\ifdefined\singcol 
    \documentclass[12pt,letterpaper,final,peerreviewca]{IEEEtran}
\else
    \documentclass[journal,final]{IEEEtran}
\fi

\usepackage{cite}
\usepackage{graphicx,color,epsfig,rotating}
\usepackage{amsfonts,amsmath,amsthm,amssymb,bbm,dsfont,stfloats}
\usepackage{hyperref}
\usepackage{xcolor}\hypersetup{linkbordercolor=green}
\usepackage{algorithm,algorithmic}
\usepackage{subfigure,enumerate}

\newtheorem{thm}{Theorem}

\newtheorem{lem}{Lemma}
\newtheorem{corol}{Corollary}

\newtheorem{assum}{Assumption}
\newtheorem{rem}{Remark}

\newcounter{storeeqcounter}
\newcounter{tempeqcounter}

\long\def\comment#1{}

\newfont{\bbb}{msbm10 scaled 700}

\newfont{\bbc}{msbm10 scaled 1100}

\newcommand{\nv}{{\pmb n}}

\newcommand{\sv}{{\pmb s}}

\newcommand{\vv}{{\pmb v}}

\newcommand{\Nm}{{\pmb N}}

\newcommand{\Cc}{{\cal C}}

\newcommand{\Nc}{{\cal N}}
\newcommand{\Oc}{{\cal O}}

\newcommand{\Rc}{{\cal R}}

\newcommand{\Lambdam}{\hbox{\boldmath$\Lambda$}}

\newcommand{\Phim}{\hbox{\boldmath$\Phi$}}

\newcommand{\Upsilonm}{\hbox{\boldmath$\Upsilon$}}
\newcommand{\Psim}{\hbox{\boldmath$\Psi$}}
\newcommand{\Thetam}{\hbox{\boldmath$\Theta$}}
\newcommand{\Omegam}{\hbox{\boldmath$\Omega$}}
\newcommand{\Xim}{\hbox{\boldmath$\Xi$}}

\newcommand{\trace}{{\hbox{tr}}}

\newcommand{\var}{{\hbox{var}}}

\renewcommand{\Re}{{\rm Re}}

\renewcommand{\vec}{{\rm vec}}

\def\pa{{\pmb a}}\def\pe{{\pmb e}}\def\pg{{\pmb g}}\def\ph{{\pmb h}}\def\pn{{\pmb n}}\def\pp{{\pmb p}}\def\pu{{\pmb u}}\def\pv{{\pmb v}}\def\pw{{\pmb w}}\def\px{{\pmb x}}\def\py{{\pmb y}}\def\pz{{\pmb z}}

\def\pA{{\pmb A}}\def\pB{{\pmb B}}\def\pC{{\pmb C}}\def\pD{{\pmb D}}\def\pF{{\pmb F}}\def\pG{{\pmb G}}\def\pI{{\pmb I}}\def\pJ{{\pmb J}}\def\pQ{{\pmb Q}}\def\pR{{\pmb R}}\def\pT{{\pmb T}}\def\pU{{\pmb U}}\def\pV{{\pmb V}}\def\pX{{\pmb X}}\def\pZ{{\pmb Z}}\def\p0{{\pmb 0}}

\makeatletter
\newcommand{\vast}{\bBigg@{3.5}}
\newcommand{\Vast}{\bBigg@{5}}
\makeatother

\def\ct{\mathsf{H}}
                         
\def\snr{{\small\textsf{SNR}}}

\def\mmse{{\small\textsf{mmse}}}

\def\Eb{{\mathbb{E}}}

\IEEEoverridecommandlockouts

\begin{document}

\title{Capacity Scaling of Massive MIMO in Strong Spatial Correlation Regimes}

\author{Junyoung Nam,~\IEEEmembership{Member,~IEEE}, Giuseppe Caire,~\IEEEmembership{Fellow,~IEEE}, M{\'e}rouane Debbah,~\IEEEmembership{Fellow,~IEEE}, \\ and H. Vincent Poor,~\IEEEmembership{Fellow,~IEEE}
\thanks{ The material in this paper was presented in part  at the IEEE International Conference on Communications (ICC),  Shanghai, China, May 2019.}
\thanks{J. Nam is with Qualcomm, San Jose, CA 95110 (e-mail: nam@ qti.qualcomm.com). } 
\thanks{G. Caire is with the Communication Systems Department, Faculty IV, Technical University of Berlin, Germany (e-mail: caire@tu-berlin.de).}
\thanks{M. Debbah is with the Large Networks and Systems Group, CentraleSup{\'e}lec, Gif-sur-Yvette, France (e-mail: merouane.debbah@centralesupelec. fr).}
\thanks{H. V. Poor is with the Department of Electrical Engineering, Princeton University, NJ (e-mail: poor@princeton.edu).}

\thanks{Copyright (c) 2019 IEEE. Personal use of this material is permitted.  However, permission to use this material for any other purposes must be obtained from the IEEE by sending a request to pubs-permissions@ieee.org.}
}

\maketitle 

\begin{abstract}

This paper investigates the capacity scaling of multicell massive MIMO systems in the presence of spatially correlated fading. In particular, we focus on the strong spatial correlation regimes where the covariance matrix of each user channel vector has a rank that scales sublinearly with the number of base station antennas, as the latter grows to infinity. We also consider the case where the covariance eigenvectors corresponding to the non-zero eigenvalues span randomly selected subspaces. For this channel model, referred to as the ``random sparse angular support'' model, we characterize the asymptotic capacity scaling law in the limit of large number of antennas. To achieve the asymptotic capacity results, \emph{statistical spatial despreading} based on the second-order channel statistics plays a pivotal role in terms of pilot decontamination and interference suppression. A remarkable result is that even when the number of users scales linearly with base station antennas, a linear growth of the capacity with respect to the number of antennas is achievable under the sparse angular support model. We also note that the achievable rate lower bound based on massive MIMO ``channel hardening'', widely used in the massive MIMO literature, yields rather loose results in the strong spatial correlation regimes and may significantly underestimate the achievable rate of massive MIMO. This work therefore considers an alternative bounding technique which is better suited to the strong correlation regimes. In fading channels with sparse angular support, it is further shown that spatial despreading (spreading) in uplink (downlink) has a more prominent impact on the performance of massive MIMO than channel hardening.

\end{abstract}


\section{Introduction}
\label{sec:intro}

Achieving ever higher spectral efficiency has always been a central problem in wireless networks. To this end, massive multiple-input multiple-output (MIMO) \cite{Mar10,MLY16}, also referred to as large-scale MIMO, is a viable technology that avoids centralized processing of multiple base station (BS) sites and yet provides unprecedented spectral efficiency, provided that every BS has a sufficiently large-scale antenna array and that uplink/downlink channel reciprocity holds despite hardware impairments. 
In order to accurately predict the performance of multicell massive MIMO, it will be important to investigate the asymptotic sum capacity in the limit of a large number of antennas, when not only channel training cost, channel uncertainty, and out-of-cell interference but spatial correlation is also taken into account. This work considers particular regimes where the covariance matrix of each user channel has a rank which scales sublinearly with the number of BS antennas, $M$.

For the large-scale antenna array regime, in which typically the number of antennas $M$ per cell is larger than the number of all active users in the homogeneous $L$-cell network with $K$ users per cell and finite $L$, we can characterize the achievable rate of massive MIMO in several ways. For a simple isotropic channel model with pilot sequence reuse factor 1, where every cell shares the same set of pilot sequences, the spectral efficiency of typical massive MIMO networks can be written by following the line of arguments in \cite{Zhe02,Mar10,Hoy13} as 
\begin{align} \label{eq:intro-2}
    \Rc_M = \bigg(1-\frac{\kappa_1}{T_c}\bigg)\kappa_1L\log \bigg(1+\frac{1}{\iota(L-1)}\bigg)  +o(1)
\end{align}
where $\Rc_M$ is the asymptotic\footnote{In fact, some dependency on $M$ is captured in the $o(1)$ factor, which vanishes in the large $M$ limit.} achievable rate as $M\rightarrow\infty$, $T_c$ is the number of channel uses of a time-frequency coherence block, $\kappa_1=\min\{M,K,\lfloor T_c/2\rfloor\}$, $\iota \in (0,1]$ is a symmetric intercell interference factor, and $o(1)$ goes to zero as $M$ grows without bound. This result clearly shows performance limits of massive MIMO due to the pilot contamination effect coming from the reuse of pilot sequences among different cells. More specifically, the multiplexing gain (or spatial degrees of freedom) defined by $\lim_{\snr\rightarrow\infty}\frac{\Cc(\snr)}{\log\snr}$, where $\Cc(\snr)$ is the capacity of a channel with $\snr$ being equivalent to the uplink/downlink sum power per cell in this work, vanishes and the power (beamforming) gain is saturated, no matter how large $M$ is. 
In order to avoid the pilot contamination problem, one may utilize the {globally orthogonal} pilot sequences across the entire network. As a straightforward consequence of \cite{Zhe02,Mar10,Huh12}, the corresponding massive MIMO network has then the capacity scaling law 
\begin{align} \label{eq:intro-3}
   \Cc_M = \bigg(1-\frac{\kappa_2}{T_c}\bigg)\kappa_2\log \left(\snr \;\frac{M}{K}\right) +o(1)
\end{align}
where $ \Cc_M$ is the asymptotic capacity in the limit of $M$, $\kappa_2=\min\{M,KL,\lfloor {T_c}/{2}\rfloor\}$, and $o(1)$ follows from the fact that every interference asymptotically vanishes in the absence of pilot contamination. 
For finite $T_c\le 2KL$,\footnote{
In the block-fading channel model, we presume a single channel use per coherence block for channel training. For a desirable channel estimation performance with noisy observations of realistic (non block-fading) channels, it is common to design an ``oversampled" allocation of pilot symbols across time and frequency grid by a factor of $O = 2$ or even more, where $O$ is an oversampling factor per time/frequency domain (e.g, \cite{Kal03,Faz03}). In this case, the pre-log factor becomes $(1-{\kappa_2'O^2/T_c})\kappa_2'$ with $\kappa_2'=\min\{KL,\lfloor \frac{T_c}{2O^2}\rfloor\}$, meaning that the condition $T_c\le 2KL$ should be replaced with $T_c\le 2KLO^2$.}  
however, the above multiplexing gain of the $L$-cell network is approximately limited by $\frac{T_c}{4}$ even if both $M$ and $K$ grow without bound, yielding that every cell only gets multiplexing gain of $\frac{T_c}{4L}$. 
Therefore, for fixed $T_c$ with large $KL$, the channel training cost immediately turns out to be a critical limiting factor that undermines the performance gain of MIMO networks \cite{Nam17}. Recently, allocating up to $\lfloor {T_c}/{2}\rfloor$ channel uses to pilots was restated by \cite{Bjo16} in the context of massive MIMO along with an analysis of the optimal number of scheduled users in terms of spectral efficiency. In fact, the scaling law of \eqref{eq:intro-3} serves as a lower bound on asymptotic achievable rate for the globally orthogonal pilot scheme, regardless of channel statistics, whereas it becomes tight when user channels are isotropic. This implies that for different and more realistic channel models, both \eqref{eq:intro-2} and \eqref{eq:intro-3} are not necessarily tight. { Meanwhile, other capacity scaling results can also be found that investigate non-coherent single-input multiple-output (SIMO) multiple access channel  \cite{CMG16} and single-user multiple-input single-output (MISO) channel with feedback of previous channel outputs \cite{Che18}.}

Since the use of orthogonal pilot sequences across the entire network is too constraining in terms of the pilot dimensionality overhead (see \eqref{eq:intro-3}) and pilot contamination without a further countermeasure yields an interference limited system (see \eqref{eq:intro-2}), several techniques to tackle the pilot contamination problem have been proposed in the literature. 
For instance, \cite{Ash12} proposed multicell cooperative precoding/combining over the entire network, and blind pilot decontamination was given by \cite{Mul14} to separate signal and interference subspaces into disjoint supports. Following \cite{Yin13, Adh13}, many pilot decontamination techniques have exploited the linear independence between the subspaces spanned by the eigenvectors of the rank-deficient channel covariance matrices of users so that one can find some useful structure of subspaces with orthogoanl (non-overlapping) angular supports. A joint angle and delay domain based pilot decontamination was also developed in \cite{YCG16,Hag17}.  In a different line of work, \cite{BHS18} recently proved that the linear independence of those subspaces is rather surprisingly not a necessary condition for the elimination of contamination with infinitely many $M$ antennas. The more general sufficient condition therein is an \emph{asymptotic} linear independence of the covariance matrices themselves other than that of their subspaces. This leads to the linear independence of all user channels that is then utilized to eliminate pilot contamination through multicell non-cooperative precoding/combining, requiring each cell to estimate all user channels in the multicell network. 


\ifdefined\singcol 
\else
\begin{figure*}
  \vspace{-3mm}
  \normalsize
 \setcounter{storeeqcounter}{\value{equation}}

\begin{align} \label{eq:intro-4}
   \Cc_M=\left\{ \begin{array}{ll}
  \big(1-\frac{1}{T_c}\big)\min\{M,K\}L\log\left(\snr \frac{M}{K}\right) +o(1) & \ \text{under strong spatial correlation} \\
  \big(1-\frac{1}{T_c}\big)KL\log\left(\snr \frac{r}{K}\right) +o(1)  & \ \text{under very strong spatial correlation}
  \end{array} \right.
\end{align}

 \addtocounter{storeeqcounter}{1} 
 \setcounter{equation}{\value{storeeqcounter}} 
  \hrulefill
\end{figure*}
\fi

Rather than any explicit pilot decontamination technique, this work focuses on the capacity scaling law in the massive MIMO network when the randomness and the sparsity of angular supports of channel covariance matrices are taken into consideration. 
Different scattering geometries of users located arbitrarily in a network make the angular support of each user channel random. This randomness also captures the fact that due to common scatterers, the angular supports of user channels are often partially overlapped. The sparsity of the angular support means that the number of significant multipaths in angular domain is much smaller than $M$.  
This arises in scenarios of limited scattering geometry, which has been observed in several channel measurement campaigns in { not only millimeter wave (mm-Wave) but also} below 6 GHz bands (see \cite{Gao15,JCZ17} and references therein), where the number of non-negligible angular components of the user channel is fairly smaller than $M$, although the covariance matrix is mathematically of full rank. In this paper, we consider particularly two strong spatial correlation regimes, in which the rank $r$ of channel covariance matrices grows sublinearly with $M$. For such correlated fading channels in the homogeneous $L$-cell network, where every BS has sufficiently large $M$ antennas and serves $K$ users with common signal-to-noise ratio (SNR) and with the same coherence block size $T_c$, we show by some extensions of the method of deterministic equivalents \cite{Bai98,Cou11a,Cou11b,Wag12} that the ergodic sum capacity behaves as  
\ifdefined\singcol 
\begin{align} \label{eq:intro-4}
   \Cc_M=\left\{ \begin{array}{ll}
  \big(1-\frac{1}{T_c}\big)\min\{M,K\}L\log\left(\snr \frac{M}{K}\right) +o(1) &  \text{under strong spatial correlation} \\
  \big(1-\frac{1}{T_c}\big)KL\log\left(\snr \frac{r}{K}\right) +o(1)  &  \text{under very strong spatial correlation}
  \end{array} \right.
\end{align}
\else
\eqref{eq:intro-4}, shown on the top of page~\pageref{eq:intro-4},  
\fi
where we used (intra-cell) non-orthogonal pilot that consumes only a single channel use per coherence block across the network as an extreme case and also assumed $\lim_{M\rightarrow\infty}\frac{K}{M}= 0$ for the former regime and $\limsup_{M\rightarrow\infty}\frac{K}{M}<\infty$ for the latter.  
Note that the above scaling law is asymptotically tight and its prelog factor is indeed the best one can ever expect through a cut-set upper bound from the perspective of either pilot-aided or non-coherent  communication with a single antenna in block fading \cite{Zhe02}, whose prelog factor is $(1-{T_c}^{-1})$.

The main differences of \eqref{eq:intro-4} and prior work can be summarized as follows:
\begin{enumerate}
\item \emph{Capacity scaling characterization}: To the best of our knowledge, it is not clear in the prior work how the sum rate scales with respect to any of $M,K,T_c$, and SNR, except for the simple case of orthogonal pilots across the whole network as in \eqref{eq:intro-3}. In particular, as mentioned earlier, the multiplexing gain of the massive MIMO network is limited by $T_c/4$ when $T_c\le 2KL$ and globally orthogonal pilots are used, meaning that $K\le \frac{T_c}{2L}$ users per cell can only be served. Otherwise, pilot contamination eventually suffocates multiplexing gain as in \eqref{eq:intro-2}. However, the capacity scaling in \eqref{eq:intro-4} is not dominated by $T_c$ any more under strong spatial correlation regimes. 

\item \emph{Scaling of $K$ with respect to $M$}: Past work has generally assumed $\lim_{M\rightarrow\infty}\frac{K}{M}= 0$, i.e., $K$ is finite or at most grows slower than $M$. For instance, if $KL>M$, the linear independence of the subspaces of covariance matrices in \cite{Yin13, Adh13} is never attainable. Likewise, if $\liminf_{M\rightarrow\infty}\frac{K}{M}>0$, then the asymptotic linear independence of covariance matrices in \cite{BHS18} does not hold any longer.  
Therefore, the results based on the linear independence of the signal subspaces or of the covariance matrices cannot capture the capacity scaling with respect to the ratio $\frac{K}{M}>0$, as both $M$ and $K$ grow large.  This poses a fundamental methodological question since in reality we are in the presence of a finite system with given $M$ and $K$. Which of the following large system analyses will produce the more meaningful prediction of its behavior: considering finite $K$ and letting $M \rightarrow \infty$ \cite{Mar10,MLY16,BHS18} or considering both $K$ and $M \rightarrow \infty$ with fixed ratio equal to the actual ratio $\frac{K}{M}$ of the practical finite system \cite{Hoy13}? We claim that the latter methodology yields a more relevant asymptotics.

\item \emph{Scaling of $T_c$ with respect to $M$}: The conventional large system analysis including \cite{Hoy13} implicitly assumes that the coherence block $T_c$ of the channel scales linearly with $K$ to accommodate the intra-cell orthogonal pilot sequences consuming $K$ channel uses.  Moreover, the semi-blind decontamination technique \cite{Mul14} can completely remove pilot contamination, provided that $T_c$ goes to infinity as well as $M$. However, the channel coherence block depends on Doppler spread due to mobility of users and on frequency selectivity due to delay spread of multipath channels, irrespectively  of $M$, and hence it is finite in practice. Meanwhile, our large system limits hold in finite $T_c$ since the use of non-orthogonal pilot turns out to be feasible in strong spatial correlation regimes.

\item \emph{Unlimited capacity as $M$ and $K$ grow with $\frac{K}{M}>0$ and finite $T_c$}: The second capacity characterization in \eqref{eq:intro-4} shows that every user can get unlimited spectral efficiency in the very strong spatial correlation regime with finite $T_c$, as long as $K$ grows no faster than $M$ and the uplink/downlink per-user transmit power $\frac{\snr}{K}$ is not vanishing. 
It turns out that spatial despreading in uplink (or spreading in downlink),  which acts as statistical spatial filters matched to the angular supports of user channel covariances, makes pilot contamination terms fade away and that an infinite sum of the vanishing contamination terms still vanishes. Moreover, overlapped angular components between any pair of the channel covariance matrices of users only occupy a vanishing portion of the total signal space dimension if their angular supports become sufficiently sparse as $M\rightarrow\infty$.

\item \emph{Beamforming architecture and complexity}: Spatial despreading (spreading) is followed by low-dimensional (i.e., $r$-dimensional) channel estimation and single-cell combining in uplink (precoding in downlink), which is asymptotically optimal in the strong correlation regimes even if each BS only estimates the low-dimensional effective channels of its own users. We will use the low-dimensional processing to prove the achievability of the large system limits in \eqref{eq:intro-4}. The low-dimensional minimum mean-squared error (MMSE) channel estimation and combining/precoding show a comparable performance to the conventional $M$-dimensional MMSE processing at finite $M$. Furthermore, spatial despreading/spreading is in fact long-term or wideband (i.e., frequency non-selective) processing, while low-dimensional combining/precoding is short-term or narrowband (frequency selective) processing. Therefore, they are amenable to the hybrid beamforming architecture (see \cite{Mol16} and references therein) that is beneficial to realize massive MIMO in practice by saving hardware cost and channel training/feedback overhead.


\end{enumerate}

{ In order to provide some intuition behind the main results, let us explain why pilot contamination vanishes even though we do not rely on any explicit decontamination technique and only use a finite pilot dimension (e.g., a single symbol) in finite coherence block $T_c$, such that there is even pilot contamination for the users in the same cell.  From a technical perspective, as long as the number of users $K$ and the rank of the channel covariance $r$ grow sublinearly with (i.e., slower than) the number of BS antennas $M$, overlapped angular support of channel covariances of different users can be managed as only a subset of Lebesgue measure zero in positive reals $\mathbb{R}^+$ under appropriate conditions, as $M\rightarrow\infty$, which leads to the almost sure convergence of pilot contamination to zero. It is mainly due to the effect of {spatial (de)spreading}, which is implicitly conducted by channel propagation itself and suppresses the contamination by a factor of $\frac{M}{r}$, that the user channels come to be ``asymptotically orthogonal'' to each other. Therefore, the effect of pilot contamination is indeed not a fundamental limiting factor for the user channels with random sparse angular support. This also holds true in the very strong correlation regime, in which $K$ grows linearly with $M$, as long as $r$ grows rather more slowly than $M$ and the received channel energy also grows slower than $M$.}

Another contribution of this work is an investigation into lower-bounding techniques in terms of massive MIMO with random sparse angular support. We begin with the fact that the well-known channel hardening effect in massive MIMO can be undermined by spatial correlation at finite $M$. In particular, the useful signal coefficient does not concentrate on its mean when the fading channels are highly correlated. As a consequence, the widely used non-coherent bound \cite{MLY16} in massive MIMO downlink, which works very well in the case of channel hardening, may significantly underestimate the achievable rate, depending on spatial correlation. Furthermore, for finite $M$, the coherent bounding technique \cite{MLY16} also widely used in uplink may suffer from channel estimation error due to imperfect CSI at the receiver (CSIR) especially when non-orthogonal pilot is employed. In this work, we further consider a lesser-known alternative non-coherent bounding technique given in \cite{Cai17} to better estimate the performance of massive MIMO in the strong spatial correlation regimes. 

The paper is organized as follows: In Sec. \ref{sec:Pre}, we describe the system model with two stochastic spatial correlation models and explain spatial (de)spreading. Sec. \ref{sec:MR} addresses our main results on the capacity scaling of massive MIMO along with their implication. In Sec. \ref{sec:ED}, we present an extension of the deterministic equivalents technique.  Sec. \ref{sec:PF} provides the proofs of the main results in Sec. \ref{sec:MR}. Sec. \ref{sec:NR} contains some numerical results.  We conclude this work in Sec. \ref{sec:Con}.   

{ \emph{Notation:} We use $\overset{a.s.}{\longrightarrow} $ for the almost sure convergence such that, for sequences $a_n$ and $b_n$, $a_n - b_n \overset{a.s.}{\longrightarrow} 0$ as $n\rightarrow\infty$,  and $\simeq$ is also used for the sake of compactness. Let $\limsup_{n\rightarrow\infty} a_n$ and $\liminf_{n\rightarrow\infty} a_n$ denote the limit superior and the limit inferior of $a_n$, respectively. For a matrix $\pA$,  $\|\pA\|_2$ and $\trace\pA$ denote the spectral norm and the trace of $\pA$, respectively. For a vector $\pa$, $\|\pa\|$ denotes the $\ell_2$ norm of $\pa$. $\mathcal{CN}(0,1)$ denotes the zero-mean circularly symmetric complex Gaussian distribution.}

\section{Preliminaries }
\label{sec:Pre}

\subsection{System Model}
\label{sec:SM}

We consider an $L$-cell time-division duplex (TDD) MIMO network where the $\ell$-th BS has $M$ antennas and serves $K_\ell$ single-antenna users. The user channel follows the frequency-flat block-fading model for which it remains constant during the coherence block of $T_c$ but changes independently every interval, where $T_c$ is the number of channel uses or the signal dimension in the time-frequency domain. The fading distribution is known at both the transmitters and the receivers. We do not require any cooperation between multiple BSs such that there is neither channel state information (CSI), data, scheduling information, nor pilot allocation information coordinated across the network through backhaul links. 
Indexing the $k$th user in BS $\ell$ by $\ell_k$, $\ph_{\ell\ell'_k}$ is the uplink channel from user $\ell'_k$ to BS $\ell$, and $\ph_{\ell'\!\ell_k}^\ct$ is the downlink channel from BS $\ell'$ to user $\ell_k$. Using the Karhunen-Lo\`eve transform, the channel vector $\ph_{\ell\ell'_k}$ can be expressed as  
\begin{align} \label{eq:SM-2}
   \ph_{\ell\ell'_k}  =\pU_{\ell\ell'_k}\Lambdam_{\ell\ell'_k} \textsf{h}_{\ell\ell'_k}
\end{align}
where $\Lambdam_{\ell\ell'_k} \in \mathbb{C}^{r_{\ell\ell'_k}\times r_{\ell\ell'_k}}$ is the diagonal matrix whose elements are the non-zero eigenvalues of the channel covariance matrix $\pR_{\ell\ell'_k}$, $\pU_{\ell\ell'_k} \in \mathbb{C}^{M\times r_{\ell\ell'_k}}$ is the eigenvector matrix of $\pR_{\ell\ell'_k}$, and $\textsf{h}_{\ell\ell'_k} \in \mathbb{C}^{r_{\ell\ell'_k} } \sim\mathcal{CN} (\p0, \pI)$ is the small-scale channel component. 
Throughout this work, it is assumed that the rank $r_{\ell\ell'_k}$ of $\pR_{\ell\ell'_k}$ is assumed to be much smaller than $M$ due to limited scattering environments. Furthermore, BS $\ell$ has a prior knowledge on the low-rank covariance matrices $\pR_{\ell\ell_k}$ of its own users and a sum of low-dimensional covariance matrices of the other-cell user channels projected by $\pU_{\ell\ell'_k}$, depending on the channel training schemes in Subsection \ref{sec:SC}.


In the MIMO uplink, the received signal vector at BS $\ell$ can be given by
\begin{align} \label{eq:SM-1}
   \py_\ell=\sum_{k}\ph_{\ell\ell_k}x_{\ell_k} +\sum_{\ell'\neq\ell}\sum_{k}\ph_{\ell\ell'_k}x_{\ell'_k} +\pz_\ell 
\end{align}
where $x_{\ell_k}$ is the input signal of user $\ell_k$ chosen from a Gaussian codebook and satisfies the equal power constraint such that $\mathbb{E}[|x_{\ell_k}|^2 ]\le P_\text{ul}$, and $\pz_\ell  \sim\mathcal{CN} (\p0, \pI)$ is the Gaussian noise at the BS antennas. Since the noise power per antenna is normalized to be unity, ${P_\text{ul}}$ can be regarded as the transmit SNR per user in uplink. 
Meanwhile, the received signal vector at  user $\ell_k$ at BS $\ell$ in the downlink can be given by
\begin{align} \label{eq:SM-4}
   y_{\ell_k}= \ph_{\ell\ell_k}^\ct &\pp_{\ell_k}d_{\ell_k}+\sum_{k'\neq k}\ph_{\ell\ell_k}^\ct \pp_{\ell_{k'}}d_{\ell_{k'}} \nonumber \\ &+\sum_{\ell'\neq\ell}\sum_{k'}\ph_{\ell'\!\ell_k}^\ct \pp_{{\ell'}\!\!_{k'}}d_{{\ell'}\!\!_{k'}} +z_{\ell_k} 
\end{align}
where $\pp_{\ell_k}$ and $d_{\ell_k}$ are the precoding vector and the input signal of user $\ell_k$ satisfying $\mathbb{E}[\|\pp_{\ell_k}d_{\ell_k}\|^2]\le P_{\text{dl},\ell}$, respectively, and $z_{\ell_k}  \sim\mathcal{CN} (0, 1)$ is the Gaussian noise. 
We assumed the equal power allocation within a cell such that $P_{\text{dl},\ell}=\frac{P_\text{dl}}{K_\ell}$, where ${P_\text{dl}}$ is the sum-power constraint per cell, and hence ${P_\text{dl}}$ becomes the normalized transmit SNR in line with the classical MIMO downlink, which is denoted by $\snr$ in this paper.

\subsection{Stochastic Spatial Correlation Models}
\label{sec:CM}


The channel covariance matrix is determined by the propagation geometry, and in particular by the distribution of the signal power over the angle domain (angular scattering function). It is well known that the propagation geometry changes on a time-scale much larger than the coherence time of the small-scale fading. This means that the channel can be considered as ``locally wide-sense stationary (WSS)''. In other words, on a relatively large-scale time span, which we call \emph{local stationarity interval} in this paper, it is seen as a snapshot taken from a WSS process with given second-order statistics (e.g., \cite{Rap02,Cho17}).  In this paper, we assume that on each such local stationarity interval, the channel covariance matrices are drawn at random from a distribution with 
$\pR_{\ell\ell'_k}=\pU_{\ell\ell'_k}\Lambdam_{\ell\ell'_k}\pU_{\ell\ell'_k}^\ct$. While the columns of $\pU_{\ell\ell'_k}$ span the large-scale angular domain subspace of multipath components, $\trace\Lambdam_{\ell\ell'_k}$ captures the large-scale fading factors such as path loss and shadow fading, i.e., the channel energy that BS $\ell$ receives from user ${\ell'_k}$. 
In the following, we introduce two stochastic models for the eigenvector matrix $\pU_{\ell\ell'_k}$ to capture the randomness of spatial correlation, which typically arises in wireless channel propagation due to arbitrary user and scattering geometry. 

{

\subsubsection{Random partial unitary model} 
A simple model for $\pU_{\ell\ell'_k}$  is a random partial unitary matrix. In this model,  $\pU_{\ell\ell'_k}$ is independently and uniformly drawn from a random partial unitary matrix whose column space is in the Grassmann manifold $\mathcal{G}(M,r_{\ell\ell'_k})$, which is the set of all $r_{\ell\ell'_k}$-dimensional subspaces in $\mathbb{C}^{M}$. Hence, the random partial unitary matrices are mutually independent for all $(\ell, \ell',k)$. 
The $i$th column of $\pU$ (denoted by $\pu_i$) is a random unit vector uniformly distributed on the $M$-dimensional complex unit sphere such that $\pu_i = \frac{\px}{\|\px\|_2}$ with covariance $\frac{1}{M}\pI_M$, where $\px \sim\mathcal{CN}(\p0,\pI_M)$. 



\subsubsection{Random partial Fourier model}
Another interesting spatial correlation model is a random partial (subsampled) Fourier matrix, motivated by the typical uniform linear array (ULA) in multiple antenna systems. Let $F_{jk}$ denote the $(j,k)$th entry of the discrete Fourier transform (DFT) matrix $\pF\in \mathbb{C}^{M\times M}$, as shown by
$$F_{jk}=\frac{1}{\sqrt{M}} e^{\jmath 2\pi jk/M}, \ j, k = 0,\ldots,M-1.$$
Suppose that $\pU_{\ell\ell'_k}$ is composed of $r_{\ell\ell'_k}$ column vectors uniformly drawn at random \emph{without replacement} from the Fourier basis functions of $\pF$ so that different users can have common basis elements, taking into account common scatterers shared by multiple users. 
The resulting unitary matrix can be represented by 
\begin{align} \label{eq:CM-1}
   \pU_{\ell\ell'_k} = \pF\pG_{\ell\ell'_k}
\end{align}
where $\pG_{\ell\ell'_k}\in\mathbb{C}^{M\times {r_{\ell\ell'_k}}}$ is the random selection matrix that chooses ${r_{\ell\ell'_k}}$ columns (angular components) {without replacement} from $M$ columns of $\pF$. 
This correlation model is a reminiscence of the angular domain representation of MIMO channels \cite{Tse05} or virtual channel representation \cite{Say02} widely-used in the literature. In particular, the random partial Fourier model can be justified by the known asymptotic behavior of channel covariance in massive MIMO with ULA \cite{Adh13}, showing that the eigenvectors of channel covariance matrices are well approximated by the columns of DFT for large $M$. 
Notice that the random partial Fourier model has much less degrees of freedom (i.e., highly structured or less randomness) than the random partial unitary model. In particular, the columns of $\pU_{\ell\ell'_k}$ in \eqref{eq:CM-1} are not statistically independent due to sampling without replacement, even if they are linearly independent.


It is important to notice that given the above random realizations of the channel covariance matrices of users, the resulting ergodic achievable rates are conditional to such realizations. Therefore they are random variables unlike most sum-rate analyses in the massive MIMO literature. One might then be interested in their distribution, in particular, 
the outage probability defined by the  distribution function of such ergodic rates conditioned on the covariances. In order to characterize the ergodic capacity scaling in this work rather than the outage capacity scaling, we make the assumption that the long-term channel energy captured by the large-scale fading factor $\trace\Lambdam_{\ell\ell'_k}$ does not change over user mobility and different scattering geometries for a given value of $M$ such that 
\begin{align} \label{eq:CM-1b}
  \trace\Lambdam_{\ell\ell'_k}= c_{\ell\ell'_k}(M), \ \forall  (\ell,\ell',k)
\end{align}
where $c_{\ell\ell'_k}(M)$ is a positive real constant over {certain local stationarity intervals}\footnote{ For an  illustrative example, 5G new radio (NR) mm-Wave systems may have OFDM symbol duration of 4.5 $\mu$sec with 240 KHz subcarrier spacing \cite{NR_TS}. In this case, $10^4$ symbols only span 45 msec, which corresponds to a local stationarity interval over which channel second order statistics does not change. With a few hundred such intervals, a pedestrian only travels 10 m at a walking speed of 1 m/sec. Therefore it may be reasonable to assume that the long-term channel energy captured by $\trace\Lambdam_{\ell\ell'_k}$ remains unchanged over such intervals since the amount of intervals does not substantially change path loss and shadowing.}. Under this deterministic $\trace\Lambdam_{\ell\ell'_k}$,  we will show later in Sec. \ref{sec:IMR} that the conditional ergodic rates converge to a deterministic limit for sufficiently large $M$, meaning that the limit does not depend any longer on a specific covariance realization but only on the distribution. As a consequence, we will focus on such deterministic limits in our asymptotic sum-rate analysis.

Our spatial correlation models that capture the random (or diverse) nature of angular components in user channels allow different angular supports among users as in \cite{Adh13,Yin13}.  
We notice here that in some other works, e.g., \cite{Hoy13,Ngo13}, antenna correlation was modeled by letting all users to have the same covariance matrix. Such model is physically less justifiable, since it implies that all users share the same multipath components with the  same strength, i.e., the users are all co-located.

Given the stochastic models on $\pU_{\ell\ell'_k}$, we next make several assumptions on the  eigenvalue matrices $\Lambdam_{\ell\ell'_k}$ with ${\trace\Lambdam_{\ell\ell'_k}}$ deterministic for the large system analysis in the random covariance matrices $\pR_{\ell\ell'_k}$. 

\begin{assum} \label{as-2} \normalfont
 For all $(\ell,\ell',k)$
\begin{subequations}
  \begin{IEEEeqnarray}{c}
   \limsup_{M\rightarrow \infty}\frac{r_{\ell\ell'_k}}{M}\;\|\Lambdam_{\ell\ell'_k}\|_2 < \infty  \label{eq:SM-3c}  \\
     \limsup_{M\rightarrow \infty} \frac{\trace\Lambdam_{\ell\ell'_k}}{M}<\infty . \label{eq:SM-3d}    
\end{IEEEeqnarray}
\end{subequations}
\end{assum}


\begin{assum}[\emph{Strong Spatial Correlation Regime}] \label{as-3} \normalfont
    The number $r_{\ell\ell'_k}$ of non-zero eigenvalues of $\pR_{\ell\ell'_k}$ grows without bound but slower than $M$ such that
\begin{subequations}
  \begin{IEEEeqnarray}{c}
     \frac{r_{\ell\ell'_k}}{M} \triangleq \alpha_{\ell\ell'_k} \xrightarrow[M\rightarrow\infty]{} 0     \label{eq:SM-32a} \\
      \liminf_{M\rightarrow \infty}\frac{\trace\Lambdam_{\ell\ell'_k}}{M}>0, \ \forall (\ell,\ell',k) .    \label{eq:SM-32b}
  \end{IEEEeqnarray}
\end{subequations}
\end{assum}

First of all, the condition \eqref{eq:SM-32a} implies in conjunction with \eqref{eq:SM-32b} that the spectral norm\footnote{For positive semidefinite $\pA$, it is simply the maximum eigenvalue of $\pA$} $\|\Lambdam_{\ell\ell'_k}\|_2$  is \emph{not necessarily} uniformly bounded with respect to $M$. The uniform boundedness is a necessary condition for the method of deterministic equivalents \cite{Sil95,Bai98,Bai07,Cou11b}. Hence, under those conditions the deterministic equivalents technique cannot directly apply any longer. We will extend the technique later on in \ref{sec:ED} to address this issue.


\begin{figure*} 
\center
  \includegraphics[scale=.9]{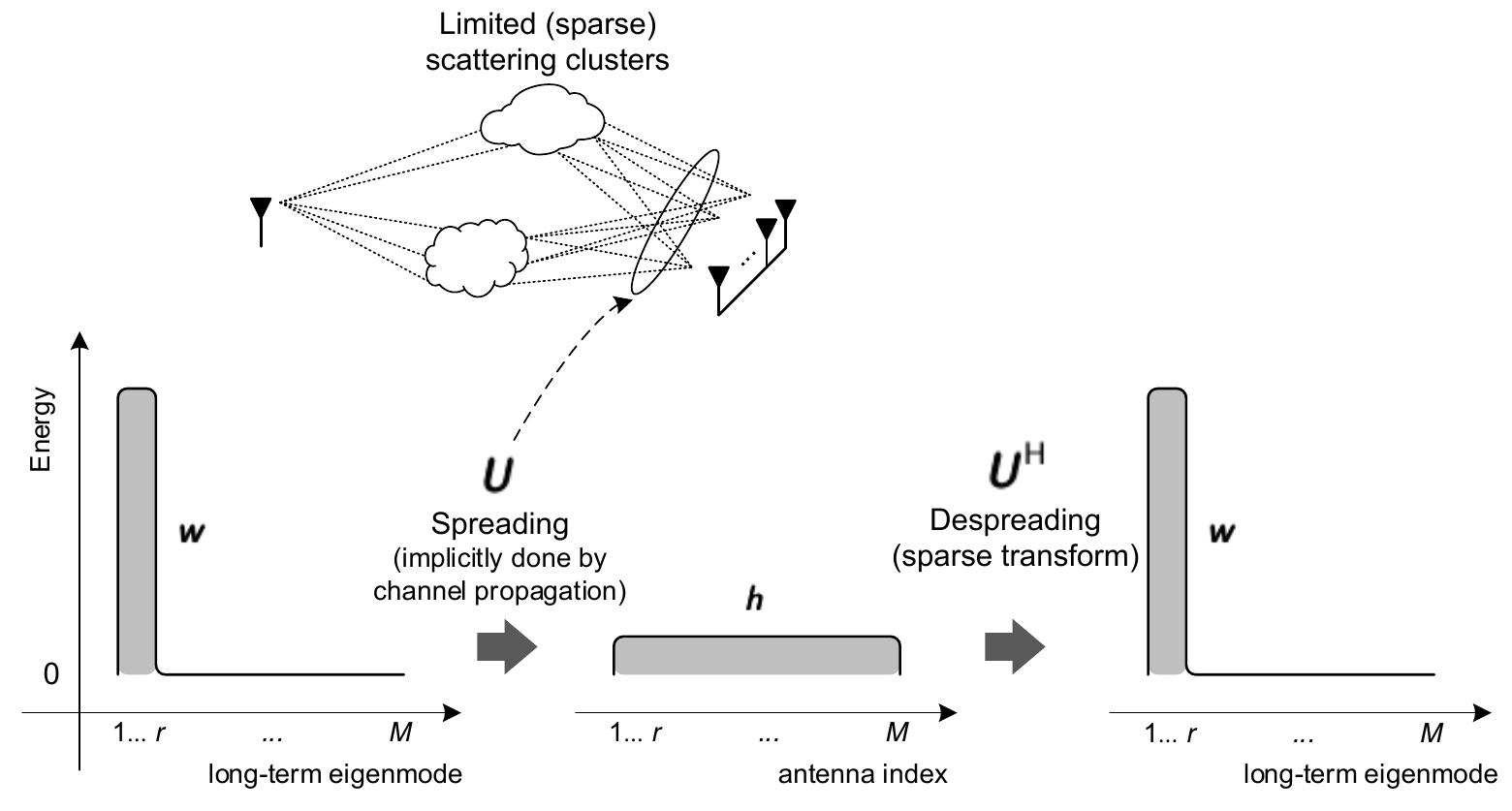}
  \caption{Change of coordinates (sparse transform of the channel vector $\ph$) in uplink.}
\vspace{-2mm} \label{fig-1}
\end{figure*} 

{
\begin{rem} \label{rem-2} \normalfont
The sublinear sparsity assumption in \eqref{eq:SM-32a} hinges on the premise that the number of angular components of each user channel grows without bound, but slower than the number of BS antennas. In the massive MIMO literature, the angular domain model is categorized into two cases: a large but finite number of angular components (e.g., \cite{Ngo13}) and infinitely many components with a constant ratio of $\alpha_{\ell\ell'_k}>0$ (e.g., \cite{Hoy13}). The former model is based on the well-known fact \cite{Tse05} that angular resolution of an array is proportional to the array aperture that should be finite in practice. In contrast, the latter can be justified for high carrier frequency systems like mm-Wave, where the angular resolution might continue to increase proportionally with the carrier frequency $f_c$. 
In the mm-Wave channels, however, it was observed (e.g., \cite{Zha10}) that the higher  $f_c$, the smaller number of multipaths that can arrive at the receivers due to higher path loss, shadowing, and blockage. This sparsity of surviving multipath components is also verified by mm-Wave propagation measurement campaigns \cite{Rap13} and also by \cite{JCZ17} even in below 6 GHz, meaning that non-negligible eigenvalues of $\pR_{\ell\ell'_k}$ may be very sparse in the large $M$ limit.\footnote{Although user terminals in mm-Wave systems have multiple or even large-scale antennas, the analog beamforming architecture turns the MIMO channel into an effective multiple-input single-output channel. Hence, our channel model with sparse angular support can cover the typical mm-Wave beamforming architecture.} 
As a matter of fact, Assumptions \ref{as-2} and \ref{as-3} model the above \emph{limited scattering} channel environments in the sense that more energy concentrates upon relatively sparse angular support of $\pR_{\ell\ell'_k}$ as $M$ increases.  
Therefore, it is particularly reasonable to assume the sublinear sparsity when the size of the antenna
array does not scale with $M$ in practice and rather carrier frequency $f_c$ scales. 
\end{rem}
}

An insufficiency of Assumption \ref{as-3} is that the large-scale fading factor $\trace\Lambdam_{\ell\ell'_k}$ grows at the same speed as $M\rightarrow\infty$, even though the rank $r_{\ell\ell'_k}$ of $\Lambdam_{\ell\ell'_k}$  grows only sublinearly with $M$. This means that despite such a sparse angular support in limited scattering clusters, the channel energy does not dissipate so as to concentrate upon the $r_{\ell\ell'_k}$-dimensional subspace spanned by the columns of $\pU_{\ell\ell'_k}$. Therefore, we also consider another spatial correlation regime, where $\trace\Lambdam_{\ell\ell'_k}$  grows no faster than $r_{\ell\ell'_k}$ as $M\rightarrow\infty$ and spatial correlation is rather stronger.

\begin{assum}[\emph{{Very Strong Spatial Correlation Regime}}] \label{as-4} \normalfont
   The number of non-zero eigenvalues $r_{\ell\ell'_k}$ grows  without bound but much more slowly than $M$ such that
\begin{subequations}
  \begin{IEEEeqnarray}{c}
     \frac{r_{\ell\ell'_k}^{4}}{M^3} \xrightarrow[M\rightarrow\infty]{} 0  \label{eq:SM-3c1} \\   
         \limsup_{M\rightarrow \infty}\frac{r_{\ell\ell'_k}^{2}}{M} <\infty  \label{eq:SM-3c2} \\
         \limsup_{M\rightarrow \infty}\frac{\trace\Lambdam_{\ell\ell'_k}}{r_{\ell\ell'_k}}<\infty , \ \forall (\ell,\ell',k).  \label{eq:SM-3c3}
  \end{IEEEeqnarray}
\end{subequations}
\end{assum}

This very strong correlation regime is of importance since it captures significant path loss, shadowing, and penetration loss in high carrier frequency bands like mm-Wave and above. Interestingly, it will be shown later on that the two  correlation regimes lead to different sum-rate scaling laws in terms of the ratio of the number of users per cell and the number of BS antennas.

\subsection{Spatial Despreading and Spreading}
\label{sec:Pre-A}

In this subsection, we introduce statistical spatial despreading in uplink (spreading in downlink) based on channel covariance matrices and explain its implication.

\begin{itemize}
  \item \emph{Sparse transform:} Under our spatial correlation models, $\ph_{\ell\ell'_k}$ has a sparse representation and 
$\pU_{\ell\ell'_k}$ serves as a sparse transformation matrix \footnote{This is also known as a sparse representation matrix and should not be confused with a sensing (or measurement) matrix in compressed sensing \cite{CRT06,Don06}.} of $\ph_{\ell\ell'_k}$ such that
\begin{align} \label{eq:PS-1}
   \pU_{\ell\ell'_k}^\ct\ph_{\ell\ell'_k}  = {\Lambdam}_{\ell\ell'_k}^{\frac{1}{2}}\textsf{h}_{\ell\ell'_k} \triangleq \pw_{\ell\ell'_k} 
\end{align}
where $\pw_{\ell\ell'_k} \in \mathbb{C}^{r_{\ell\ell'_k}}\sim \mathcal{CN} (\p0, \Lambdam_{\ell\ell'_k})$ is the projected {effective channel} vector, whose dimension is much lower than that of the original vector $\ph_{\ell\ell'_k}$ in $\mathbb{C}^{M}$.  
   \item \emph{Energy concentration:} 
In the above transformation, the energy of $\ph_{\ell\ell'_k}$ is preserved and concentrated on $\pw_{\ell\ell'_k}$ in (\ref{eq:PS-1}) since $ \trace\pR_{\ell\ell'_k}=  \trace{\Lambdam}_{\ell\ell'_k}$, implying ${\pU_{\ell\ell'_k}^{\bot}}^\ct\ph_{\ell\ell'_k}=\p0$, where ${\pU_{\ell\ell'_k}^{\bot}}$ spans the null space of $\pU_{\ell\ell'_k}$. The energy concentration is simply due to the classical Parseval theorem implying that the change of coordinates (basis) preserves inner products (energy).
\end{itemize}

The sparse transform naturally gives rise to an interesting interpretation of spatial despreading as follows.  
One can regard $\{\pU_{\ell\ell'_k}, \forall \ell,\ell',k\}$ as ``statistical spatial matched filters" or ``random spreading sequences" in the classical uplink CDMA \cite{Ver99} with asynchronous users since the sparse transform (multiplying $\ph_{\ell\ell'_k}$ by $\pU_{\ell\ell'_k}^\ct$) in uplink and its counterpart  (multiplying $\pw_{\ell\ell'_k}$ by $\pU_{\ell\ell'_k}$) in downlink are a reminiscence of \emph{despreading} and \emph{spreading}, respectively. Hence the sparse transform and its counterpart will be referred to as spatial despreading and {spreading}, respectively. Unlike CDMA, the spatial (de)spreading is not controllable at the cost of bandwidth, but it depends on the propagation channel. In particular, under the assumptions made before, the channel with limited scattering offers an unbounded spatial spreading gain (energy concentration) as $M \rightarrow \infty$ without incurring any bandwidth cost.
Fig. \ref{fig-1} illustrates how {spatial spreading} is implicitly conducted by channel propagation and how  {spatial despreading} based on the second-order channel statistics $\pU_{\ell\ell'_k}$ is done by the change of coordinates. 

For the homogeneous network where all users have the same $r_{\ell\ell'_k}=r$, the symmetric spatial spreading gain is defined as 
\begin{align} \label{eq:SM-10}
   \zeta =\frac{M}{r} .
\end{align}
Similar to the uplink CDMA with random spreading signatures, for large $M$, one can then intuitively expect that spatial despreading can suppress interference power and, more importantly, pilot contamination by the factor $\zeta$ in the homogeneous network. This effect will be addressed later on in (\ref{eq:MR-9b}) of Sec. \ref{sec:PF}.

Once spatial despreading is performed upon the received signal $\py_\ell $ in \eqref{eq:SM-1}, the transformed vector $\py_{\ell_k}$ for user $\ell_k$ is given by
\begin{align} \label{eq:SM-5}
   \py_{\ell_k} &= \pU_{\ell\ell_k}^\ct\py_\ell \nonumber \\
   &= \pw_{\ell\ell_k}x_{\ell_k} +\sum_{k'\neq k} \pw_{\ell_k\ell_{k'}}x_{\ell_{k'}} +\sum_{\ell'\neq\ell}\sum_{k'}\pw_{\ell_k{\ell'}\!\!_{k'}}x_{{\ell'}\!\!_{k'}} +\pz_{\ell_k}
\end{align}
where $\pz_{\ell_k}=\pU_{\ell\ell_k}^\ct\pz_\ell $ and 
\begin{align} \label{eq:SM-5b}
  \pw_{\ell_k{\ell'}\!\!_{k'}}=\pU_{\ell\ell_k}^\ct\pU_{\ell{\ell'}\!\!_{k'}} \pw_{\ell{\ell'}\!\!_{k'}}, \ \forall (\ell,k,\ell',k') .
\end{align}
The subsequent receiver processing applies to the transformed vector $\py_{\ell_k}$ instead of $\py_\ell$. 
Meanwhile, in order to leverage the sparsity in spatially correlated fading channels in downlink, we let the BS perform {spatial spreading} such that $\pp_{\ell_k}=\pU_{\ell\ell_k}\pg_{\ell_k}, \forall \ell, k,$ where $\pg_{\ell_k} \in \mathbb{C}^{r_{\ell\ell_k}}.$ We can then express \eqref{eq:SM-4} as 
\begin{align} \label{eq:SM-4b}
   y_{\ell_k}=\pw_{\ell\ell_k}^\ct \pg_{\ell_k}d_{\ell_k} +\sum_{(\ell',k')\neq(\ell,k)}\pw_{{\ell'}\!\!_{k'}\ell_k}^\ct \pg_{{\ell'}\!\!_{k'}}d_{{\ell'}\!\!_{k'}} +z_{\ell_k} .
\end{align}


It should be noted that spatial despreading in uplink (or spreading in downlink) by $\pU_{\ell\ell_k}$ does not incur any loss of optimality from the single-user perspective, but it is suboptimal from the perspective of MU-MIMO  because spatial despreading projects the $M$-dimensional signal space onto the $r_{\ell\ell_k}$-dimensional subspace, in which multiuser combining (or precoding) is performed based on $\{\pw_{\ell\ell_k}, \pw_{\ell_k{\ell'}\!\!_{k'}}\}$. As a matter of fact, it is only asymptotically optimal under Assumption \ref{as-3} (or \ref{as-4}), which will be shown by the main capacity scaling result in Theorem \ref{thm-3}. Nevertheless, the purpose of representing the uplink/downlink received signal in the above forms \eqref{eq:SM-5} and \eqref{eq:SM-5b}  is three-fold: 1) to explicitly show the role of spatial spreading/despreading, 2) to separate the effect of channel hardening through $\pw_{\ell\ell_k}$ and that of spatial despreading through $\pU_{\ell\ell_k}$, and 3) to study the asymptotic performance of the resulting $r_{\ell\ell_k}$-dimensional channel estimation and combining/precoding. In particular, the effect of channel hardening arises when   
$\frac{\|\pw_{\ell\ell_k}\|^2}{\mathbb{E}[\|\pw_{\ell\ell_k}\|^2]} \xrightarrow{a.s.} 1$ for sufficiently large ${r_{\ell\ell_k}}$ as $M\rightarrow\infty$. The almost sure convergence can be given by the well-known trace lemma \cite[Lem. 2.7]{Bai98} (see also \cite[Thm. 3.4]{Cou11b})}, although the elements of $\pw_{\ell\ell_k}$ are non-identically distributed.  Therefore, channel hardening depends directly on the dimension ${r_{\ell\ell_k}}$ of the effective channel $\pw_{\ell\ell_k}$ in our rank-deficient spatial correlation models.



\subsection{Low-Dimensional Channel Estimation}
\label{sec:SC}

For channel training through uplink pilot signals, we consider both orthogonal pilot per cell and non-orthogonal pilot over the entire network. In each case, the corresponding MMSE channel estimate is described in the following. 

\subsubsection{(Intra-Cell) Orthogonal Pilot Scheme}
\label{sec:SC1}


{Let us consider a {pilot-aided} multicell MIMO system, where the uplink/downlink system uses a certain amount of channel uses to allow the BSs to estimate their associated user channel vectors.} For the conventional orthogonal pilot scheme \cite{Mar10,MLY16}, users in each cell send a set of mutually orthogonal pilot sequences, which are shared by $L$ cells in the network, so that user $k$ in every cell uses the same pilot sequence, resulting in pilot contamination. The received pilot signal for user $k$ in cell $\ell$ can then be expressed as  
\begin{align} \label{eq:CM-3}
   \bar{\sv}_{\ell_k}= \sum_{\ell'=1}^L\ph_{\ell\ell'_k} +\frac{1}{\sqrt{\rho_p }}\pz_\ell 
\end{align}
where $\rho_p = \varrho_p P_\text{ul}$ with $\varrho_p$ the power boosting factor (i.e., power gap between the training phase and the communication phase). This requires the overall training cost of $\max_\ell  K_\ell $ channel uses in half duplex TDD, where we assume synchronized cells such that in a given time slot all BSs schedule transmissions in the same direction to avoid severe inter-cell interference. In general, the maximum number of users per cell is limited by the uplink pilot dimension in this orthogonal scheme \cite{Mar10}.   
After spatial despreading for the same purpose as \eqref{eq:SM-5}, we will make use of the projected pilot signal  $\sv_{\ell_k}$ represented as
\begin{align} \label{eq:Pre-2b}
   \sv_{\ell_k}= \pU_{\ell\ell_k}^\ct\bar{\sv}_{\ell_k} =\pw_{\ell\ell_k}  +\sum_{\ell'\neq\ell}\pw_{\ell_k{\ell'}\!\!_{k}} +\frac{1}{{\sqrt{\rho_p }}}\pz_{\ell_k}.
\end{align} 


Given the noisy observation $\sv_{\ell_k}$ and the prior knowledge on $\pU_{\ell} =[\pU_{\ell\ell_1},\ldots,\pU_{\ell\ell_{K_\ell}}]$ 
 and the sum, $\sum_{\ell'\neq\ell}\pU_{\ell\ell'_k}^\ct \pR_{\ell\ell'_k}\pU_{\ell\ell'_k}$, of low-dimensional covariance matrices of the other-cell user channels after spatial despreading by $\pU_{\ell\ell'_k}^\ct$, and conditioned on a realization of $\underline{\pU}=\{\pU_{\ell\ell'_{k}}, \forall(\ell,\ell',k)\}$,  
the MMSE estimate $\hat{\pw}_{\ell\ell_k}$ of the {effective channel} $\pw_{\ell\ell_k}$ is given by
\begin{align} \label{eq:Pre-3}
   \hat{\pw}_{\ell\ell_k} &= \mathbb{E}\big[\pw_{\ell\ell_k}\sv_{\ell_k}^\ct\big|\underline{\pU}\big] \mathbb{E}\big[\sv_{\ell_k}\sv_{\ell_k}^\ct\big|\underline{\pU}\big]^{-1} \sv_{\ell_k} \nonumber \\
   &=  \Lambdam_{\ell\ell_k}\Xim_{\ell\ell_k} \sv_{\ell_k}  
\end{align} 
where $\Xim_{\ell\ell_k}  \triangleq \big(\Lambdam_{\ell\ell_k} +  \sum_{\ell'\neq\ell}\tilde{\pR}_{\ell_k{\ell'}\!\!_{k'}} +\rho_p^{-1}\pI_{r_{\ell\ell_k}}\big)^{-1}$ with  
$ \tilde{\pR}_{\ell_k{\ell'}\!\!_{k'}}  \triangleq \pU_{\ell\ell_k}^\ct\pR_{\ell{\ell'}\!\!_{k'}} \pU_{\ell\ell_k}  \in \mathbb{C}^{r_{\ell\ell_k}\times r_{\ell\ell_k}}$ being the covariance of $\pw_{\ell_k{\ell'}\!\!_{k'}}$. The sum of  low-dimensional covariance matrices  $\Xim_{\ell\ell_k}$ can rather be estimated  by an empirical covariance matrix based on ${\sv}_{\ell_k}$ instead of the prior knowledge since the fading channel statistics are ergodic and WSS, as long as the scattering geometry of users remains unchanged over the local stationarity interval. The distribution of $\hat{\pw}_{\ell\ell_k}$ is $\mathcal{CN}(\p0, \Phim_{\ell\ell_k}),$
where $\Phim_{\ell\ell_k}= \Lambdam_{\ell\ell_k}\Xim_{\ell\ell_k}\Lambdam_{\ell\ell_k}.$


The effective channel $\pw_{\ell\ell_k}$ of user $\ell_k$ can then be written 
as
\begin{align} 
   \pw_{\ell\ell_k} = \hat{\pw}_{\ell\ell_k} +\pn_{\ell\ell_k} \nonumber
\end{align} 
where $\pn_{\ell\ell_k} \sim \mathcal{CN} (\p0, \Nm_{\ell\ell_k})$ is conditionally independent of $\hat{\pw}_{\ell\ell_k}$ given $\pU_{\ell}$ by the orthogonality property of the MMSE estimate and the joint Gaussianity of $\hat{\pw}_{\ell\ell_k}$ and $\pn_{\ell\ell_k}$.
From (\ref{eq:Pre-3}), the conditional error covariance matrix  is given by 
\begin{align} 
   \mathbb{E}\big[ \pn_{\ell\ell_k}\pn_{\ell\ell_k}^\ct|\underline{\pU}\big] &= \mathbb{E}\big[\pw_{\ell\ell_k}\pw_{\ell\ell_k}^\ct\big] -\mathbb{E}\big[\hat{\pw}_{\ell\ell_k}\pw_{\ell\ell_k}^\ct|\underline{\pU}\big]  \nonumber \\
   &=  \Lambdam_{\ell\ell_k}  - \Lambdam_{\ell\ell_k}\Xim_{\ell\ell_k} \Lambdam_{\ell\ell_k} . \nonumber
\end{align}  


Notice that spatial despreading gives us the $r_{\ell\ell_k}$-dimensional MMSE channel estimation for $\pw_{\ell\ell_k}$ rather than the $M$-dimensional one. In fact, spatial despreading (spreading) is implicitly done in the full-dimensional MMSE channel estimation and combining (precoding) based on $\bar{\sv}_{\ell_k}$ in \eqref{eq:CM-3}. 
The following result shows that the low-dimensional channel estimate $\hat{\pw}_{\ell\ell_k}$ after spatial despreading is asymptotically equivalent to the full-dimensional channel estimate based on $\bar{\sv}_{\ell_k}$.  

\begin{lem}\label{lem-4} \normalfont
Under Assumption \ref{as-3}, $\sv_{\ell_k}$ in (\ref{eq:Pre-2b}) is an asymptotically sufficient statistic conditioned on $\pU_\ell$ to estimate the effective channel $\pw_{\ell\ell_k}$ from $\bar{\sv}_{\ell_k}$ as $M\rightarrow\infty$.
\end{lem}

\begin{IEEEproof}
Refer to Appendix \ref{app-1b}.
\end{IEEEproof}
\vspace{2mm}

\subsubsection{(Intra-Cell) Non-Orthogonal Pilot Scheme}
\label{sec:SC2}

It is known \cite{Zhe02,Nam17} that the conventional orthogonal pilot scheme may significantly limit the sum-rate performance of MU-MIMO systems unless $\max_\ell  K_\ell $ is smaller than $T_c/2$. To overcome this limiting factor, we also consider the non-orthogonal pilot scheme. 
This pilot signal $\sv_{\ell_k}'$ is non-orthogonal over intra cell as well as inter cell such that it consumes a single channel use\footnote{For more general non-orthogonal pilot design, one may use Welch bound equality frames \cite{Sar98}, where the cross-correlation coefficient between pilot sequences shared by different cells in the network is non-zero but the same. We can then design  non-orthogonal pilot sequences with training cost $K'$ such that $1\le K'< \max_\ell  K_\ell $ (e.g., see \cite{SZL15} in the context of massive MIMO).} per $L$-cell network per coherence block, given by 
\begin{align} \label{eq:SC2-1}
   \sv_{\ell_k}'= \pw_{\ell\ell_k}  +\sum_{(\ell',k')\neq(\ell,k)}\pw_{\ell_k{\ell'}\!\!_{k'}} +\frac{1}{\sqrt{\rho_p }}\pz_{\ell_k}.  
\end{align}
Similar to the orthogonal pilot scheme, we have the MMSE channel estimate of $\pw_{\ell\ell_k}$ given $\sv_{\ell_k}'$ as follows.
\begin{align} \label{eq:SC2-3}
   \check{\pw}_{\ell\ell_k} =  \Lambdam_{\ell\ell_k}\Xim'_{\ell\ell_k} \sv'_{\ell\ell_k}  
\end{align} 
where $\Xim_{\ell\ell_k}'  \triangleq \big(\Lambdam_{\ell\ell_k} +  \sum_{(\ell',k')\neq(\ell,k)}\tilde{\pR}_{\ell_k{\ell'}\!\!_{k'}} +\rho_p^{-1}\pI_{r_{\ell\ell_k}}\big)^{-1}$.
The distribution of $\check{\pw}_{\ell\ell_k}$ is given by $\mathcal{CN}(\p0, \Phim'_{\ell\ell_k}),$
where $\Phim'_{\ell\ell_k}= \Lambdam_{\ell\ell_k}\Xim'_{\ell\ell_k}\Lambdam_{\ell\ell_k}.$
As in Lemma \ref{lem-4}, one can show that $\sv'_{\ell\ell_k}$ is an asymptotically sufficient statistic to estimate the effective channel $\pw_{\ell\ell_k}$ from $\sv'_\ell$ under Assumption \ref{as-3}. 

A natural question arises on how we estimate channel covariance matrices in the non-orthogonal pilot scheme. While non-orthogonal pilot can give a better sum-rate performance by reducing the training overhead for the estimation of effective channels when $T_c$ is relatively small, extra ``long-period" orthogonal pilot would be necessary for the channel covariance estimation. As $T_c $ is typically much smaller than the local stationarity interval over which the channel process is WSS \cite[Chap. 4]{Rap02}, the corresponding training overhead for channel covariances will be rather small relative to ``short-period" orthogonal pilot $\sv_{\ell_k}$. In the practical scenario that one should estimate channel covariance matrices from contaminated pilot sequences, \cite{Hag17} proposed algorithms that exploit sparse angle-delay supports between users to separate the covariance estimates of the different users of interest.

\subsection{Low-Dimensional Multiuser Combining and Precoding}

We will only describe single-cell MMSE combining and precoding vectors based on the channel estimates $\hat{\pw}_{\ell\ell_k}$ in \eqref{eq:Pre-3} for the orthogonal pilot scheme, not requiring cell $\ell$ to estimate the effective channels $\pw_{\ell{\ell'}\!\!_{k'}}$ of users $k'$ in other cells $\ell'\neq\ell$. Replacing $\hat{\pw}_{\ell\ell_k}$ with 
$\check{\pw}_{\ell\ell_k}$ in \eqref{eq:SC2-3}, we can have the MMSE combining and precoding vectors for non-orthogonal pilot. 
Since we restrict our attention to the noncooperative multiuser combining/precoding and the decoder that take the intercell interference as additional additive noise, the resulting sum rates in this work is achieved by treating interference as noise \cite{GNA15}. A sum-rate gap of such trivial (intercell) interference managements including time-division multiplexing from several { capacity} upper bounds was studied and turned out to be not significant at practical values of SNR for two-user \cite{Nam17b} and $K$-user \cite{Nam17c} Gaussian interference channels.

\subsubsection{Uplink}

Spatial despreading in \eqref{eq:SM-5} indicates that we restrict the full-dimensional linear combining vector to $\pU_{\ell\ell_k}\pv_{\ell_k}\in \mathbb{C}^{M}$, where $\pv_{\ell_k}$ is the ${r_{\ell\ell_k}}$-dimensional combining vector for user $\ell_k$.  
Given $\hat{\pw}_{\ell_1},\ldots,\hat{\pw}_{\ell_{K_\ell}}$ and letting $\hat{\pw}_{\ell_k{\ell'}\!\!_{k'}} \triangleq \pU_{\ell\ell_k}^\ct\pU_{\ell{\ell'}\!\!_{k'}} \hat{\pw}_{\ell{\ell'}\!\!_{k'}}, \forall (\ell',k')$, the single-cell MMSE combining vector can be written as 
\begin{align} \label{eq:MR-15}
   \pv_{\ell_k}^\mmse &= \Upsilonm _{\ell_k} \hat{\pw}_{\ell\ell_k}
\end{align}
where $$ \Upsilonm _{\ell_k}  = \bigg({\hat{\pw}}_{\ell\ell_k}{\hat{\pw}}_{\ell\ell_k}^\ct +\sum_{k'\neq k}{\hat{\pw}}_{\ell_k{\ell}_{k'}}{\hat{\pw}}_{\ell_k{\ell}_{k'}}^\ct +\pz_{\ell_k} + {P_\text{ul}}^{-1}\pI_{r_{\ell\ell_k}}\bigg)^{-1} $$
where $\pz_{\ell_k}$ is a system design parameter and typically defined as $\pz_{\ell_k} = \sum_{\ell'\neq\ell}\sum_{k}\Xim_{\ell\ell'_k}+\sum_{\ell',k'}\Nm_{\ell_k{\ell'}\!\!_{k'}}$.
The above low-dimensional vector $\pv_{\ell_k}^\mmse$ is given based on the received vector $\py_{\ell_k}$ in \eqref{eq:SM-5}, to which spatial despreading of $\pU_{\ell\ell_k}^\ct$ applied.

\subsubsection{Downlink}

For downlink, the low-dimensional MMSE precoding vector can be written as
\begin{align} \label{eq:MR-15b}
   &\pg_{\ell_k}^\mmse = \nonumber \\ &\bigg({\hat{\pw}}_{\ell\ell_k}{\hat{\pw}}_{\ell\ell_k}^\ct +\sum_{k'\neq k}{\hat{\pw}}_{\ell_{k'}{\ell}_{k}}{\hat{\pw}}_{\ell_{k'}{\ell}_k}^\ct +\pZ'_{\ell\ell_k} + {P_{\text{dl},\ell}}^{-1}\pI_{r_{\ell\ell_k}}\bigg)^{-1} \hat{\pw}_{\ell\ell_k}
\end{align}
where $\pZ'_{\ell\ell_k} = \sum_{\ell'\neq\ell}\sum_{k}\big(\Lambdam_{\ell\ell'_k} +  \sum_{\ell'\neq\ell}\Lambdam_{{\ell'}\!\!_{k'}\ell_k} +\rho_p^{-1}\pI_{r_{\ell\ell_k}}\big)^{-1}+\sum_{\ell',k'}\Nm_{{\ell'}\!\!_{k'}\ell_k}$.   
The final precoding vector is followed by spatial spreading such that $\pp_{\ell_k}=\pU_{\ell\ell_k}\pg_{\ell_k}^\mmse$. One can also derive a more general form of the above MMSE precoding, i.e., in a regularized zero-forcing fashion.

By the data processing inequality, the performance of the above ${r_{\ell\ell_k}}$-dimensional combining (or precoding) vector is upper-bounded by that of the conventional ($M$-dimensional) MMSE combining (precoding) vector based on $\py_{\ell}$ in \eqref{eq:SM-1} ($y_{\ell_k}$ in \eqref{eq:SM-4}). We will see that the performance of $\pv_{\ell_k}^\mmse$ is asymptotically equivalent to the large-dimensional MMSE combining vector 
as $M\rightarrow\infty$, although $\pv_{\ell_k}^\mmse$ shows a performance degradation at finite $M$. This is also the case with $\pg_{\ell_k}^\mmse$ in downlink. 

A practical issue of the conventional full-dimensional MMSE combining/precoding is prohibitive computational complexity in the multicell massive MIMO network, where the $M\times M$ covariance matrix of intercell interference should be taken into account to mitigate the intercell interference (e.g., see \cite[Eqn. (11)]{Hoy13}).  The full-dimensional MMSE processing generally shows the complexity of $O(M^3)$ complex multiplications and $O(M^2)$  complex divisions per cell due to matrix inversion. Meanwhile, the low-dimensional processing involves  $O(K r^3)$ complex multiplications and $O(K r^2)$  complex divisions per cell in the homogeneous network, where every cell has $K$ users with the same rank of ${r_{\ell\ell_k}}=r$. { This low computational complexity also holds for the low-dimensional MMSE channel estimation in Sec. \ref{sec:SC}}. Consequently, the proposed MMSE processing may have much lower complexity in strong spatial correlation regimes, where $r\ll M$.


\section{Main Results}
\label{sec:MR}

In this section, we provide capacity scaling results in massive MIMO under the system model and assumptions in Sec. \ref{sec:Pre}. To this end, we consider three lower bounds on the achievable uplink rate based on an extension of the deterministic equivalents technique \cite{Sil95,Bai98,Bai07,Cou11a,Cou11b,Wag12}. While the first lower bound is based on coherent demodulation and decoding, the others are  non-coherent. The results are then extended to downlink.

\subsection{Uplink Capacity Scaling in the Strong Spatial Correlation Regime }



\subsubsection{Coherent Lower Bound}

We first consider the standard lower bound on the achievable rate proposed by Hassibi and Hochwald \cite{Has03}, based on the worst-case uncorrelated additive noise lemma. This lemma has been widely used for coherent detection with imperfect CSI in the vast literature of limited feedback single/multiple-user MIMO systems (e.g., see \cite{Cai10} and reference therein). In this paper, the channel estimate of either \eqref{eq:Pre-3} or \eqref{eq:SC2-3} is available at the BS for coherent detection in the uplink. 

Using the above bounding technique and the channel estimate in \eqref{eq:Pre-3}, the ergodic achievable rate of user $\ell_k$ is lower-bounded by 
\ifdefined\singcol 
\begin{align} \label{eq:MR-9}
   &\Rc_{\text{ul},\ell_k}^\text{(1)} = \mathbb{E} \left[\log\left(1+ \frac{|\pv_{\ell_k}^\ct\hat{\pw}_{\ell\ell_k}|^2}{\mathbb{E} \Big[|\pv_{\ell_k}^\ct\nv_{\ell\ell_k}|^2 +\sum_{(\ell',k')\neq(\ell,k)}|\pv_{\ell_k}^\ct\pw_{\ell_k{\ell'}\!\!_{k'}}|^2 +\frac{1}{P_\text{ul}}|\pv_{\ell_k}^\ct\pz_{\ell_k}|^2  \big|\; \hat{\pw}_{\ell}, \underline{\pU} \Big]}\right) \Bigg|\; \underline{\pU}\right]
\end{align}
\else
\eqref{eq:MR-9}, shown on the top of page~\pageref{eq:MR-9}, 
\fi
where { $\hat{\pw}_{\ell}=[\hat{\pw}_{\ell_1},\ldots,\hat{\pw}_{\ell_{K_\ell}}]$}. Notice that the above outer expectation is taken over $\{\pw_{\ell\ell'_{k}},\forall(\ell,\ell',k)\}$ conditioned on $\underline{\pU}$,\footnote{One might take the outer expectation also over $\underline{\pU}$. In order to attain the ergodicity with respect to the channel covariances, we will need codewords that span hundreds of local stationary intervals much longer than small-fading coherence block. However, this is not very realistic in practical systems.} implying that $\Rc_{\text{ul},\ell_k}^\text{(1)}$ is a random variable conditional to a particular covariance realization of $\underline{\pU}$ since it is a function of $\underline{\pU}$, which is itself random. As mentioned earlier, it will be shown that such ergodic rates $\Rc_{\text{ul},\ell_k}^\text{(1)}$ converge to a deterministic limit under the stochastic spatial correlation models in Sec. \ref{sec:CM}. 
The derivation of \eqref{eq:MR-9} follows from the MMSE decomposition of the useful signal channel and from the worst-case uncorrelated additive noise argument. In this subsection, it is assumed that $K_\ell$ is a finite constant or scales sublinearly with $M$ such that $\lim_{M\rightarrow\infty}\frac{K_\ell}{M}= 0$ for all $\ell$, as usual in the massive MIMO literature. We will address the case where $\limsup_{M\rightarrow\infty}\frac{K}{M}<\infty$ in Sec. \ref{sec:SC3}.
Using \eqref{eq:MR-9}, we obtain the following lower bound on the asymptotic capacity with respect to $M$.

\ifdefined\singcol 
\else
\begin{figure*}
  \vspace{-3mm}
  \normalsize
 \setcounter{storeeqcounter}{\value{equation}}

\begin{align} \label{eq:MR-9}
   &\Rc_{\text{ul},\ell_k}^\text{(1)} = \mathbb{E} \left[\log\left(1+ \frac{|\pv_{\ell_k}^\ct\hat{\pw}_{\ell\ell_k}|^2}{\mathbb{E} \Big[|\pv_{\ell_k}^\ct\nv_{\ell\ell_k}|^2 +\sum_{(\ell',k')\neq(\ell,k)}|\pv_{\ell_k}^\ct\pw_{\ell_k{\ell'}\!\!_{k'}}|^2 +\frac{1}{P_\text{ul}}|\pv_{\ell_k}^\ct\pz_{\ell_k}|^2  \big|\; \hat{\pw}_{\ell}, \underline{\pU} \Big]}\right) \Bigg|\; \underline{\pU}\right]
\end{align}

 \addtocounter{storeeqcounter}{1} 
 \setcounter{equation}{\value{storeeqcounter}} 
  \hrulefill
\end{figure*}
\fi

\begin{thm}\label{thm-2} \normalfont
For large $M$ with $\lim_{M\rightarrow\infty}\frac{K}{M}= 0$ and Assumptions \ref{as-2} and \ref{as-3} with the orthogonal pilot scheme, the ergodic sum capacity of MIMO uplink is lower-bounded by
\begin{align} \label{eq:MR-1}
   \Cc^\text{ul}_M \ge  \sum_{\ell=1}^L\sum_{k=1}^{\kappa_3}\Big(1-\frac{\kappa_3}{T_c}\Big)\log (P_\text{ul} \trace\Lambdam_{\ell\ell_k}) +o(1). 
\end{align}
where $\kappa_3={\min\{\max_\ell K_\ell,\lfloor \frac{T_c}{2}\rfloor\}}$.
\end{thm}


\begin{IEEEproof}
Refer to Sec. \ref{app-3}.
\end{IEEEproof}
\vspace{2mm}

For the above large system analysis, we extended the standard technique of deterministic equivalents in a few aspects in Appendix \ref{app-1}: 1) the unbounded spectral norm of channel covariance matrices with respect to $M$ and 2) the partial random Fourier correlation model for which the trace lemma in \cite{Bai98}, crucial for deterministic equivalents, is not applicable. The latter is due to the fact that the entries of the columns of $\pU_{\ell\ell'_k}$ are not i.i.d. any longer and that the column vectors are also not independent of each other. 


Suppose the homogeneous network such that 
\begin{gather} 
   K_\ell=K, \ P_\text{ul}=P_{\text{dl},\ell}=\frac{\snr}{K}, \ {r_{\ell\ell'_k}}=r,\ \forall  (\ell,\ell',k). \label{eq:MR-45}
\end{gather}
In this homogeneous network, it follows from \eqref{eq:MR-1} that the coherent lower bound yields the multiplexing gain per user of $1-\frac{\kappa_1}{T_c}$. Also multiplexing gain per cell is given by $\big(1-\frac{\kappa_1}{T_c}\big)\kappa_1$ for the typical (orthogonal) pilot scheme.  
Accordingly, an intriguing question arises: \emph{Is the sum-rate scaling law with the multiplexing gain of \eqref{eq:MR-1} optimal in massive MIMO under Assumption \ref{as-3}?} Such  multiplexing gain per cell is indeed limited by $\frac{T_c}{4}$ when $K\ge\frac{T_c}{2}$ for relatively small $T_c$ \cite{Nam17}. To answer this question, we need to consider the non-orthogonal pilot $\sv_{\ell_k}'$ in Subsection \ref{sec:SC}, whose training cost is only a single channel use across the $L$-cell network.  In this case, unfortunately, the  lower bound  $\Rc_{\text{ul},\ell_k}^\text{(1)}$ in \eqref{eq:MR-9} is not amenable to large system analysis because the trace lemma \cite[Lem. 2.7]{Bai98} is not applicable. Furthermore, this coherent bound  is expected to suffer from significant channel estimation error in the non-orthogonal pilot scheme. We thus turn our attention to non-coherent bounding techniques.

\ifdefined\singcol 
\else
\begin{figure*}
  \vspace{-3mm}
  \normalsize
 \setcounter{storeeqcounter}{\value{equation}}

\begin{align} \label{eq:NC-1}
   &\Rc_{\text{ul},\ell_k}^\text{(2)} = \log\left(1+  \frac{\big|\mathbb{E}[\pv_{\ell_k}^\ct{\pw}_{\ell\ell_k}| \underline{\pU}]\big|^2}{\frac{1}{P_\text{ul}} +\var[\pv_{\ell_k}^\ct{\pw}_{\ell\ell_k}| \underline{\pU}] +\sum_{(\ell',k')\neq(\ell,k)}\mathbb{E} \big[|\pv_{\ell_k}^\ct\pw_{\ell_k{\ell'}\!\!_{k'}}|^2| \underline{\pU} \big]}\right)
\end{align}

  \hrulefill
\end{figure*}
\fi

\subsubsection{Non-Coherent Lower Bound}


Marzetta \cite{Mar06} proposed a non-coherent bound based on separating the useful signal coefficient into a deterministic part and a random fluctuation part, not requiring the coherent detection. The bounding technique may be traced back to M\'edard \cite{Med00}. 
Following this technique, \cite{Jos11} developed a lower bound on the achievable rate of the MIMO downlink, where the fading distribution is assumed to be known, but the channel estimates are unavailable at the receivers unless additional dedicated downlink pilots are provided. Although this bound is accordingly very natural to use in downlink, it is applicable in uplink as well (e.g., \cite{Ngo11}), where the uplink pilot is used to only construct the coherent combining vector $\pv_{\ell_k}$ rather than coherent demodulation. Using this bounding technique, we have the ergodic achievable rate 
\ifdefined\singcol 
\begin{align} \label{eq:NC-1}
   &\Rc_{\text{ul},\ell_k}^\text{(2)} = \log\left(1+  \frac{\big|\mathbb{E}[\pv_{\ell_k}^\ct{\pw}_{\ell\ell_k}| \underline{\pU}]\big|^2}{\frac{1}{P_\text{ul}} +\var[\pv_{\ell_k}^\ct{\pw}_{\ell\ell_k}| \underline{\pU}] +\sum_{(\ell',k')\neq(\ell,k)}\mathbb{E} \big[|\pv_{\ell_k}^\ct\pw_{\ell_k{\ell'}\!\!_{k'}}|^2| \underline{\pU} \big]}\right).
\end{align}
\else
in \eqref{eq:NC-1}, shown on the top of page~\pageref{eq:NC-1}.
 \addtocounter{storeeqcounter}{1} 
 \setcounter{equation}{\value{storeeqcounter}} 
\fi
We can then obtain the following main result of this work.

\begin{thm}\label{thm-3} \normalfont
For large $M$ with $\lim_{M\rightarrow\infty}\frac{K}{M}= 0$ and Assumptions \ref{as-2} and \ref{as-3} with the non-orthogonal pilot scheme, the sum 
capacity of MIMO uplink behaves as
\begin{align} \label{eq:MR-1b}
   \Cc^\text{ul}_M = \big(1-T_c^{-1}\big) \sum_{\ell=1}^L\sum_{k=1}^{K_\ell}\log (P_\text{ul}  \trace\Lambdam_{\ell\ell_k}) +o(1)  .
\end{align}
\end{thm}

\begin{IEEEproof}
Refer to Sec. \ref{app-5}.
\end{IEEEproof}

The factor $P_\text{ul}\trace\Lambdam_{\ell\ell_k}$ in the right-hand side (RHS) of \eqref{eq:MR-1b} represents the post-processing SNR in uplink. 
For the homogeneous network in \eqref{eq:MR-45} with $\trace\Lambdam_{\ell\ell_k}=M$ and $\trace\Lambdam_{\ell\ell'_k}=\iota M$ for all $(\ell,\ell',k)$, the first scaling law in (\ref{eq:intro-4}) directly follows from \eqref{eq:MR-1b}. 
Compared to (\ref{eq:intro-2}) and (\ref{eq:intro-3}), this scaling function shows multiplexing gain is not limited any longer by $\frac{T_c}{4}$ for $T_c\le {2LK}$ in the strong correlation regime. 
This capacity scaling is asymptotically tight, as mentioned in the introduction, and shows the maximum multiplexing gain of $(1-{T_c}^{-1})LK$ from the perspective of either pilot-aided or non-coherent communications.


The lower bound $\Rc_{\text{ul},\ell_k}^\text{(2)}$ predicts well the achievable rate of massive MIMO when the useful signal coefficient $\pv_{\ell_k}^\ct{\pw}_{\ell\ell_k}$ hardens,  i.e., behaves almost deterministically. However, in fading channels with strong spatial correlation, where the dimension $r_{\ell\ell_k}$ of the effective channel $\pw_{\ell\ell_k}$ could be much smaller than $M$ (i.e., lack of channel hardening), $\Rc_{\text{ul},\ell_k}^\text{(2)}$ suffers from the self-interference due to the non-negligible variance term $\var[\pv_{\ell_k}^\ct{\pw}_{\ell\ell_k}| \underline{\pU}]$ in \eqref{eq:NC-1} unless $r_{\ell\ell_k}$ becomes sufficiently large. 
As a consequence, the non-coherent bound $\Rc_{\text{ul},\ell_k}^\text{(2)}$ may substantially underestimate the sum-rate performance of massive MIMO, which will be examined through numerical results in Sec. \ref{sec:NR}. 
This is also particularly evident in the large but finite antenna regime. 

\subsubsection{Alternative Non-Coherent Lower Bound}

We consider another non-coherent bounding technique very recently derived in \cite{Cai17} for the single-cell downlink scenario. The bound can be traced back to a mutual information decomposition given in the tutorial paper by Biglieri, Proakis, and Shamai \cite[Eqn. 3.3.27]{BPS98} to study non-coherent block fading channels.
Assuming Gaussian inputs, linear combining/precoding vector, and treating interference as noise, we have the following upper bound on the ergodic achievable rate \cite{Cai17}
\begin{align} \label{eq:SC-1c}
  \Rc_{\text{ul},\ell_k}^\text{ub} = \mathbb{E} \left[\log\left(1+  \frac{\big|\pv_{\ell_k}^\ct{\pw}_{\ell\ell_k}\big|^2}{\frac{1}{P_\text{ul}} +\sum_{(\ell',k')\neq(\ell,k)}|\pv_{\ell_k}^\ct\pw_{\ell_k{\ell'}\!\!_{k'}}|^2 } \right)\bigg|\; \underline{\pU}\right] 
\end{align}
which is a max-min bound such that the maximum is over the coding/decoding strategy of user ${\ell_k}$ and the minimum is over all input distributions of the other users. The third lower bound used in this work is given by \cite{Cai17}
\begin{align} \label{eq:SC-1}
   &\Rc_{\text{ul},\ell_k}^\text{(3)} = \Rc_{\text{ul},\ell_k}^\text{ub} \nonumber \\ & \ \ \ -\frac{1}{T_c}\sum_{(\ell',k')\neq(\ell,k)} \log\Big(1+ {P_\text{ul}}\var\big[\pv_{\ell_k}^\ct\pw_{\ell_k{\ell'}\!\!_{k'}}\big| \underline{\pU}\big]\Big).
\end{align}
The second term in the RHS of \eqref{eq:SC-1} consists of the prelog factor $\frac{\sum_\ell K_\ell-1}{T_c}$ and the variances of coherent interference $\var[\pv_{\ell_k}^\ct\pw_{\ell_k{\ell'}\!\!_{k'}}| \underline{\pU}]$ multiplied by the transmit power ${P_\text{ul}}$ inside the logarithm.  On one hand, this bound comes very close to the max-min upper bound when coherence block is sufficiently large or coherent interference is limited. The latter is the case with our main scenario under the sublinear sparsity assumption, where interference is significantly suppressed by spatial despreading.  On the other hand, a negative feature of this bound is that when the first term $\Rc_{\text{ul},\ell_k}^\text{ub}$ is interference limited at finite $M$, ${P_\text{ul}}$ grows to infinity, and/or $T_c$ does not scale well with $M$, the bound goes to even below zero. Therefore, another lower bound with a complementary behavior can be found in \cite[Lemma 4]{Cai17}. One can prove that the non-coherent lower bound $\Rc_{\text{ul},\ell_k}^\text{(3)}$ achieves the same capacity scaling as Theorem \ref{thm-3}. A sketch of the proof will be given in Sec. \ref{sec:FP-C}. 

\ifdefined\singcol 
\else
\begin{figure*}
  \vspace{-3mm}
  \normalsize
  \setcounter{tempeqcounter}{\value{equation}} 
\begin{align} 
   &\Rc_{\text{dl},\ell_k}^\text{(1)} = \log\left(1+  \frac{\big|\mathbb{E}[{\pw}_{\ell\ell_k}^\ct\pg_{\ell_k}| \underline{\pU}]\big|^2}{\frac{1}{P_{\text{dl},\ell}} +\var[{\pw}_{\ell\ell_k}^\ct\pg_{\ell_k}| \underline{\pU}] +\sum_{(\ell',k')\neq(\ell,k)}\mathbb{E} \big[|\pw_{{\ell'}\!\!_{k'}\ell_k}^\ct \pg_{{\ell'}\!\!_{k'}}|^2| \underline{\pU}\big]} \right) \label{eq:SCD-2}
\\
  & \Rc_{\text{dl},\ell_k}^\text{(2)} = \mathbb{E} \left[\log\left(1+  \frac{\big|{\pw}_{\ell\ell_k}^\ct\pg_{\ell_k}\big|^2}{\frac{1}{P_{\text{dl},\ell}} +\sum_{(\ell',k')\neq(\ell,k)}| \pw_{{\ell'}\!\!_{k'}\ell_k}^\ct \pg_{{\ell'}\!\!_{k'}}|^2 } \right)\bigg|\; \underline{\pU}\right]  -\frac{1}{T_c}\sum_{(\ell',k')\neq(\ell,k)} \log\left(1+ {P_{\text{dl},\ell}}\var[\pw_{{\ell'}\!\!_{k'}\ell_k}^\ct \pg_{{\ell'}\!\!_{k'}}| \underline{\pU}]\right) \label{eq:SCD-3}
\end{align}
 \addtocounter{tempeqcounter}{2} 
\setcounter{equation}{\value{tempeqcounter}} 
  \hrulefill
\end{figure*}
\fi

\subsection{Downlink Capacity Scaling in the Strong Spatial Correlation Regime }
\label{sec:SCD}

Unless we employ downlink dedicated (precoded) pilot signals consuming additional $K_\ell$ channel uses for the communication phase, the useful signal coefficient necessary for coherent detection is unknown so that we cannot use $\Rc_{\text{ul},\ell_k}^\text{(1)}$ for downlink. On the contrary, we can utilize both non-coherent bounds $\Rc_{\text{ul},\ell_k}^\text{(2)}$ and $\Rc_{\text{ul},\ell_k}^\text{(3)}$ for downlink as well to derive the capacity scaling result under sparse angular support models. Given the downlink received signal in \eqref{eq:SM-4b} and conditioned on $\underline{\pU}$, the lower bounds are given  by
\ifdefined\singcol 
\begin{align} 
   &\Rc_{\text{dl},\ell_k}^\text{(1)} = \log\left(1+  \frac{\big|\mathbb{E}[{\pw}_{\ell\ell_k}^\ct\pg_{\ell_k}| \underline{\pU}]\big|^2}{\frac{1}{P_{\text{dl},\ell}} +\var[{\pw}_{\ell\ell_k}^\ct\pg_{\ell_k}| \underline{\pU}] +\sum_{(\ell',k')\neq(\ell,k)}\mathbb{E} \big[|\pw_{{\ell'}\!\!_{k'}\ell_k}^\ct \pg_{{\ell'}\!\!_{k'}}|^2| \underline{\pU}\big]} \right) \label{eq:SCD-2}
\\
  & \Rc_{\text{dl},\ell_k}^\text{(2)} = \mathbb{E} \left[\log\left(1+  \frac{\big|{\pw}_{\ell\ell_k}^\ct\pg_{\ell_k}\big|^2}{\frac{1}{P_{\text{dl},\ell}} +\sum_{(\ell',k')\neq(\ell,k)}| \pw_{{\ell'}\!\!_{k'}\ell_k}^\ct \pg_{{\ell'}\!\!_{k'}}|^2 } \right)\bigg|\; \underline{\pU}\right]  -\frac{1}{T_c}\sum_{(\ell',k')\neq(\ell,k)} \log\left(1+ {P_{\text{dl},\ell}}\var[\pw_{{\ell'}\!\!_{k'}\ell_k}^\ct \pg_{{\ell'}\!\!_{k'}}| \underline{\pU}]\right). \label{eq:SCD-3}
\end{align}
\else
\eqref{eq:SCD-2} and \eqref{eq:SCD-3} on the top of page~\pageref{eq:SCD-2}.
\fi
Using the above bounds under Assumptions \ref{as-2} and \ref{as-3} with the orthogonal pilot scheme, we can first derive the achievable ergodic sum rate in MIMO downlink 
\begin{align} \label{eq:SCD-1}
   \Cc^\text{dl}_M \ge  \sum_{\ell=1}^L\sum_{k=1}^{\kappa_3}\Big(1-\frac{{\kappa_3}}{T_c}\Big)\log (P_{\text{dl},\ell} \trace\Lambdam_{\ell\ell_k}) +o(1) 
\end{align}
where  $P_{\text{dl},\ell}\trace\Lambdam_{\ell\ell_k}$ is the post-processing SNR of user $\ell_k$ in the downlink. The above sum-rate characterization follows from the same footsteps in the proof of Theorem \ref{thm-2}. Similar to Theorem \ref{thm-3}, we also have the following result.

\begin{thm}\label{thm-3b} \normalfont
For large $M$ with $\lim_{M\rightarrow\infty}\frac{K}{M}= 0$ and Assumptions \ref{as-2} and \ref{as-3} with the non-orthogonal pilot scheme, the sum capacity of MIMO downlink behaves as
\begin{align} \label{eq:DL-2}
   \Cc^\text{dl}_M = \big(1-T_c^{-1}\big) \sum_{\ell=1}^L\sum_{k=1}^{K_\ell}\log (P_{\text{dl},\ell} \trace\Lambdam_{\ell\ell_k}) +o(1).
\end{align}
\end{thm}

In order to prove the above result, it suffices to use either $\Rc_{\text{dl},\ell_k}^\text{(1)}$ or $\Rc_{\text{dl},\ell_k}^\text{(2)}$ with the transmit maximal ratio (or conjugate) precoding, i.e., $\pg_{\ell_k} =\check{\pw}_{\ell\ell_k}$. The proof is omitted for the sake of brevity, but  follows the footsteps and arguments in the proofs of Theorems \ref{thm-2} and \ref{thm-3}. A main difference is that we sometimes use Lemmas \ref{cor-1c} and \ref{lem-1d} instead of Lemmas \ref{lem-1b} and \ref{lem-1e}, respectively. As expected, under the channel model assumptions of this paper the downlink has the same capacity scaling of the uplink.  


\subsection{Very Strong Spatial Correlation Regime}
\label{sec:SC3}

It is important to notice that in order to obtain \eqref{eq:MR-1b} (or the first scaling law in \eqref{eq:intro-4}) in Theorem \ref{thm-3} and \eqref{eq:DL-2} in Theorem \ref{thm-3b} in the strong correlation regime, we have assumed that $L$ is finite and $K$ is also finite or $\lim_{M\rightarrow\infty}\frac{K}{M}= 0$, which is implicit in the massive MIMO literature \cite{Mar10,BHS18,Hoy13,MLY16}, together with $T_c$ growing linearly with $K$. 
Rather, we will see that under the very strong correlation regime in Assumption \ref{as-4}, the linear sum capacity scaling with respect to $M$ at finite SNR can still be achieved as long as $\limsup_{M\rightarrow\infty}\frac{K}{M}<\infty$, i.e., $K$ does not grow faster than $M$, even though the coherence block size $T_c$ does not scale with $K$. 
To show this, we will use $\Rc_{\text{ul},\ell_k}^\text{ub}$ in  \eqref{eq:SC-1c} and the naive channel estimate $\sv_{\ell_k}'$ in \eqref{eq:SC2-1} based on non-orthogonal pilot instead of the MMSE estimate $\check{\pw}_{\ell\ell_k}$ for the sake of exposition, since the former channel estimate essentially captures the pilot contamination term and the inter/intra-cell interference term. With $\vv_{\ell_k}=\sv_{\ell_k}'$ (ignoring normalization), the overall interference term in \eqref{eq:SC-1c} for user $\ell_k$ in uplink is given by
\begin{align} \label{eq:IMR-1}
   &\frac{1}{M^2} \sum_{(\ell',k')\neq(\ell,k)}|\pv_{\ell_k}^\ct\pw_{\ell_k{\ell'}\!\!_{k'}}|^2  = \frac{1}{M^2} \times \nonumber \\ & \sum_{(\ell',k')\neq(\ell,k)}\bigg|\Big(\pw_{\ell\ell_k}  +\sum_{(j,m)\neq(\ell,k)}\pw_{\ell_kj_m} +\frac{1}{\sqrt{\rho_p }}\pz_{\ell_k}\Big)^\ct \pw_{\ell_k{\ell'}\!\!_{k'}}\bigg|^2 .
\end{align}
We investigate the asymptotic behaviors of three main terms in the above equation. 
\begin{itemize}
\item{Coherent pilot contamination term (${(j,m)=(\ell',k')}$):}  
In the limit of $M$, we can derive the following almost sure convergence
\begin{align} \label{eq:MR-1c}
  \frac{\|\pw_{\ell_k{\ell'}\!\!_{k'}}\|^2}{M} = &\; \frac{\big|\pw_{\ell{\ell'}\!\!_{k'}}^\ct \pU_{\ell{\ell'}\!\!_{k'}}^\ct \pU_{\ell\ell_k}\pU_{\ell\ell_k}^\ct\pU_{\ell{\ell'}\!\!_{k'}} \pw_{\ell{\ell'}\!\!_{k'}} \big|^2}{M}\nonumber \\ & \ \ \xrightarrow{a.s.} \; \frac{r}{M^2}\trace\Lambdam_{\ell{\ell'}\!\!_{k'}}  
\end{align}
where $\pw_{\ell_k{\ell'}\!\!_{k'}}$ is given in \eqref{eq:SM-5b}. The convergence follows from Lemma \ref{lem-a10} since $\frac{\|\pw_{\ell{\ell'}\!\!_{k'}}\|^2}{M} \xrightarrow{a.s.}\frac{1}{M}\trace\Lambdam_{\ell{\ell'}\!\!_{k'}}$ by the trace lemma and  $\pU_{\ell{\ell'}\!\!_{k'}}^\ct \pU_{\ell\ell_k}\pU_{\ell\ell_k}^\ct\pU_{\ell{\ell'}\!\!_{k'}} \xrightarrow{a.s.}\frac{r}{M}\pI_r$ by Lemma \ref{lem-1b} (or \ref{lem-1e}) for the random partial unitary correlation model (or the random partial Fourier model). Note that the latter convergence shows the role of spatial despreading in uplink, which is more crucial than the former (channel hardening) to obtain the capacity scaling results in this paper. Using the dominated convergence theorem, we have 
\begin{align} \label{eq:IMR-2}
   \frac{1}{M^2} \sum_{(\ell',k')\neq(\ell,k)}\big|\pw_{\ell_k{\ell'}\!\!_{k'}} ^\ct \pw_{\ell_k{\ell'}\!\!_{k'}}\big|^2 \; \xrightarrow{a.s.} \; \frac{LK}{M} \cdot \frac{r^2 \trace\Lambdam_{\ell{\ell'}\!\!_{k'}}^2}{M^3} 
\end{align}
which vanishes under Assumption \ref{as-4} as long as $\limsup_{M\rightarrow\infty}\frac{K}{M}<\infty$ and $L$ is finite. 





\item{Inter/intra-cell interference term:}  Although interference does not matter in the conventional massive MIMO because of finite $K$ and the law of large number in the limit of $M$, this is not necessarily the case with $\limsup_{M\rightarrow\infty}\frac{K}{M}<\infty$. 
For the residual interference term after spatial despreading, $|\pw_{\ell\ell_k} ^\ct \pw_{\ell_k{\ell'}\!\!_{k'}}|^2 = |\pw_{\ell\ell_k}^\ct \pU_{\ell\ell_k}^\ct\pU_{\ell{\ell'}\!\!_{k'}} \pw_{\ell{\ell'}\!\!_{k'}}|^2, \forall (\ell',k')\neq(\ell,k)$, ``common" angular components between the channel covariance matrices of any pair of users are relatively very small such that 
$\lim_{r\rightarrow\infty}\frac{\Delta_{\ell_k,{\ell'}\!\!_{k'}}}{r}=0$,
where $\Delta_{\ell_k,{\ell'}\!\!_{k'}}$ is the number of the similar angular components shared by users $\ell_k$ and ${\ell'}\!\!_{k'}$ at BS $\ell$. 
Specifically, $\pU_{\ell\ell_k}^\ct\pU_{\ell{\ell'}\!\!_{k'}} \xrightarrow{a.s.}\frac{1}{\sqrt{M}}\pJ_r$ as a straightforward corollary of Lemma \ref{lem-1b} (or \ref{lem-1e}), where $\pJ_r$ is the $r$-dimensional all-ones matrix, and $\frac{|\textsf{h}_{\ell\ell_k} ^\ct \textsf{h}_{\ell{\ell'}\!\!_{k'}}|^2}{r^2} \xrightarrow{a.s.}0$  by Lemma \ref{lem-2}, where $\textsf{h}_{\ell{\ell'}\!\!_{k'}}$ is defined in \eqref{eq:SM-2}. Similar to \eqref{eq:IMR-2}, we have 
\begin{align} \label{eq:IMR-5}
   \frac{1}{M^2}& \sum_{(\ell',k')\neq(\ell,k)}\big|\pw_{\ell\ell_k} ^\ct \pw_{\ell_k{\ell'}\!\!_{k'}}\big|^2 \; \xrightarrow{a.s.} \nonumber \\ &\; \frac{LK}{M} \cdot \frac{r^2 \trace\Lambdam_{\ell\ell_k}\trace\Lambdam_{\ell{\ell'}\!\!_{k'}}}{M^2}\cdot\frac{|\textsf{h}_{\ell\ell_k} ^\ct \textsf{h}_{\ell{\ell'}\!\!_{k'}}|^2}{r^2} 
\end{align}
which also vanishes under Assumption \ref{as-4}, particularly, \eqref{eq:SM-3c2} and \eqref{eq:SM-3c3}.

\item{Noncoherent pilot contamination term (${(j,m)\neq(\ell',k')}$):}  Although the amount of these terms $|\pw_{\ell_kj_m} ^\ct \pw_{\ell_k{\ell'}\!\!_{k'}}|^2  = |\pw_{\ell j_m}^\ct \pU_{\ell j_m}^\ct \pU_{\ell\ell_k}\pU_{\ell\ell_k}^\ct\pU_{\ell{\ell'}\!\!_{k'}} \pw_{\ell{\ell'}\!\!_{k'}} |^2$ is extremely large due to non-orthogonal pilot, the infinite sum of quickly vanishing terms eventually goes to zero under our channel models and assumptions. 

\end{itemize}

Note that since the deterministic approximation of the normalized signal power $\frac{\|\pw_{\ell\ell_k}\|^2}{M}$ as well as \eqref{eq:IMR-2} and \eqref{eq:IMR-5} depends on  realizations of  $\{\trace\Lambdam_{\ell\ell'_k}\}$, the resulting SINR of user $\ell_k$ is conditional to random channel covariance matrices of all users in the network. Under the deterministic $\trace\Lambdam_{\ell\ell'_k}$ assumption in \eqref{eq:CM-1b}, however, the conditional SINR converges to a deterministic limit by Lemma \ref{lem-a10} for sufficiently large $M$, thus allowing the ergodic rate analysis in this work. 
Using the same argument as the above derivations, we have the following corollary of Theorems \ref{thm-3} and \ref{thm-3b}. 

\begin{thm}\label{cor-1} \normalfont
For large $M$ with $\limsup_{M\rightarrow\infty}\frac{K}{M}<\infty$ under Assumptions \ref{as-2} and \ref{as-4} with the non-orthogonal pilot scheme, the sum capacities of massive MIMO uplink and downlink asymptotically behave as \eqref{eq:MR-1b} and \eqref{eq:DL-2}, respectively. For downlink, the sum power $P_\text{dl}$ is assumed to scale linearly with $K$. 
\end{thm}

Using the homogeneous network in \eqref{eq:MR-45} with $\trace\Lambdam_{\ell\ell_k}=r$ and $\trace\Lambdam_{\ell\ell'_k}=\iota r$ for all $(\ell,\ell',k)$ in the very strong correlation regime, the second scaling law in (\ref{eq:intro-4}) immediately follows from the above result.  
For the consistency (or duality) with uplink, where the uplink per-user  transmit power $P_\text{ul}$ is fixed and the uplink total power automatically scales with $K$, we assumed that the downlink sum power $P_\text{dl}$ also scales with $K$. Otherwise, the post-processing SNR $\frac{P_\text{dl}}{K}\trace\Lambdam_{\ell\ell_k}$ will vanish as $K\rightarrow\infty$.

\subsection{Interpretation of Main Results}
\label{sec:IMR}

Let us consider the homogeneous network in \eqref{eq:MR-45} with $\iota=1$. 
The main results in Theorems \ref{thm-3} and \ref{thm-3b} show the role of spatial despreading and spreading in strong spatial correlation regimes because the achievability proof is based on the corresponding low-dimensional channel estimation and combining/precoding. We have not invoked any explicit pilot decontamination technique such as covariance-aided coordination with non-overlapping angular supports \cite{Yin13}, semi-blind estimation \cite{Mul14}, or multicell precoding and combining \cite{Ash12,Bjo16}. As a matter of fact, our assumptions on strong correlation regimes are rather restrictive or favorable in the sense that each covariance matrix occupies a vanishing fraction of dimensions with respect to the total signal space dimensions $M$ so that a sort of coordination based on covariances is implicitly achieved without information exchange among cells. Accordingly, our asymptotic results in this paper may be regarded as a direct consequence of the channel models and their assumptions.  
However, since we leverage the asymptotic sum-rate scaling law only as a tool to provide insight into the role of spatial spreading and despreading in terms of $M$, $K$, $T_c$, and $r$, it is important to see if our asymptotic results are translated into a good prediction in the performance behavior of finite systems with strong spatial correlation. %
To this end, the linear sum-rate scaling of our results with respect to either SNR or $M$ (equivalently, $K$ for fixed $\frac{K}{M}>0$)  in \eqref{eq:MR-1} will be shown valid through numerical examples in Figs. \ref{fig-3a} and \ref{fig-6} for \emph{moderate} values of $r$ with respect to finite $M$, as the large system analysis in general works quite precisely even for rather small system dimensions \cite{Hoy13,Huh12}. 
It is also pointed out that any advanced precoding technique like block diagonalization \cite{Nam12,Adh13} over channel covariances is not required to realize the sum-rate performance predicted by the asymptotic capacity scaling under our channel models. We have shown that single-cell linear precoding (combining) based on spatial spreading (despreading) is sufficient to achieve the optimal capacity scaling laws  \eqref{eq:intro-4} in the limit of $M$, although multicell precoding can significantly outperform single-cell processing at finite $M$.

Theorem \ref{cor-1} indicates that \emph{every user can achieve unlimited spectral efficiency} in multicell massive MIMO even when the number of users per cell scales linearly with the number of BS antennas. The result is only achievable under the very strong correlation regime, which  stipulates a much larger value of $M$ for $r$ to be sufficiently large relative to the strong correlation regime, e.g., to guarantee the asymptotic orthogonality of ${|\textsf{h}_{\ell\ell_k} ^\ct \textsf{h}_{\ell{\ell'}\!\!_{k'}}|^2}/{r^2}$ in the RHS of \eqref{eq:IMR-5}. We also used the non-orthogonal pilot scheme whose training overhead is not a function of $K$ and in general much less than $T_c$, no matter how large $K$ is. 
This somewhat surprising argument may appear counterintuitive, since it has been presumed in the literature that each data stream requires at least one orthogonal pilot dimension within its own cell, in order to allow the receiver (uplink) or the transmitter (downlink) to separate the streams and provide an interference-free channel. Therefore, the common intuition suggests that the number of data streams is upper-bounded by $\min\{M,K,T_c/2\}$. While this is indeed the case for isotropic fading or, more in general, for channel covariances with rank $r = \alpha M$, with fixed $\alpha > 0$, it is not the case under the strong correlation regimes, where $r$ grows without bound but slower than $M$ (see \eqref{eq:SM-3c1} and \eqref{eq:SM-3c2}) so that $\alpha$ vanishes in the limit of $M$ { and the channel energy captured by $\trace\Lambdam_{\ell{\ell'}\!\!_{k'}}$ also grows slower than $M$ (see \eqref{eq:SM-3c3})}, thus making non-orthogonal pilot based channel training feasible. 

Finally it deserves to mention the assumption on the downlink sum power $P_\text{dl}$ in Theorem \ref{cor-1}. Due to limited downlink transmit power per cell in practice, the scaling of $P_\text{dl}$ with respect to $K$ makes the downlink system power-limited for large $K$. 
Consequently, the second scaling law in \eqref{eq:intro-4} is more feasible in uplink rather than in downlink. Furthermore, the result does not mean that spatial multiplexing per cell is represented by $(1-{T_c}^{-1}) K$ since we only have $\log(\alpha\snr)$ with $\alpha$ vanishing in the limit of $M$.

\section{An Extension of Deterministic Equivalents}
\label{sec:ED}

As mentioned earlier, it is essential to extend Theorem 1 in \cite{Cou11a,Wag12} to the case  of \eqref{eq:SM-3c} and \eqref{eq:SM-32a}, where the spectral norm of $\Lambdam_{\ell\ell_k}$ grows without bound as $M\rightarrow \infty$. 
In this section, we develop the following extension of deterministic equivalents based on Corollary \ref{corol-a1} in Appendix \ref{app-1}. 

\begin{thm} \label{thm-a1}
\normalfont Let  
\begin{align} \label{eq:ED-1}
     \pB_N = \pX_N \pX_N^\ct + \pA_N
\end{align}
where 
\begin{enumerate}[1)]
\item $N , N_1,\ldots,N_n$, $n$, and $M_0,\ldots,M_n$ grow with ratios $a_0 = N/n$, $a_i = N/N_i$, and $b_i=N/M_i$, respectively, such that $0 < \lim\inf_N a_i \le \lim\sup_N a_i  < \infty $ and $0 \le \lim\inf_N b_i \le \lim\sup_N b_i \le 1, \forall i$, and $0 <\lim\inf_N\frac{b_i}{b_j}\le \lim\sup_N\frac{b_i}{b_j}<\infty, \forall i\neq j$;
\item $\pX_N \in \mathbb{C}^{N\times n}$ has independent columns $\px_i = \Psim_i\py_i$, where $\py_i \in \mathbb{C}^{N_ i}$ has either i.i.d. entries of zero mean, variance $\frac{1}{N}$, and $4 + \epsilon$ moment of order $\Oc(\frac{1}{N^{2+\epsilon/2}})$, or uncorrelated entries  of zero mean, variance $1/n$, $|y_{n,i}|^2=1/n,$ where $y_{n,i}$ is the $i$th entry of $\py_n$ and $i=1,\ldots,n,$ and eighth order moment of order $\Oc(\frac{1}{n^4})$;
\item $\Psim_i \in \mathbb{C}^{N\times N_ i}$ are deterministic and the spectral norm of $\Thetam_i=\Psim_i\Psim_i^\ct$ is not necessarily uniformly bounded with respect to $N$, but there exist $M_i\ge N, i=1,\ldots,n$ satisfying 
$$\limsup_{N\rightarrow \infty} \|b_i{\Thetam}_i\|_2 < \infty, \forall i ;$$
\item $\pA_N \in \mathbb{C}^{N\times N}$ is non-negative Hermitian;
\item and let $\pQ_N \in \mathbb{C}^{N\times N}$ be non-negative Hermitian with not necessarily  uniformly bounded spectral norm with respect to $N$, but there exist $M_0\ge N$ satisfying $\limsup_{N\rightarrow \infty} \|b_0{\pQ}_N\|_2 < \infty$. 
\end{enumerate}
Define $$m_N(z)\triangleq \frac{1}{N}\trace\;\pQ_N\big(\pB_N -z\pI_N\big)^{-1}$$ and $\beta_i\triangleq\frac{M_i}{M_\text{max}}$ with $M_\text{max}=\min_{i=0,1,\ldots,n}M_i$.

For $z\in \mathbb{C}\setminus\mathbb{R}^+$, as $N\rightarrow\infty$ with ratios $a_i$, $b_i$,  and $\beta_i$,$\forall i$, we have 
\begin{align} \label{eq:ED-2}
     \frac{1}{\beta_0} m_N(z) - \frac{1}{\beta_0 N}\trace\;{\pQ}_N\pT_N(z) \overset{a.s.}{\longrightarrow} 0
\end{align}
where 
\begin{align} \label{eq:ED-3b}
\pT_N(z)=\left(\frac{1}{N}\sum_{j=1}^n\frac{{\Thetam}_j}{1+e_{N,j}(z)}+\pA_N-z\pI_N\right)^{-1}
\end{align}
and $e_{N,1}(z),\ldots,e_{N,n}(z)$ form the unique functional solution of
\begin{align} \label{eq:ED-3}
     e_{N,i}(z)= \frac{1}{\beta_i N}\trace\;{\Thetam}_i\pT_N(z) 
\end{align}
which is the Stieltjes transform of a nonnegative finite measure on $\mathbb{R}^+$ and given by $e_{N,i}(z)=\lim_{t\rightarrow\infty} e_{N,i}^{(t)}(z)$ where $e_{N,j}^{(0)}(z)= -\frac{1}{z}$ and for $t\ge 1$
\begin{align} \label{eq:ED-4}
     e_{N,i}^{(t)}(z)= \frac{1}{\beta_iN}\trace\;{\Thetam}_i\left(\frac{1}{N}\sum_{j=1}^n\frac{{\Thetam}_j}{1+e_{N,j}^{(t-1)}(z)}+\pA_N-z\pI_N\right)^{-1} .
\end{align}
\end{thm}

\begin{IEEEproof}
See Appendix \ref{app-4}.
\end{IEEEproof}


It is clear that if all the related matrices have uniformly bounded spectral norm, the above result reduces to Theorem 1 in \cite{Cou11a,Wag12}. The contribution of our result is that we relax the condition of uniformly bounded spectral norms into the more general conditions 1) and 3). Accordingly, the method of deterministic equivalents is still valid and hence we could obtain the capacity scaling law in the previous section.


It is pointed out that, similar to Theorem \ref{thm-a1}, we can naturally extend other results requiring random or deterministic matrices with uniformly bounded spectral norm in large random matrix theory. 
For instance, a generalization of \cite[Thm. 2]{Hoy13} (see also \cite[Thm. 2]{Wag12}) is given next and will be used in this work. 

\begin{thm} \label{thm-4} \normalfont
Let $\Omegam_N \in \mathbb{C}^{N\times N}$ be non-negative Hermitian with not necessarily  uniformly bounded spectral norm with respect to $N$, but there exist $M_c\ge N$ satisfying $\limsup_{N\rightarrow \infty} \|\frac{N}{M_c}\Omegam_N\|_2 < \infty$, and also let $\vartheta\triangleq\frac{M_0M_c}{M_\text{max}^2}$ and $\vartheta_{jk}\triangleq\frac{M_jM_k}{M_\text{max}^2}$. Under the same conditions as in Theorem \ref{thm-a1}, we have 
\begin{align} \label{eq:ED-6}
     \frac{1}{\vartheta N} \trace\;\pQ_N\big(&\pB_N -z\pI_N\big)^{-1}\Omegam_N\big(\pB_N -z\pI_N\big)^{-1} \nonumber \\ 
     &- \frac{1}{\vartheta N}\trace\;{\pQ}_N\pT'_N(z) \xrightarrow[N\rightarrow\infty]{a.s.} 0
\end{align}
where $\pT'_N(z)$ is defined as
\begin{align}
\pT'_N(z)=&\;\pT_N(z)\Omegam_N\pT_N(z) \nonumber \\ &+ \pT_N(z)\frac{1}{N}\sum_{j=1}^n\frac{{\Thetam}_j e'_{N,j}(z)}{(1+e_{N,j}(z))^2}\pT_N(z) \nonumber
\end{align}
with  $\pe'(z)=[e'_{N,1}(z),\ldots,e'_{N,n}(z)]$ is given by 
\begin{align} 
     \pe'(z)= \big(\pI_n-\pJ(z)\big)^{-1}\pv(z) \nonumber
\end{align}
where $\pJ(z)$ and $\pv(z)$ are defined as
\begin{align} 
     [\pJ(z)]_{jk}&= \frac{1}{\vartheta_{jk}N}\frac{\trace\;{\Thetam}_j \pT_N(z) {\Thetam}_k \pT_N(z)}{N(1+e_{N,j}(z))^2}, \forall j,k \nonumber \\
     [\pv(z)]_j&= \frac{1}{\vartheta N}{\trace\;{\Thetam}_j \pT_N(z) \Omegam_N\pT_N(z)}, \forall j\nonumber .
\end{align}
\end{thm}

Finally, the above result directly applies to the random partial Fourier model, in which i.i.d. sequences are replaced with uncorrelated constant modulus sequences defined in Corollary \ref{corol-a0}.

\section{Proofs of Mains Results}
\label{sec:PF}

We first derive a deterministic equivalent of the SINR $\gamma_{\text{ul},\ell_k}^\mmse$ of the low-dimensional MMSE combining in uplink. Our analysis is based on \eqref{eq:SM-5} to highlight the role of spatial despreading. 
Based on Theorems \ref{thm-a1} and \ref{thm-4}, the deterministic equivalent $\bar{\gamma}_{\text{ul},\ell_k}^\mmse$ of the SINR $\gamma_{\text{ul},\ell_k}^\mmse$ of the MMSE detector is given by the following result.



\begin{lem}\label{thm-1} \normalfont
Under Assumption \ref{as-2} and the orthogonal pilot scheme in Subsection \ref{sec:SC1}, we have $\gamma_{\text{ul},\ell_k}^\mmse \xrightarrow[M\rightarrow\infty]{} \bar{\gamma}_{\text{ul},\ell_k}^\mmse$, where $\bar{\gamma}_{\text{ul},\ell_k}^\mmse$ is given by 
\begin{align} \label{eq:MR-9b}
   \bar{\gamma}_{\text{ul},\ell_k}^\mmse &=   \frac{\delta_{\ell_k}^2}{\frac{1}{{P_\text{ul}}M}\mu_{\ell_k} +\sum_{\ell'\neq\ell} \alpha_{\ell\ell_k}^2{|\nu_{\ell\ell'_k} |^2}  +\frac{1}{M}\sum_{\ell'\neq\ell, k'\neq k}\mu_{\ell_k{\ell'}\!\!_{k'}}}  
\end{align}
with 
$   \delta_{\ell_k} =  \frac{1}{M}\trace\; {\Phim}_{\ell\ell_k}\pT_{\ell_k}  , \;
   \mu_{\ell_k} = \frac{1}{M}\trace\; {\Phim}_{\ell\ell_k} \pT'_{\ell_k} , \;
   \nu_{\ell\ell'_k} = \frac{1}{r_{\ell\ell_k}M} \trace\;\Lambdam_{\ell\ell'_k} \Xim_{\ell\ell_k}\Lambdam_{\ell\ell_k} \pT_{\ell_k} , \;
   \mu_{\ell_k{\ell'}\!\!_{k'}} = \frac{\trace\Lambdam_{\ell{\ell'}\!\!_{k'}}}{M}\;\mu_{\ell_k}   \nonumber 
$, where 
\begin{enumerate}[1)]
\item  $\pT_{\ell_k}=\pT_N(\frac{1}{{P_\text{ul}}{r_{\ell\ell_k}}})$ are given by Theorem \ref{thm-a1} with $N={r_{\ell\ell_k}}$, $M_k=M, b_k=\frac{1}{M}$ (i.e., $\beta_k=1$), $\pQ_N ={r_{\ell\ell_k}}{\Phim}_{\ell\ell_k}$, $\Thetam_k={r_{\ell\ell_k}}{\Phim}_{\ell\ell_k}$, $\pA_N=\frac{1}{{r_{\ell\ell_k}}}\pz_{\ell_k}$, and $e_k=\delta_{\ell_k}, \forall i$; 
\item $\pT'_{\ell_k}=\pT'_N(\frac{1}{{P_\text{ul}}{r_{\ell\ell_k}}})$ is given by Theorem \ref{thm-4} with $\Omegam_N=\pI_{r_{\ell\ell_k}}, \forall i$.
\end{enumerate}
\end{lem}

\begin{IEEEproof} 
A high-level sketch of the proof is as follows. Given the random realization of $\{\pU_{\ell\ell'_{k}}\}$, the SINR $\gamma_{\text{ul},\ell_k}^\mmse$ converges to a conditional random variable that does not depend on the particular realization of effective channels $\{\pw_{\ell\ell'_{k}}\}$ in the limit of $M$.  Using Lemma \ref{lem-a10}, we can show that the conditional random variable in fact converges to a \emph{deterministic} limit independent of the specific covariance realization under the distributions of $\pU_{\ell\ell'_{k}}$ in Section \ref{sec:CM}. For the details, see Appendix \ref{app-2}. 
\end{IEEEproof}

Notice that $\bar{\gamma}_{\text{ul},\ell_k}^\mmse$ is a limiting value that does not depend on the particular realization of small-scale fading channels $\{\pw_{\ell\ell'_{k}}\}$ in the limit of $M$, but conditional to the random realization of $\{\pU_{\ell\ell'_{k}}\}$ through $\Xim_{\ell\ell_k}$ defined in \eqref{eq:Pre-3} and also through ${\Phim}_{\ell\ell_k}$. 

Using the standard technique in the literature  \cite{Sil95,Bai98,Bai07,Cou11a,Cou11b,Wag12,Hoy13} and some new tools herein, the above result can be naturally extended to other cases such as other lower bounds and receive algorithms like the multicell MMSE detector. We skip the details because our focus is the asymptotic capacity scaling of massive MIMO rather than deterministic approximation of different receive algorithms.

\subsection{Proof of Theorem \ref{thm-2}}
\label{app-3}

We begin with the brief derivation of \eqref{eq:MR-9} using the standard arguments in \cite{Lap02,Med00,Has03,Cai10}. 
Given $\sv_{\ell}=[\sv_{\ell_1},\ldots,\sv_{\ell_{K_\ell}}]$ in the training phase and conditioned on a realization of $\underline{\pU}$ on the channel second-order statistics of all users in the network, the individual rate $\Rc_{\text{ul},\ell_k}$ of user $\ell_k$ can be lower-bounded by  
\begin{align} 
   \Rc_{\text{ul},\ell_k} &= I(x_{\ell_k};\py_{\ell},\sv_{\ell} |\underline{\pU}) \nonumber \\
   &\ge  h(x_{\ell_k})- \mathbb{E} \Big[\log\Big(\pi e\; \mathbb{E} \big[|x_{\ell_k}-\xi\pv_{\ell_k}^\ct \py_{\ell_k}|^2\big|\; \hat{\pw}_{\ell}, \underline{\pU} \big]\Big)\Big] \nonumber
\end{align}
where we used the data processing inequality, the independence of $x_{\ell_k}$ and $(\hat{\pw}_{\ell},\underline{\pU})$, and the fact that Gaussian distribution maximizes the conditional distribution for a given covariance. 

The received signal vector $\py_{\ell_k}$ in (\ref{eq:SM-5}) of user $\ell_k$ can be rewritten as
\begin{align} 
   \py_{\ell_k} &= \hat{\pw}_{\ell\ell_k}x_{\ell_k} +{{\pn}_{\ell\ell_k}x_{\ell_k} +\sum_{(\ell',k')\neq(\ell,k)}\pw_{\ell_k{\ell'}\!\!_{k'}}x_{\ell_k{\ell'}\!\!_{k'}} +\pz_{\ell_k}}. \nonumber
\end{align} 
Based on the fact that $\hat{\pw}_{\ell\ell_k}x_{\ell_k}$ and $\pz'_{\ell\ell_k}={\pn}_{\ell\ell_k}x_{\ell_k} +\sum_{(\ell',k')\neq(\ell,k)}\pw_{\ell_k{\ell'}\!\!_{k'}}x_{\ell_k{\ell'}\!\!_{k'}} +\pz_{\ell_k}$  are uncorrelated, the worst-case uncorrelated noise argument \cite{Lap02,Has03} implies that the worst-case $\pz'_{\ell\ell_k}$ is Gaussian with variance of $\mathbb{E} [{\pz'_{\ell\ell_k}}{\pz'_{\ell\ell_k}}^{\ct}]$. 
{Letting $\xi\pv_{\ell_k}^\ct \py_{\ell_k}$ equal to the linear MMSE estimate of $x_{\ell_k}$ given $\pv_{\ell_k}^\ct \py_{\ell_k}$ and $(\hat{\pw}_{\ell}, \underline{\pU})$ to minimize $\mathbb{E} \big[|x_{\ell_k}-\xi\pv_{\ell_k}^\ct \py_{\ell_k}|^2\big|\; \hat{\pw}_{\ell}, \underline{\pU} \big]$ and using 
the orthogonality principle, for any $\pv_{\ell_k}$, the achievable rate of user $\ell_k$ is lower-bounded by \eqref{eq:MR-9}.}

For the MMSE detector in (\ref{eq:MR-15}), we now use the deterministic equivalent of the corresponding SINR in Lemma \ref{thm-1}. 
Let $\varphi_{\text{ul}}=P_\text{ul}M<\infty$. 
As $M\rightarrow\infty$, we have  $N={r_{\ell\ell_k}}$, $M_k=M, b_k=\frac{1}{M}$ (i.e., $\beta_k=1$), $\pQ_N ={r_{\ell\ell_k}}{\Phim}_{\ell\ell_k}$, $\Thetam_k={r_{\ell\ell_k}}{\Phim}_{\ell\ell_k}$, $\pA_N=\frac{1}{{r_{\ell\ell_k}}}\pz_{\ell_k}$, and $e_k=\delta_{\ell_k}, \forall i$. Based on Lemma \ref{lem-a10}, we can further obtain the following asymptotic equivalence   
\begin{align} \label{eq:MR-9c}
\Xim_{\ell\ell_k} \simeq \bigg(\Lambdam_{\ell\ell_k} +  \Big(\sum_{\ell'\neq\ell} \frac{\trace\Lambdam_{\ell{\ell'}\!\!_{k}}}{M} +\rho_p^{-1}\Big)\pI_{r_{\ell\ell_k}}\bigg)^{-1} \simeq \Lambdam_{\ell\ell_k}^{-1}
\end{align} 
where the first step follows from Lemmas \ref{lem-1b} and the second step is due to Assumption \ref{as-3} (i.e., the eigenvalues of $\Lambdam_{\ell\ell_k}$ grow without bound as $M\rightarrow\infty$) with finite $L$ and $M$. 

From \eqref{eq:ED-3b}, we have 
\begin{align} \label{eq:MR-9d}
\pT_{\ell_k}^{-1}&= \frac{1}{r_{\ell\ell_k}}\sum_{j=1}^{K_\ell}\frac{{r_{\ell\ell_j}}{\Phim}_{\ell\ell_j}}{1+\delta_{\ell\ell_j}(\varphi_{\text{ul}}^{-1})}+\frac{1}{{r_{\ell\ell_k}}}\pz_{\ell_k}-\varphi_{\text{ul}}^{-1}\pI_{r_{\ell\ell_k}}\nonumber \\
  &\overset{(a)}{\simeq} \frac{{\Phim}_{\ell\ell_k}}{1+\delta_{\ell_k}(\varphi_{\text{ul}}^{-1})}+\frac{1}{{r_{\ell\ell_k}}}\pz_{\ell_k}-\varphi_{\text{ul}}^{-1}\pI_{r_{\ell\ell_k}} \nonumber \\
  &\overset{(b)}{\simeq} \frac{\Lambdam_{\ell\ell_k}}{1+\delta_{\ell_k}(\varphi_{\text{ul}}^{-1})}+\frac{1}{r_{\ell\ell_k}}\Lambdam_{\ell\ell_k}^{-1}-\varphi_{\text{ul}}^{-1}\pI_{r_{\ell\ell_k}}\nonumber \\
  &\overset{(c)}{\simeq} \varphi_{\text{ul}}^{-1}\pI_{r_{\ell\ell_k}}  
\end{align} 
where $(a)$ follows from the fact that $\Phim_{\ell\ell_j}= \Lambdam_{\ell_j\ell_k}\Xim_{\ell\ell_j}\Lambdam_{\ell_j\ell_k} \simeq 0$ as $\alpha_{\ell\ell_j}\rightarrow0$ for $j\neq k$ by Lemma \ref{lem-1b}, where $\Lambdam_{\ell_j\ell_k}$ is defined in \eqref{eq:Pre-3}, $(b)$ follows from \eqref{eq:MR-9c}, and in $(c)$ we used $\delta_{\ell_k}(\varphi_{\text{ul}}^{-1})\simeq\frac{P_\text{ul}}{r_{\ell\ell_k}}\trace^2\Lambdam_{\ell\ell_k}$ through \eqref{eq:ED-4}.
Similarly, we can have 
\begin{align} \label{eq:MR-9e}
\pT'_{\ell_k}&= \pT_{\ell_k}\pT_{\ell_k} +\pT_{\ell_k}\frac{1}{r_{\ell\ell_k}}\sum_{j=1}^{K_\ell}\frac{{r_{\ell\ell_j}}{\Phim}_{\ell\ell_j} \delta'_{\ell\ell_j}(\frac{1}{{P_\text{ul}}{r_{\ell\ell_j}}})}{(1+\delta_{\ell\ell_j}(\frac{1}{{P_\text{ul}}{r_{\ell\ell_j}}}))^2}\pT_{\ell_k} \nonumber \\ &\simeq\varphi_{\text{ul}}^{2}\pI_{r_{\ell\ell_k}}  .
\end{align} 
Substituting \eqref{eq:MR-9c}--\eqref{eq:MR-9e} into \eqref{eq:MR-9b},  \eqref{eq:MR-32} 
\ifdefined\singcol 
  \begin{IEEEeqnarray}{ll}
   \bar{\gamma}_{\text{ul},\ell_k}^\mmse
   &\simeq   \frac{\big(\frac{\varphi_{\text{ul}}}{M}\trace\Lambdam_{\ell\ell_k}\big)^2}{\frac{\varphi_{\text{ul}}}{M}\trace\Lambdam_{\ell\ell_k}\!\!+\!\!\sum_{\ell'\neq\ell} \alpha_{\ell\ell_k}^2{\big(\frac{\varphi_{\text{ul}}}{M}\trace\Lambdam_{\ell\ell'_k}\big)^2}  \!\!+\!\frac{1}{M}\sum_{\ell',k'}\frac{\varphi_{\text{ul}}}{M^2}\trace\Lambdam_{\ell\ell'_k}\trace\Lambdam_{\ell\ell_k}} \nonumber \\
   &\simeq  \frac{\varphi_{\text{ul}}}{M}\trace\Lambdam_{\ell\ell_k}  \nonumber \\
   &=\ P_\text{ul}\trace\Lambdam_{\ell\ell_k}.     \label{eq:MR-32}
  \end{IEEEeqnarray}
\else
\addtocounter{equation}{1}%
\setcounter{storeeqcounter}%
 {\value{equation}}%
on the top of Page~\pageref{eq:MR-32} immediately follows under Assumption \ref{as-3}.   

\begin{figure*}[!t]
  \normalsize
  \setcounter{tempeqcounter}{\value{equation}} 
  \begin{IEEEeqnarray}{ll}
\setcounter{equation}{\value{storeeqcounter}} 
   \bar{\gamma}_{\text{ul},\ell_k}^\mmse
   &\simeq   \frac{\big(\frac{\varphi_{\text{ul}}}{M}\trace\Lambdam_{\ell\ell_k}\big)^2}{\frac{\varphi_{\text{ul}}}{M}\trace\Lambdam_{\ell\ell_k}\!\!+\!\!\sum_{\ell'\neq\ell} \alpha_{\ell\ell_k}^2{\big(\frac{\varphi_{\text{ul}}}{M}\trace\Lambdam_{\ell\ell'_k}\big)^2}  \!\!+\!\frac{1}{M}\sum_{\ell',k'}\frac{\varphi_{\text{ul}}}{M^2}\trace\Lambdam_{\ell\ell'_k}\trace\Lambdam_{\ell\ell_k}} \simeq  \frac{\varphi_{\text{ul}}}{M}\trace\Lambdam_{\ell\ell_k}  =\ P_\text{ul}\trace\Lambdam_{\ell\ell_k}.     \label{eq:MR-32}
  \end{IEEEeqnarray}
  \setcounter{equation}{\value{tempeqcounter}} 
  \hrulefill
  \vspace*{4pt}
\end{figure*}

\fi


The corresponding ergodic achievable rate $\Rc_{\text{ul},\ell_k}$ of user $\ell_k$ is given by 
\begin{align} \label{eq:MR-13}
   \Rc_{\text{ul},\ell_k} 
   &\ge \mathbb{E} \Big[\log\big(1+\gamma_{\text{ul},\ell_k}^\mmse \big)\big| \underline{\pU} \Big]  \\
   &\simeq  \log\big(1+\bar{\gamma}_{\text{ul},\ell_k}^\mmse\big)  \nonumber 
\end{align}
where the expectation is over $\{\pw_{\ell{\ell'}\!\!_{k}}\}$, and we used the dominated convergence theorem. Although the RHS of \eqref{eq:MR-13}  is conditional to the realization of $ \underline{\pU}$, $\Rc_{\text{ul},\ell_k}$ eventually converges to the deterministic limit by Lemma \ref{lem-a10}. The prelog factor  in \eqref{eq:MR-1} of scheduled users is simply given by finding $\kappa'$ that maximizes multiplexing gain in the quadratic function $(1-\frac{\kappa'}{T_c}){\kappa'}$ for cell $\ell$.
This completes the achievability.

\begin{rem} \normalfont
One can make an important observation from \eqref{eq:MR-32} on the role of spatial despreading. The coherent  pilot contamination term eventually approximated as a non-zero finite deterministic value $\nu_{\ell\ell'_k}$ multiplied by the ratio ${\alpha_{\ell\ell_k}}$ vanishes under the sublinear sparsity in Assumption \ref{as-3}. 
{ Due to its ``non-coherent" nature (i.e., $\ph_i ^\ct\ph_j$ terms with $i\neq j$), the interference term disappears much faster than pilot contamination through spatial despreading. For illustrative purposes, suppose that $r=\sqrt{M}$ in the homogeneous network. When $M=100$, \eqref{eq:MR-32} shows that spatial despreading can suppress pilot contamination and interference in the uplink by a factor of $\zeta=10$ in \eqref{eq:SM-10}. Meanwhile, $r=10$ is not large enough to see the channel hardening effect including asymptotic orthogonality of small-scale or fast fading channel components of different users by the law of large number.} 
Therefore, spatial despreading plays the central role to eliminate pilot contamination and interference. This is even prominent when one cannot take advantage of channel hardening due to small $r_{\ell\ell'_k}$, as will be also shown through numerical results in Sec. \ref{sec:NR}.
\end{rem}

\subsection{Proof of Theorem \ref{thm-3}}
\label{app-5}

It suffices to consider the matched filter (MF) receiver to derive the asymptotic capacity scaling in \eqref{eq:MR-1b}, although the MF would perform inferior to the MMSE receiver at finite $M$. This is because the effect of spatial despreading remains essentially the same regardless of MF and MMSE. 
We accordingly set $\pv_{\ell_k}=\check{\pw}_{\ell\ell_k}$, where $\check{\pw}_{\ell\ell_k}$ is given by \eqref{eq:SC2-3}, and denote the resulting SINR by ${\gamma}_{\text{ul},\ell_k}^\text{MF}$.

For the MF receiver, it is straightforward to show that 
\begin{align} 
   \frac{1}{M^2}\big|\mathbb{E}[\check{\pw}_{\ell\ell_k}^\ct{\pw}_{\ell\ell_k}\big|\underline{\pU}]\big|^2 &\overset{(a)}{\simeq} \bigg(\frac{1}{M} \trace\Phim'_{\ell\ell_k}\bigg)^2\nonumber  \\
   \frac{1}{M^2}\var[\check{\pw}_{\ell\ell_k}^\ct{\pw}_{\ell\ell_k}\big|\underline{\pU}] &\simeq 0 \nonumber 
\end{align}
where $(a)$ follows from Corollary \ref{corol-a1} and Lemma \ref{lem-2}. In what follows, we first consider the random partial unitary model for $\pU_{\ell\ell_k}$ and focus on the dominating components in the interference power term $\sum_{(\ell',k')\neq(\ell,k)}\frac{1}{M^2}\mathbb{E} \big[|\check{\pw}_{\ell\ell_k}^\ct\pw_{\ell_k{\ell'}\!\!_{k'}}|^2\big|\underline{\pU}\big] $ that incur the effect of pilot contamination. 
From  
\begin{align} 
   \mathbb{E} \big[|\check{\pw}_{\ell\ell_k}^\ct\pw_{\ell_k{\ell'}\!\!_{k'}}|^2 \big|\underline{\pU}\big] 
   = &\; \mathbb{E} \bigg[\Big|\big(\pw_{\ell\ell_k} +\sum_{(j,m)\neq(\ell,k)}\pw_{\ell_k{j}_{m}} + \nonumber \\ &\ {\rho_p}^{-1/2}\pz_{\ell_k}\big)^\ct \Xim'_{\ell\ell_k}\Lambdam_{\ell\ell_k}\pw_{\ell_k{\ell'}\!\!_{k'}}\Big|^2\Big|\underline{\pU}\bigg]   \nonumber
\end{align}
we extract the pilot contamination component (i.e., ${(j,m)=(\ell',k')}$), which leads to  
\begin{align} \label{eq:NC-3}
   \mathbb{E} \big[|&\pw_{\ell_k{\ell'}\!\!_{k'}}^\ct\Xim'_{\ell\ell_k}\Lambdam_{\ell\ell_k}\pw_{\ell_k{\ell'}\!\!_{k'}}|^2\big|\underline{\pU}\big] \nonumber \\
   &= \mathbb{E} \big[|\pw_{\ell{\ell'}\!\!_{k'}}^\ct {\pU_{\ell{\ell'}\!\!_{k'}}^\ct \pU_{\ell\ell_k}\Xim'_{\ell\ell_k}\Lambdam_{\ell\ell_k} \pU_{\ell\ell_k}^\ct\pU_{\ell{\ell'}\!\!_{k'}}} \pw_{\ell{\ell'}\!\!_{k'}}|^2\big|\underline{\pU}\big] \nonumber \\
   &\overset{(a)}{\simeq} \frac{1}{M^2}\mathbb{E} \Big[ \big|\trace \big(\Xim'_{\ell\ell_k}\Lambdam_{\ell\ell_k}\big)\pw_{\ell{\ell'}\!\!_{k'}}^\ct \pU_{\ell{\ell'}\!\!_{k'}}^\ct\pU_{\ell{\ell'}\!\!_{k'}}\pw_{\ell{\ell'}\!\!_{k'}} \big|^2 \big|\underline{\pU}\Big] \nonumber \\
   &\simeq \frac{1}{M^2}\mathbb{E} \Big[ \big|\trace \big(\Xim'_{\ell\ell_k}\Lambdam_{\ell\ell_k}\big)\pw_{\ell{\ell'}\!\!_{k'}}^\ct\pw_{\ell{\ell'}\!\!_{k'}} \big|^2 \Big] \nonumber \\
   &\overset{(b)}{\simeq}  \frac{1}{M^2}\Big(\trace\Lambdam_{\ell{\ell'}\!\!_{k'}}\Big)^2 \Big(\trace \Xim'_{\ell\ell_k}\Lambdam_{\ell\ell_k} \Big)^2\nonumber \\
   &=  {r^2_{\ell\ell_k}}\Big(\frac{1}{M}\trace\Lambdam_{\ell{\ell'}\!\!_{k'}} \frac{1}{r_{\ell\ell_k}}\trace\Xim'_{\ell\ell_k}\Lambdam_{\ell\ell_k}\Big)^2 \nonumber \\
   &=  M^2\alpha^2_{\ell\ell_k}\psi^2_{{\ell'}\!\!_{k'}}
\end{align}
where $\psi_{{\ell'}\!\!_{k'}} = \frac{1}{r_{\ell\ell_k}M}\trace\Lambdam_{\ell{\ell'}\!\!_{k'}} \trace\;\Xim'_{\ell\ell_k}\Lambdam_{\ell\ell_k}$.
In $(a)$, we used the independence of $\pU_{\ell\ell_k}$ and $\pU_{\ell{\ell'}\!\!_{k'}}$ and the direct consequence of Lemma \ref{lem-1b} that $\pU_{\ell\ell_k}\Xim'_{\ell\ell_k}\Lambdam_{\ell\ell_k} \pU_{\ell\ell_k}^\ct=\frac{\trace \Xim'_{\ell\ell_k}\Lambdam_{\ell\ell_k}}{M}\pI_M$, and $(b)$ follows from the trace lemma. We can see that $0<\liminf_{M\rightarrow \infty}\psi^2_{{\ell'}\!\!_{k'}}\le \limsup_{M\rightarrow \infty}\psi^2_{{\ell'}\!\!_{k'}}<\infty$ under Assumption \ref{as-2}.

Neglecting the interference components vanishing in the limit of $M$, under Assumptions \ref{as-2} and \ref{as-3}, we have $${\gamma}_{\text{ul},\ell_k}^\text{MF} \xrightarrow[M\rightarrow\infty]{a.s.} \bar{\gamma}_{\text{ul},\ell_k}^\text{MF}$$
where 
\begin{align} \label{eq:NC-4}
   &\bar{\gamma}_{\text{ul},\ell_k}^\text{MF} =  \frac{(\frac{1}{M} \trace\Xim'_{\ell\ell_k})^2}{\frac{1}{P_\text{ul}M}\frac{1}{M} \trace\Xim'_{\ell\ell_k} +\sum_{(\ell',k')\neq(\ell,k)}\alpha^2_{\ell\ell_k}\psi^2_{{\ell'}\!\!_{k'}}} .
\end{align}
The above pilot contamination term $\sum_{(\ell',k')\neq(\ell,k)}\alpha^2_{\ell\ell_k}\psi^2_{{\ell'}\!\!_{k'}}$ goes to zero under Assumption \ref{as-3} because $K_\ell$, $L$, and $\psi_{{\ell'}\!\!_{k'}}$ are all finite. 
Similar to \eqref{eq:MR-9c}, we can then have a deterministic approximation of $\Xim'_{\ell\ell_k}$ and eventually have 
\begin{align} \label{eq:NC-6}
   \bar{\gamma}_{\text{ul},\ell_k}^\text{MF} 
   &\simeq P_\text{ul}\trace\Lambdam_{\ell\ell_k}. 
\end{align}
For the random partial Fourier model, it suffices to use a direct combination of Corollaries \ref{corol-a0} and \ref{corol-a1} and then Lemma \ref{lem-1e} instead of Lemma \ref{lem-1b}.

{For the converse proof, we use a simple cut-set upper bound \cite{Cov06} on the sum rate of the pilot-aided MIMO uplink, where a cut divides the BSs from the users. Inspired by \cite{Zhe02}, the work in \cite{Nam17} provided the capacity upper bound of a pilot-aided single-cell frequency-division duplex (FDD)  MIMO downlink system, in which the BS perfectly obtains CSIT through a delay-free and error-free feedback channel. Likewise, assuming that CSIR is perfectly acquired by the BS through a contamination-free and noise-free uplink training channel, 
the per-user (or per-link) achievable rate is upper-bounded by the  full CSI  capacity of SIMO channel with a single channel training cost over the coherence block $T_c$ such that  
\begin{align} \label{eq:NC-7}
   \Rc_{\text{ul},\ell_k} &\le \Big(1-\frac{1}{T_c}\Big)\mathbb{E} \Big[ \log\big(1+P_\text{ul}\ph_{\ell\ell_k}^\ct\ph_{\ell\ell_k}\big) \Big]\nonumber \\
   &= \Big(1-\frac{1}{T_c}\Big)\log P_\text{ul}\trace\Lambdam_{\ell\ell_k} +o(1) . 
\end{align}
Following the above footsteps and using the non-coherent communication argument in \cite{Zhe02}, one can easily show the same capacity scaling as \eqref{eq:MR-1b}.}




\subsection{Sketch of Achievability Proof of $\Rc_{\text{ul},\ell_k}^\text{(3)}$}
\label{sec:FP-C}

Following along the lines and arguments of Thoerems \ref{thm-2} and \ref{thm-3}, and noticing \eqref{eq:MR-1c}, it is straightforward to show that $\Rc_{\text{ul},\ell_k}^\text{ub}$ in  \eqref{eq:SC-1c} behaves as $(1-{T_c}^{-1})\log P_\text{ul}\trace\Lambdam_{\ell\ell_k} +o(1)$ . 
Since the first term $\Rc_{\text{ul},\ell_k}^\text{ub}$ in \eqref{eq:SC-1} is the upper bound of all three lower bounds in this paper as in \eqref{eq:MR-1c}, 
we only need to see if 
\begin{align} \label{eq:FP-12}
   \var[\pv_{\ell_k}^\ct\pw_{\ell_k{\ell'}\!\!_{k'}}| \underline{\pU}] = &\; \mathbb{E}\big[|\pv_{\ell_k}^\ct\pw_{\ell_k{\ell'}\!\!_{k'}}|^2\big| \underline{\pU}\big] - \big|\mathbb{E}[\pv_{\ell_k}^\ct\pw_{\ell_k{\ell'}\!\!_{k'}}| \underline{\pU}] \big|^2 \nonumber \\ & \xrightarrow[M\rightarrow\infty]{} 0 .
\end{align}
With $\pv_{\ell_k}=\check{\pw}_{\ell\ell_k}$, the deterministic approximation of $\mathbb{E} \big[|\check{\pw}_{\ell\ell_k}^\ct\pw_{\ell_k{\ell'}\!\!_{k'}}|^2 \big|\underline{\pU}\big] $ is given in \eqref{eq:NC-3}. 
Following the same steps as \eqref{eq:NC-3} and using that 
\begin{align} 
   \pw_{\ell_k{\ell'}\!\!_{k'}}^\ct\Xim'_{\ell\ell_k}&\Lambdam_{\ell\ell_k}\pw_{\ell_k{\ell'}\!\!_{k'}}  
   \nonumber\\ &= \pw_{\ell{\ell'}\!\!_{k'}}^\ct {\pU_{\ell{\ell'}\!\!_{k'}}^\ct \pU_{\ell\ell_k}\Xim'_{\ell\ell_k}\Lambdam_{\ell\ell_k} \pU_{\ell\ell_k}^\ct\pU_{\ell{\ell'}\!\!_{k'}}} \pw_{\ell{\ell'}\!\!_{k'}}\nonumber \\
   &\simeq \frac{1}{M}    \trace \big(\Xim'_{\ell\ell_k}\Lambdam_{\ell\ell_k}\big)\pw_{\ell{\ell'}\!\!_{k'}}^\ct\pw_{\ell{\ell'}\!\!_{k'}}  \nonumber
\end{align}
where we used Lemma \ref{lem-1b} (or \ref{lem-1e}) for the random partial unitary model (or partial Fourier model), we have $\check{\pw}_{\ell\ell_k}^\ct\pw_{\ell_k{\ell'}\!\!_{k'}} \xrightarrow[]{a.s.} M\alpha_{\ell\ell_k}\psi_{{\ell'}\!\!_{k'}}$.
Applying the continuous mapping theorem and the dominated convergence theorem gives rise to \eqref{eq:FP-12}. This concludes that $\Rc_{\text{ul},\ell_k}^\text{(3)}$ achieves the scaling law in \eqref{eq:MR-1b}.

\section{Numerical Results}
\label{sec:NR}

For numerical examples in this section, we only consider the homogeneous scenario that has $L$ cells serving $K$ users each with inter-cell interference factor $\iota={0.2}$ and use the single-cell MMSE combining or precoding.  The symmetric geometry of users is assumed such that we normalize channel covariance matrices to satisfy $\trace{\pR_{\ell\ell'_k}}=M$  for all $(\ell,\ell',k)$ and $r_{\ell\ell_k}=r$   for all $(\ell,k)$. We assume that the number $r_{\ell\ell'_k}$ of non-zero eigenvalues  of the covariance matrices of channels from other cells $\ell'\neq \ell$ is equal to $r/2$ to take into account the fact that spatial correlation depends on the distance between sender and receiver. The larger the distance, the smaller multipath components survive after several specular reflections and diffusion. We used pilot power gap (boosting) of ${\varrho_p}=2$ (i.e., 3 dB). The random partial Fourier correlation model in Sec. \ref{sec:CM} is only considered. Under the random partial unitary correlation model, the sum-rate performance shows degradation compared to the Fourier model, since the eigenmodes of channel covariance matrices of different users are not orthogonal but linearly independent of each other. 
In what follows, we evaluate all uplink and downlink lower bounds in this paper for finite $M$ and SNR to verify the asymptotic sum-rate scaling results.

\begin{figure}
\hspace{-6mm}  
  \includegraphics[scale=.5]{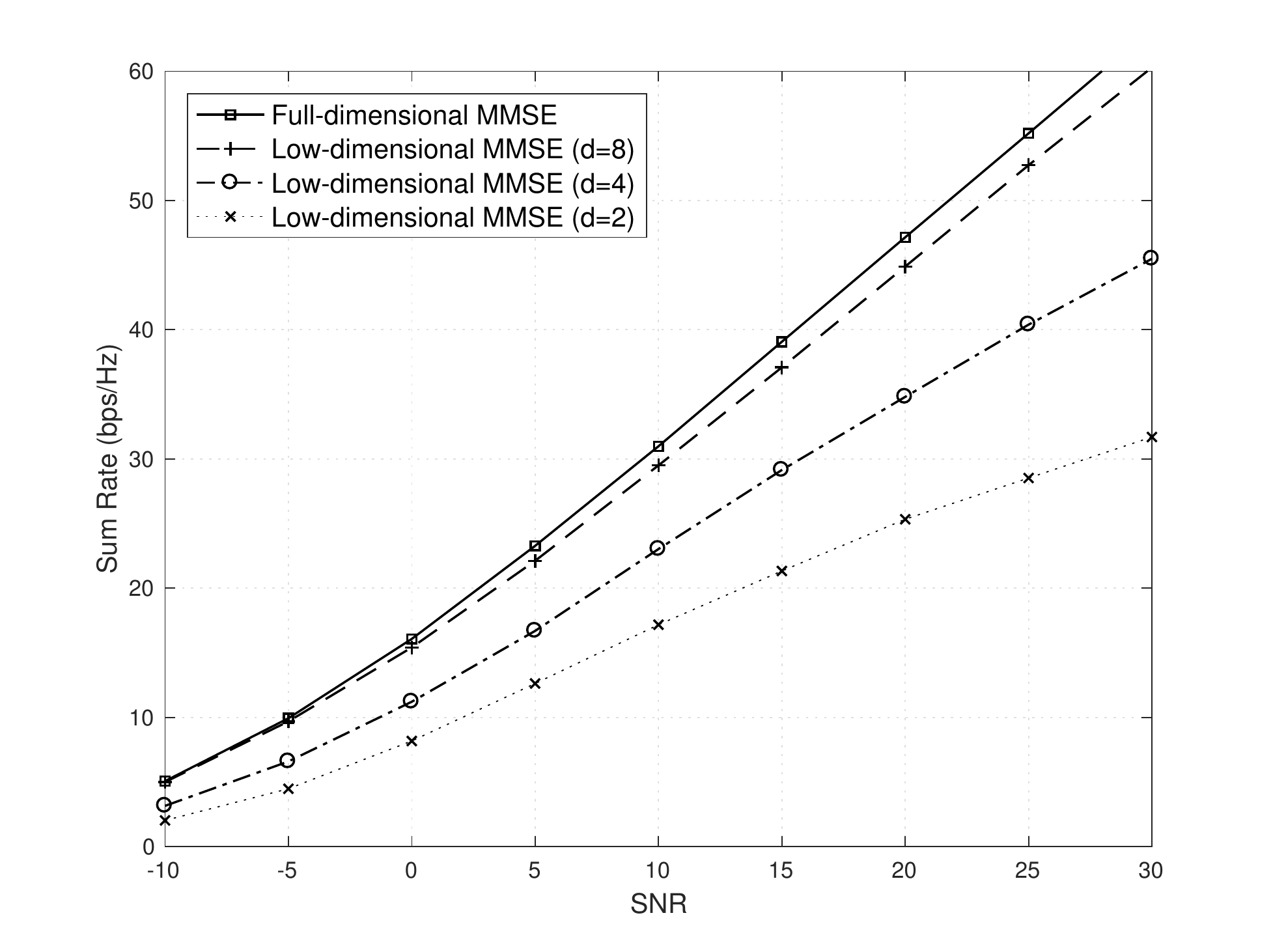}
  \caption{Sum-rate curves of different precoding schemes versus SNR (dB) in MIMO downlink, where $M=100,  L=4, K=5, r=8, T_c=500$. The random partial Fourier model in \eqref{eq:CM-1} was used for channel covariance matrices. For low-dimensional MMSE precoding, we used \eqref{eq:MR-15b}. }\label{fig-2}
\end{figure}

In Fig. \ref{fig-2}, we investigate the lower bound on the per-cell achievable rate  in \eqref{eq:SCD-3} of different MMSE precoding schemes in downlink, where $r=8$. While the `full-dimensional MMSE' represents the conventional $M$-dimensional MMSE channel estimation and precoding scheme based on $\bar{\sv}_{\ell_k}$ in \eqref{eq:CM-3} without spatial (de)spreading, the `low-dimensional MMSE' is given by \eqref{eq:Pre-3} and \eqref{eq:MR-15b}, where $d$ indicates the number of Fourier coefficients used out of $r$ ones under the Fourier correlation model for spatial (de)spreading. The $8$-dimensional ($d=8$) channel estimation and precoding are shown to cause only graceful performance degradation\footnote{ In a dense user scenario, e.g., $K=20$, the performance degradation will be more significant especially at high SNR.}, compared to the $100$-dimensional vector operation. However, significant performance loss is observed when we only utilize $d(<r)$ Fourier coefficients uniform randomly chosen from the $r$ elements  of angular support that have equal path gain. This implies that one should make use of as many  dominant elements of angular support as possible for spatial (de)spreading in practical systems to realize the capacity scaling.


\begin{figure*}
\vspace{-3mm}
\center  
  \includegraphics[scale=.7]{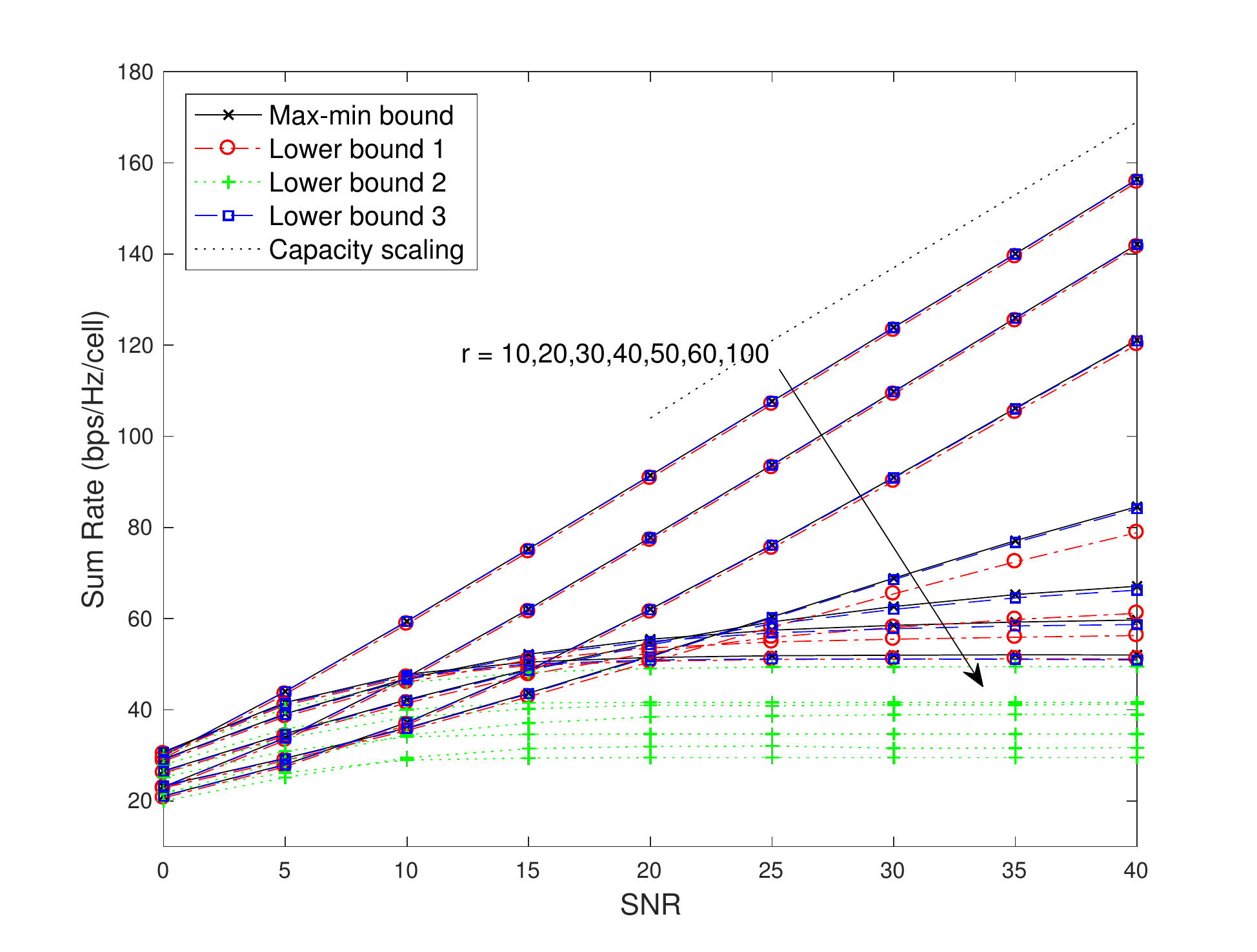}
\vspace{-5mm}
  \caption{The impact of spatial despreading on sum-rate scaling of MIMO uplink with respect to the sparsity of angular support ($r$), where $M=200$, $L=4, K=10, T_c=500$. The max-min bound is given by \eqref{eq:SC-1c}. While `Lower bound 1' is the coherent bound in \eqref{eq:MR-9}, `Lower bound 2' and `Lower bound 3' indicate the noncoherent ones in \eqref{eq:NC-1} and  \eqref{eq:SC-1}, respectively. While $r=100, 60, 10, 20, 50, 30, 40$ from top to bottom for Lower bound 2, $r=10, 20, 30, 40,50,60,100$ for the rest.  The {asymptotic} capacity scaling  is given by \eqref{eq:MR-1} with $o(1)=0$.}\label{fig-3a}
\vspace{-3mm}
\end{figure*}

Fig. \ref{fig-3a} shows how strong spatial correlation we need to achieve  linear sum-rate scaling with respect to SNR (dB) in terms of the number of non-zero eigenvalues (or multipath components in angular domain) of channel covariance matrices in uplink. We used the orthogonal pilot sequences. 
$\Rc_{\text{ul},\ell_k}^\text{(1)}$ (`lower bound 1') in \eqref{eq:MR-9} and $\Rc_{\text{ul},\ell_k}^\text{(3)}$ (`lower bound 3') in \eqref{eq:SC-1} turn out to yield a linear growth of the ergodic sum rate with SNR (dB) for both orthogonal and non-orthogonal pilot schemes. In contrast, $\Rc_{\text{ul},\ell_k}^\text{(2)}$ (`lower bound 2') in \eqref{eq:NC-1} does not show linear growth due to lack of  hardening of the effective channels $\pw_{\ell\ell_k}$, whose dimension is $r$. The coherent lower bound suffers from channel estimation error represented by the parallel shift of capacity versus SNR curves, also known as power offset. 
It can be further seen that almost the same multiplexing gain as \eqref{eq:MR-1} is achievable up to $r=30$. At low SNR, both interference suppression and pilot decontamination effects of spatial despreading are diluted by noise, and the sum-rate performance depends more on channel hardening of $\pw_{\ell\ell_k}$ than spatial despreading of $\pU_{\ell\ell_k}$. Hence, the larger $r$ turns out rather beneficial in the low SNR regime.


\begin{figure}
\hspace{-6mm}
  \includegraphics[scale=.5]{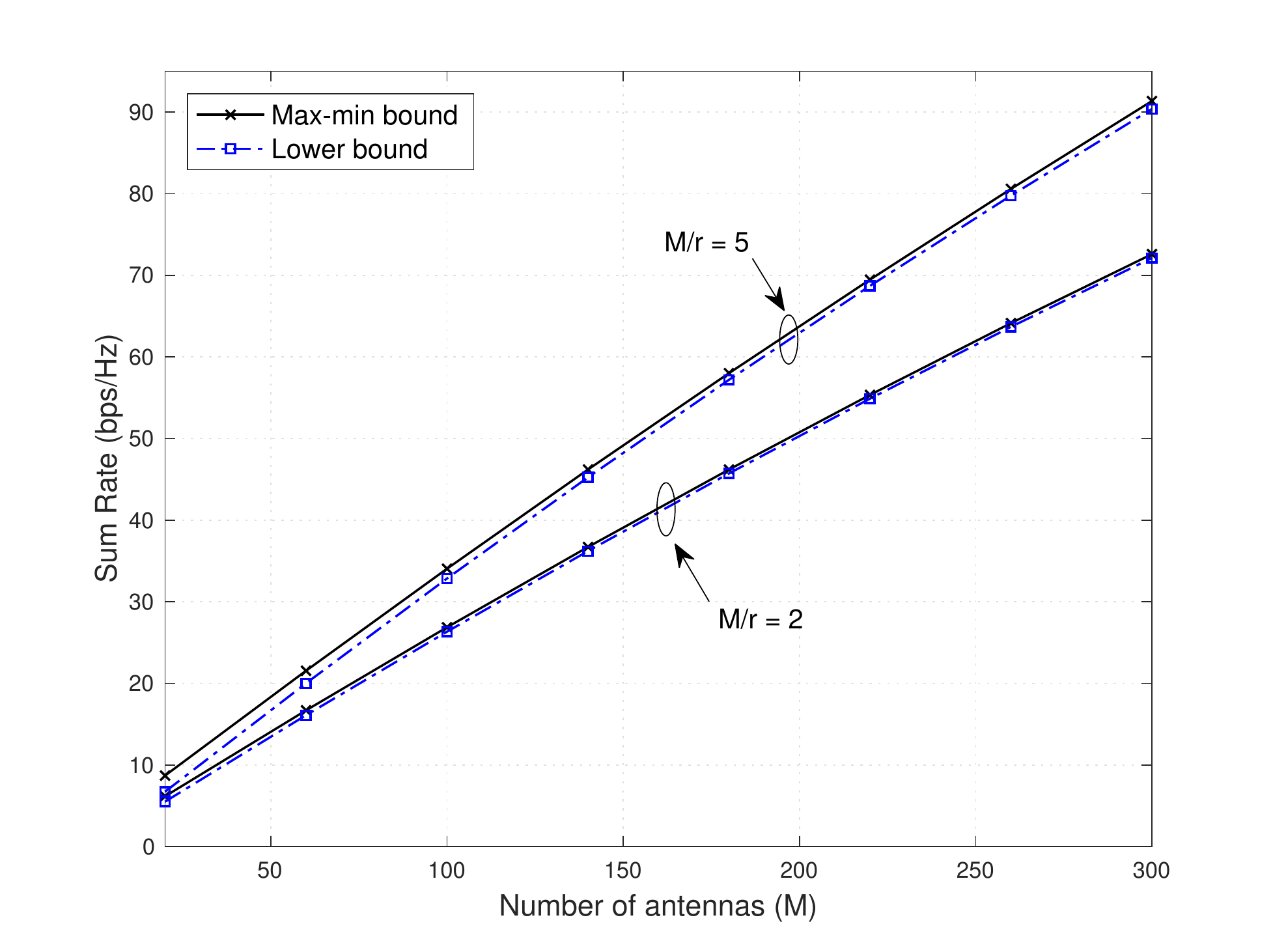}
  \caption{The impact of spatial spreading on sum-rate scaling with respect to $M$ in MIMO downlink with the ratios $\frac{M}{K}=5$ and $\zeta=\frac{M}{r}$ fixed, where the `Lower bound' is given by \eqref{eq:SCD-3}.}\label{fig-6}
\end{figure}

Fig. \ref{fig-6} verifies the scalability of the sum-rate scaling of MIMO downlink in \eqref{eq:SCD-1} with respect to $M$ with the ratios $\frac{M}{K}$ and $\zeta=\frac{M}{r}$ fixed, where $M$ is in the linear scale, and $L=7$. At  SNR $=10$ dB, we observe that given the fixed ratios of $M,K,r$, the sum rate scales almost \emph{linearly} with $M$. This implies that the effect of spatial spreading scales well with respect to both $K$ and $r$ as well as $M$, although $r$ is not much smaller than $M$. Furthermore, this linear growth is observed even for $M<KL$.

\begin{figure}
\hspace{-6mm}
  \includegraphics[scale=.5]{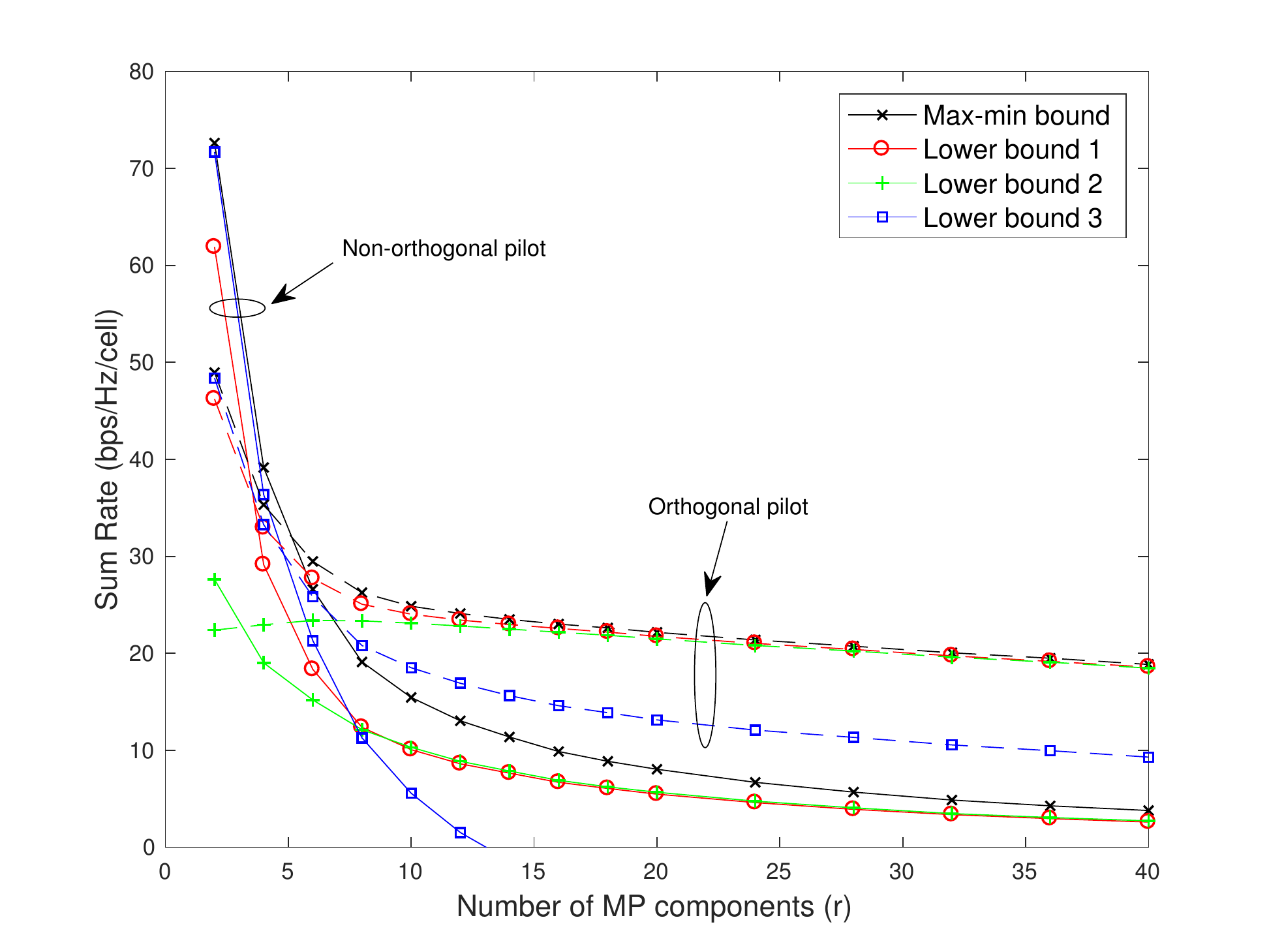}
  \caption{The performance of the non-orthogonal pilot scheme with respect to $r$ for small coherence block size $T_c=50$, where $M=100, L=7$, $K=20$, and SNR$=20$ dB.  }\label{fig-4}
\end{figure}

\begin{figure}
\hspace{-6mm}
  \includegraphics[scale=.5]{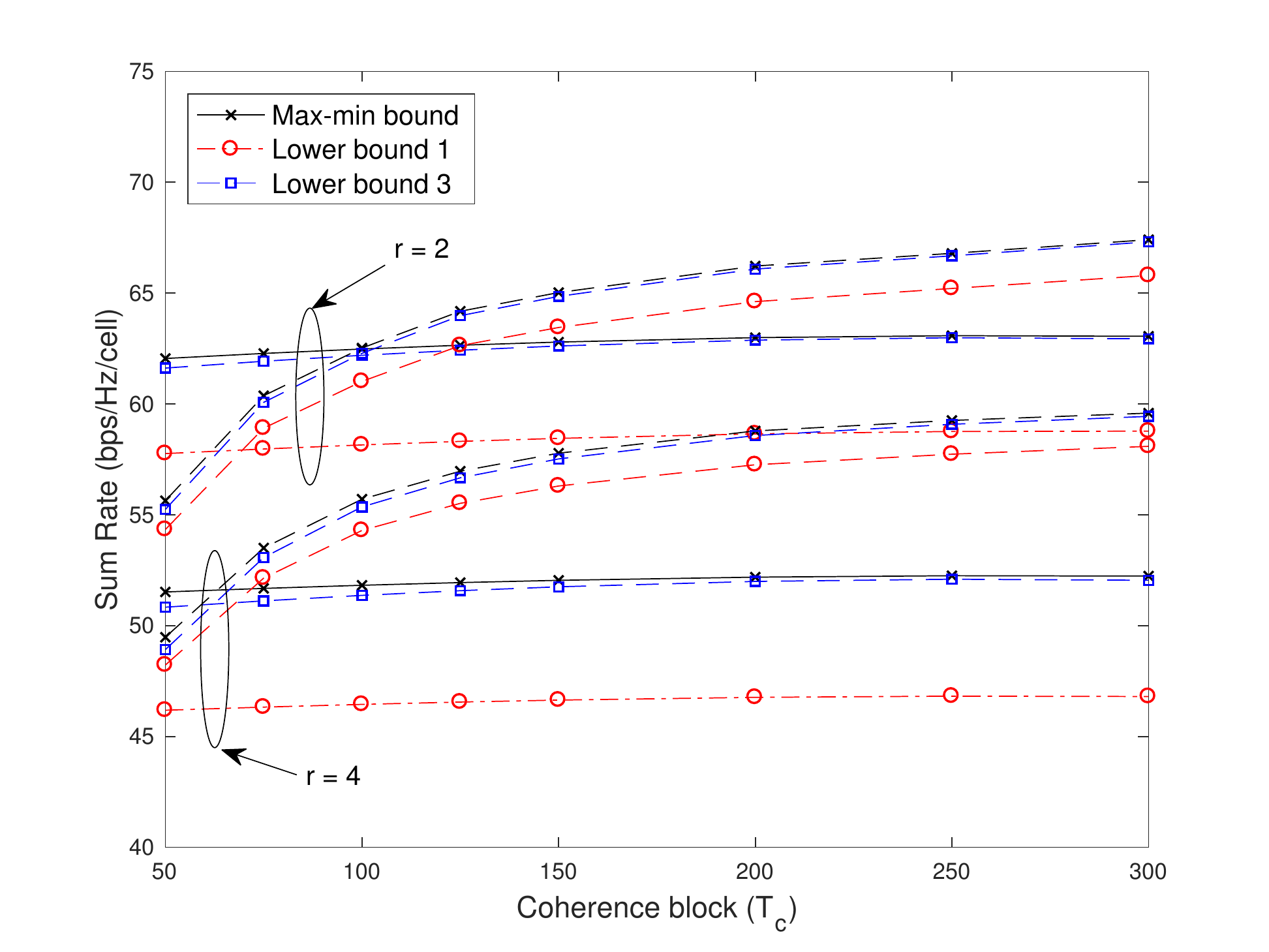}
  \caption{The impact of the coherence block size $T_c$ on both orthogonal and non-orthogonal pilot schemes, where $M=100, L=7$, $K=10$, and SNR$=20$ dB. While the solid lines indicate the non-orthogonal pilot scheme, the dashed lines represent the orthogonal pilot sequences.  }\label{fig-5}
\end{figure}

In Fig. \ref{fig-4}, we see that the non-orthogonal pilot scheme in Subsection \ref{sec:SC2} that consumes only a single channel use per network per coherence block can be beneficial in terms of the per-cell sum rate in uplink, even though the network is dense such that $L=7$ and $K=20$ for $M=100$. For small coherence block size of $T_c=50$, the non-orthogonal pilot helps only in fading channels with strong spatial correlation. As mentioned earlier, $\Rc_{\text{ul},\ell_k}^\text{(3)}$ has a shortcoming of sum-rate underestimation for this small $T_c$ unless $r$ is sufficiently small. 
Fig. \ref{fig-5} shows the impact of the coherence block size $T_c$ in the strong spatial correlation cases of $r=4,8.$ For $r=4$, the non-orthogonal pilot scheme turns out useful in a small to moderate range of $T_c$. In addition, the coherent lower bound $\Rc_{\text{ul},\ell_k}^\text{(1)}$ widely used in uplink suffers from considerable channel estimation error due to imperfect CSIR in case of non-orthogonal pilot. A general non-orthogonal pilot scheme based on Welch bound equality frames is expected to outperform the naive scheme in \eqref{eq:SC2-1}.

\section{Conclusion}
\label{sec:Con}

Channel hardening has been traditionally considered as an essential source of massive MIMO gain. This is not necessarily the case with strong spatial correlation under the random sparse angular support models. Rather, one can observe that the effect of spatial (de)spreading is indeed central to achieve the ultimate capacity scaling laws in this work. Although the exact capacity scaling of massive MIMO is achieved under the sublinear sparsity assumption in this work, the effect of spatial (de)spreading is shown to be still valid at finite $M$ with not-so-sparse angular support.  
Some important implications of this work can be summarized as follows. 
\begin{itemize}
\item Once the multiple antenna channels in the network satisfy a certain sparsity level of angular support, one can incorporate spatial (de)spreading into system designs such as non-orthogonal pilot and low-dimensional channel estimation and precoding (or combining) instead of orthogonal pilot and full-dimensional MMSE processing to realize the very promising sum-rate performance of massive MIMO in this paper. 
\item The potential of serving as many users as the number of large-scale BS antennas through interference-free links in strong spatial correlation regimes would be of importance particularly in uplink, where the per-cell sum power scales linearly with the number of users. Our results have shown that one may \emph{simultaneously} provide ``super-massive connectivity" and very high per-user data rate for the next-generation wireless network with very high carrier frequencies, where meeting the uplink data requirement is sometimes more challenging than downlink. 
\item The three lower bounds on the achievable rate of multicell MU-MIMO that we considered show mutually complementary behaviors. In order to better understand and predict the performance of massive MIMO, therefore, one should carefully select a proper bound, depending on the channel and system parameters such as sparsity of angular support, coherence block size, and orthogonal/non-orthogonal pilot sequences.  
\item Finally, our large system analysis holds true in finite coherence block and even when the number of users per cell scales linearly with the number of BS antennas. Meanwhile, large system analysis in the literature implicitly has assumed that coherence block grows unbounded and the number of users is finite or grows slower than the number of antennas. 
\end{itemize}

For further study, we are investigating the performance behavior of low-dimensional precoding/combining based on spatial spreading/despreading in more realistic spatial correlation models based on arbitrary angle of arrivals of users, whose subspaces are not mutually orthogonal but linearly independent of each other with high probability. 
{ 
The effect of spatial (de)spreading has an important implication in practical massive MIMO systems such as 5G NR, which is based on downlink training and feedback like FDD settings even if the NR system is mainly on TDD spectrum. For both below 6GHz and mm-Wave bands, 5G NR massive MIMO presumes hybrid beamforming due to power consumption and implementation cost. Since spatial (d)spreading corresponds to wideband/analog beamforming matched to space-frequency channel covariances of users in practical systems, our results strongly suggest that given $r$ dominant eigen-directions or DFT columns of users, a massive MIMO system should be designed to use as many wideband beamforming dimensions (i.e., $d\le r$ RF chains) as possible to realize the very promising spectral efficiency in this work, as shown in Fig. \ref{fig-2}. In other words, even if an mm-Wave channel has asymptotically unlimited capacity, no one can leverage such promising benefit with insufficient wideband beamforming dimensions.  Although having $d$ RF chains entails the cost of high power consumption and implementation cost at the BS, it may be justified from a viewpoint of system performance enhancement. 
We plan to address performance benefit of having spatial spreading matched to space-frequency channel covariances of users in a follow-up paper under the 5G NR channel model \cite{NR_ch}.}


\appendices

\section{Useful Lemmas}
\label{app-1}

We collect here some known or new lemmas to be used throughout this work. 
Silverstein and Bai derived the following well-known result in large random matrix theory, which is pivotal in the method of deterministic equivalents.

\begin{lem}\label{lem-a1} \normalfont
(Trace Lemma \cite[Lem. 2.7]{Bai98}, \cite[Thm. 3.4]{Cou11b}) 
Let $\px_1, \px_2, \ldots,$ with $\px_n \in \mathbb{C}^n$, be random vectors with i.i.d. entries of zero mean, variance $1/n$, and eighth order moment of order $\Oc(\frac{1}{n^4})$. Let $\pA_1, \pA_2,\ldots,$ with $\pA_n \in \mathbb{C}^{n\times n}$, be a series of matrices with uniformly bounded spectral norm with respect to $n$, independent of $\px_n$. Then
  \begin{align} \label{eq:app-1}
     \px_n^\ct \pA_n \px_n - \frac{1}{n}\; \trace{\pA_n} \overset{a.s.}{\longrightarrow} 0
  \end{align}
as $n\rightarrow\infty$.
\end{lem}

In this paper, we provide a direct generalization of the above result for the two cases: one is the case where $\px_1, \px_2, \ldots,$ are infinite sequences whose entries are \emph{uncorrelated} with each other and not necessarily identically distributed; the other is that the spectral norm of $\pA_n$ is \emph{not necessarily uniformly bounded}. 

\begin{corol}\label{corol-a0} \normalfont 
Let $\py_1, \py_2, \ldots,$ with $\py_n \in \mathbb{C}^n$, be random vectors with uncorrelated entries  of zero mean, variance $1/n$, $|y_{n,i}|^2=1/n,$ where $y_{n,i}$ is the $i$th entry of $\py_n$ and $i=1,\ldots,n$. With $\pA_1, \pA_2,\ldots,$ in Lemma \ref{lem-a1}, we have 
  \begin{align} \label{eq:app-1b}
     \py_n^\ct \pA_n \py_n - \frac{1}{n}\; \trace{\pA_n} \overset{a.s.}{\longrightarrow} 0
  \end{align}
as $n\rightarrow\infty$.
\end{corol}

\begin{IEEEproof}
It follows from the Markov inequality that for any $\epsilon>0$  
\begin{align} \label{eq:app-1b1}
     \Pr\left(\Big|\py_n^\ct \pA_n \py_n - \frac{1}{n}\; \trace{\pA_n}\Big| >\epsilon\right) \le \frac{\Eb\big[\big|\py_n^\ct \pA_n \py_n - \frac{1}{n}\; \trace{\pA_n}\big|^4\big]}{\epsilon^4} .\end{align}
In order to show that $\Eb\big[\big|\py_n^\ct \pA_n \py_n - \frac{1}{n}\; \trace{\pA_n}\big|^4\big] \le \frac{C_1}{n^2}$, where $C_1>0$ is a constant independent of $n$ and $\pA_n$, following the footsteps in \cite[Lem. 3.1]{Sil95} and \cite[Thm. 3.4]{Cou11b}, we begin with  
\begin{align} \label{eq:app-1b2}
     &\Eb\left[\Big|\py_n^\ct \pA_n \py_n - \frac{1}{n}\; \trace{\pA_n}\Big|^4\right] \nonumber \\ &\le 8
     \left( \Eb\Big[\sum_i A_{n,ii}(|y_{n,i}|^2-n^{-1})\Big]^4 + \Eb\Big[\sum_{i\neq j} A_{n,ij}y_{n,i}^*y_{n,j}\Big]^4\right) \nonumber \\
     &= 8 \left( \Eb\Big[\sum_{i\neq j} A_{n,ij}y_{n,i}^*y_{n,j}\Big]^4\right) \nonumber \\
     &= \Oc\big(\frac{1}{n^2}\big)
\end{align}
where we used $|y_{n,i}|^2=1/n, \forall i$. The last equality follows from the facts: 1) each term in the sum is finite since $y_{n,i}$ has the { eighth order moment of $\frac{1}{n^4}$} and $\pA_n$ has uniformly bounded norm, and 2) such terms amount to $\Oc(n^2)$ since the nonzero contribution to the sum arises if $y_{n,i}$ appears an even number of times due to $\Eb[y_i^*y_j]=0, \forall i\neq j$. Plugging \eqref{eq:app-1b2} into \eqref{eq:app-1b1}, we have 
\begin{align} 
  \sum_{n=1}^\infty\Pr\left(\Big|\py_n^\ct \pA_n \py_n - \frac{1}{n}\; \trace{\pA_n}\Big| >\epsilon\right) < \infty . \nonumber
\end{align}
The almost sure convergence in \eqref{eq:app-1b} immediately follows from the first Borel-Cantelli lemma \cite{Bil95}. 
\end{IEEEproof} 

Although the above constant modulus requirement of $|y_{n,i}|^2$  is a strong condition, some popular structured random matrices like random partial Fourier and Hadamard matrices \cite{Can06,Tsa06} satisfy the condition. 

\begin{corol} \label{corol-a1} \normalfont
Let $\px_1, \px_2, \ldots$  be as in Lemma \ref{lem-a1}. Let $\pA_1, \pA_2,\ldots,$ with $\pA_n \in \mathbb{C}^{n\times n}$, be a series of matrices independent of $\px_n$. If there exists $m_n\ge n$ such that 
\begin{align} \label{eq:app-2}
    \limsup_{n\rightarrow \infty}\frac{n}{m_n}\;\|\pA_n\|_2 < \infty 
\end{align}
then
  \begin{align} \label{eq:app-3}
     \frac{n}{m_n}\px_n^\ct \pA_n \px_n - \frac{1}{m_n}\; \trace{\pA_n} \overset{a.s.}{\longrightarrow} 0
  \end{align}
as $n\rightarrow\infty$.
\end{corol}

\begin{IEEEproof}
Even though the spectral norm of $\pA_n$ is not uniformly bounded across $n$, one may find a sequence $m_n\ge n$ to satisfy (\ref{eq:app-2}), which implies that the spectral norm of $\frac{n}{m_n}\pA_n$ is uniformly bounded with respect to $n$. In this case, $m_n$ and $n$ increase at the ratio $\frac{n}{m_n}$, where $m_n$ may grow  faster than $n$ such that $\frac{n}{m_n}\rightarrow 0$.  
By noticing $\frac{n}{m_n}\px_n^\ct \pA_n \px_n =\px_n^\ct \pB_n \px_n$, where $\pB_n =\frac{n}{m_n}\pA_n$, and by Lemma \ref{lem-a1}, for any $\epsilon>0$, there exists $n_0$ such that for all $n\ge n_0$, 
$     \left|\px_n^\ct \pB_n \px_n - \frac{1}{n}\; \trace{\pB_n}\right| 
     < \epsilon.$ 
\end{IEEEproof}

We can also combine the above two corollaries so as to make the trace lemma still valid in the more general case where $\px_1, \px_2, \ldots,$ are uncorrelated random vectors with constant modulus and/or the spectral norms of $\pA_1, \pA_2,\ldots,$ are not necessarily uniformly bounded. 
The following lemma is a generalization of \cite[Thm. 3.7]{Cou11b}. 

\begin{lem}\label{lem-2} \normalfont
Let $\px_n \in \mathbb{C}^m$ and $\py_n \in \mathbb{C}^n$, be two independent random vectors whose entries satisfy the conditions in either Lemma \ref{lem-a1} or Corollary \ref{corol-a0}, with a constant ratio $\frac{m}{n}$ as $p\rightarrow\infty$, where $p=\min(m,n)$. Also let $\pA \in \mathbb{C}^{m\times n}$, independent of $\px_n$ and $\py_n$.
As $p$ goes to infinity, we have  
\begin{align} \label{eq:Pre-10}
    \px^\ct \pA\; \py  \overset{a.s.}{\longrightarrow} 0 . 
\end{align}   
\end{lem}

\begin{IEEEproof}
Following the line of arguments in \cite[Lem. 3.1]{Sil95} and \cite[Thm. 3.7]{Cou11b}, it is not difficult to show that $\mathbb{E}\big[|\px^\ct \pA\; \py|^8\big]<\frac{c}{p^4}$ for some $c>0$ independent of $p$. Then, (\ref{eq:Pre-10}) follows from the Markov inequality, Borel-Cantelli lemma, and the Tonelli theorem \cite{Bil95,Ton09}.
\end{IEEEproof}

{ The above results require that a random vector either have i.i.d. entries or satisfy the conditions in  Lemma \ref{lem-a1}. Let us consider the third type of random vector $\pu_n$, which is a column of an $n\times n$ Haar  distributed unitary matrix. 
Even though the random unit vector $\pu_n$ does not have i.i.d. entries, it can be rewritten as $\pu_n = \frac{\px_n}{\|\px_n\|}$, where $\px_n$ has i.i.d. zero-mean Gaussian distributed entries with unit variance and $\|\px_n\|$ is $\chi^2$ distributed with $2n$ degrees of freedom. It can then be shown \cite[Sec. VI]{DHL03} that $\pu_n^\ct \pA_n \pu_n$ and $\frac{1}{\|\px_n\|^2}\px_n^\ct \pA_n \px_n$ have the same asymptotic behavior, where $\pA_n$ has uniformly bounded spectral norm with respect to $n$, independent of $\pu_n$. As a consequence, the above trace lemma and its variants directly apply to the random partial unitary correlation model in \ref{sec:CM}.}

The following lemma is about the convergence of a mixture of two independent random measures with the distributions considered in this paper. 

\begin{lem}\label{lem-a10} \normalfont 
Let $\px_1, \px_2, \ldots,$ with $\px_n \in \mathbb{C}^n$, be random vectors whose entries satisfy the conditions in either Lemma \ref{lem-a1} or Corollary \ref{corol-a0}. Let $\Lambdam_1, \Lambdam_2,\ldots,$ with $\Lambdam_n \in \mathbb{C}^{n\times n}$, be a series of random matrices with uniformly bounded spectral norm with respect to $n$, independent of $\px_n$ and convergent such that $\Lambdam_n-\mathring{\Lambdam}_n \overset{a.s.}{\longrightarrow} 0$. Then as $n\rightarrow\infty$
  \begin{align} \label{eq:app-30}
     \px_n^\ct \Lambdam_n \px_n - \frac{1}{n}\; \trace{\mathring{\Lambdam}_n} \overset{a.s.}{\longrightarrow} 0.
  \end{align}
\end{lem}

\begin{IEEEproof}
It suffices to show that\begin{align} %
   &\Pr\Big(\lim_n \px_n^\ct \Lambdam_n \px_n = \frac{1}{n}\; \trace{\mathring{\Lambdam}_n}\Big) 
    \nonumber \\ &\ge \Pr\Big(\lim_n\px_n^\ct \Lambdam_n \px_n = \frac{1}{n}\; \trace{\mathring{\Lambdam}_n},\; \lim_n \px_n^\ct \mathring{\Lambdam}_n \px_n = \frac{1}{n}\; \trace{\mathring{\Lambdam}_n} \Big) \nonumber \\
   &= \Pr\Big(\lim_n\px_n^\ct \big(\Lambdam_n-\mathring{\Lambdam}_n\big) \px_n = 0,\; \lim_n \px_n^\ct \mathring{\Lambdam}_n \px_n = \frac{1}{n}\; \trace{\mathring{\Lambdam}_n} \Big) \nonumber \\
   &\overset{(a)}{\ge} \Pr\Big(\lim_n \Lambdam_n - \mathring{\Lambdam}_n=0,\; \lim_n \px_n^\ct \mathring{\Lambdam}_n \px_n = \frac{1}{n}\; \trace{\mathring{\Lambdam}_n} \Big)    = 1 \label{eq:app-31}
\end{align}
where $(a)$ follows from the continuous mapping theorem since $\px_n^\ct \big(\Lambdam_n-\mathring{\Lambdam}_n\big) \px_n$ is a linear function of $\vec{(\Lambdam_n- \mathring{\Lambdam}_n)}$ and also $\px_n$ and $\Lambdam_n- \mathring{\Lambdam}_n$ are independent, and the second event in $\Pr(\cdot)$ in the RHS of \eqref{eq:app-31} is almost sure by Lemma \ref{lem-a1} and Corollary \ref{corol-a0}. 
Then the almost sure convergence in \eqref{eq:app-30} immediately follows due to the fact that the intersection of two almost sure events is also almost sure.  
\end{IEEEproof} 

Notice that the continuous mapping theorem does not directly apply to the above lemma since $\px_n$ does not converge to a random vector. 
In what follows, we derive some matrix analogs of the trace lemma for random matrices $\pX$ rather than random vectors $\px$. For the first results, we use the random partial unitary model in \ref{sec:CM}. 

\begin{lem}\label{lem-1b} \normalfont
Let $\pU \in \mathbb{C}^{p\times m}$ and $\pV\in \mathbb{C}^{p\times n}$ be { $m$ and $n$ columns of two independent $p\times p$ Haar distributed unitary matrices, respectively, where $p>\max(m,n)$.} Also let $\pD\in \mathbb{C}^{n\times n}$ be an arbitrary matrix with $\limsup_{p\rightarrow \infty}\|\pD\|_2 < \infty 
$, independent of $\pU$ and $\pV$. Then, as $p$ grows without bound, we have 
\begin{align} \label{eq:Pre-14}
    \pU^\ct \pV \pD \pV^\ct \pU  - \frac{\trace\pD}{p}\;\pI_m \overset{a.s.}{\longrightarrow} 0.
\end{align}   
\end{lem}

\begin{IEEEproof}
We denote the $i$th columns of $\pU$ and $\pV$ as $\pu_i$ and $\pv_i$, respectively. Since $\pu_i$ and $\pv_i$ 
 are taken from two independent Haar distributed unitary matrices, respectively, we can apply Lemmas \ref{lem-a1} and \ref{lem-2}, as mentioned eariler.  
The matrix multiplication $\pU^\ct \pV\pD \pV^\ct \pU$ unfolds as
\begin{align} 
   [\pU^\ct \pV \pD\pV^\ct \pU]_{k,\ell} = \sum_{i=1}^n\sum_{j=1}^n d_{ij}\pu_k^\ct\pv_i\pv_j^\ct\pu_\ell, \ \forall (k,\ell).
    \nonumber
\end{align}   
The diagonal and off-diagonal entries can be written respectively as
\begin{align}  \label{eq:Pre-11}
   \sum_{i,j} d_{ij}\pu_k^\ct\pv_i\pv_j^\ct\pu_k &= \sum_i d_{ii}|\pv_i^\ct\pu_k|^2 +\pu_k^\ct\bigg(\sum_{i\neq j}d_{ij}\pv_i\pv_j^\ct\bigg)\pu_k  \nonumber 
   \\
   \sum_{i,j} d_{ij}\pu_k^\ct\pv_i\pv_j^\ct\pu_\ell  &=  \nonumber \\ \pu_k^\ct\bigg(\sum_i d_{ii}&\pv_i\pv_i^\ct\bigg)\pu_\ell +\pu_k^\ct\bigg(\sum_{i\neq j}d_{ij}\pv_i\pv_j^\ct\bigg)\pu_\ell. 
\end{align}
 For $k=\ell$, we have 
\begin{align}  
   \sum_i d_{ii}|\pv_i^\ct\pu_k|^2 &+\pu_k^\ct\bigg(\sum_{i\neq j}d_{ij}\pv_i\pv_j^\ct\bigg)\pu_k \nonumber \\
   &\overset{(a)}{\simeq} \frac{\trace\sum_i d_{ii}\pv_i\pv_i^\ct}{p} + \frac{\trace\sum_{i\neq j} d_{ij}\pv_i\pv_j^\ct}{p}\nonumber\\
   &= \frac{\sum_i d_{ii}\pv_i^\ct\pv_i}{p} + \frac{\sum_{i\neq j} d_{ij}\pv_j^\ct\pv_i}{p}\nonumber\\
   &= \frac{\trace\pD}{p} \nonumber
\end{align}  
where $(a)$ follows from Lemma \ref{lem-a1}. For $k\neq\ell$, we obtain by Lemma \ref{lem-2}
\begin{align}  
    \pu_k^\ct\bigg(\sum_i d_{ii}\pv_i\pv_i^\ct\bigg)\pu_\ell +\pu_k^\ct\bigg(\sum_{i\neq j}d_{ij}\pv_i\pv_j^\ct\bigg)\pu_\ell 
   \simeq 0 \nonumber
\end{align}  
which yields (\ref{eq:Pre-14}).
\end{IEEEproof}

For finite dimensional matrices $\pU, \pV, \pD$, we can have the following similar result.

\begin{lem}\label{cor-1c} \normalfont
Let $\pU \in \mathbb{C}^{p\times m}$ and $\pV\in \mathbb{C}^{p\times n}$ be { $m$ and $n$ columns of two independent $p\times p$ Haar distributed unitary matrices, respectively, where $p>\max(m,n)$.} Also let $\pD\in \mathbb{C}^{n\times n}$ be a random matrix, independent of $\pU$ and $\pV$. Then we have 
\begin{align} \label{eq:Pre-1b}
   \mathbb{E}\left[ \pU^\ct \pV \pD \pV^\ct \pU  \right] = \frac{\trace\mathbb{E}[\pD]}{p}\;\pI_m .
\end{align}   
\end{lem}

\begin{IEEEproof}
Notice that $\pu_i$ and $\pv_i$ are two independent random unit vectors uniformly distributed on the $2p$-dimensional complex unit sphere such that $\pu_i = \frac{\px}{\|\px\|_2}$ with covariance $\frac{1}{p}\pI_p$, where $\px \sim\mathcal{CN}(\p0,\pI_p)$, and so $\pv_i$ is, and that $|\pv_i^\ct\pu_k|^2$ is a Beta-distributed random variable with parameter $(1,p-1)$. By the independence assumption on $\pU, \pV, \pD$ and the law of total expectation, the proof of \eqref{eq:Pre-1b} immediately follows similar to Lemma \ref{lem-1b}. 
\end{IEEEproof}

For the random partial Fourier model for $\pU$ in \ref{sec:CM}, we have the following two results analogous to the above lemmas. 

\begin{lem}\label{lem-1e} \normalfont
Let $\pU \in \mathbb{C}^{p\times m}$ and $\pV\in \mathbb{C}^{p\times n}$ be independent random orthonormal matrices whose columns have the same entries as $\py_n$ in Corollary \ref{corol-a0}, where $p\ge\max(m,n)$. Also let $\pD\in \mathbb{C}^{n\times n}$ be an arbitrary matrix with $\limsup_{p\rightarrow \infty}\|\pD\|_2 < \infty$, independent of $\pU$ and $\pV$. Then we have 
\begin{align} \label{eq:Pre-1e}
    \pU^\ct \pV \pD \pV^\ct \pU  - \frac{\trace\pD}{p}\;\pI_m \overset{a.s.}{\longrightarrow} 0.
\end{align}   
\end{lem}

\begin{IEEEproof}
The proof immediately follows from the same footsteps in the proof of Lemma \ref{lem-1b} using Corollary \ref{corol-a0} and Lemma \ref{lem-2}. 
\end{IEEEproof}

Note that sampling \emph{without} replacement in the random partial Fourier model guarantees the orthonormality between columns so that the above result holds for the channel covariance model.

\begin{lem}\label{lem-1d} \normalfont
Given the same  $\pU, \pV, \pD$ as in Lemma \ref{lem-1e}, we have 
\begin{align} \label{eq:Pre-1c}
   \mathbb{E}\left[ \pU^\ct \pV \pD \pV^\ct \pU \right] \longrightarrow \frac{\trace\mathbb{E}[\pD]}{p}\;\pI_m .
\end{align}   
\end{lem}

\begin{IEEEproof}
Following the same reasoning and procedure as the proofs of Lemmas \ref{lem-1b} and \ref{lem-1e}, it suffices to be aware of the distribution of the binary random variable 
$$ \big|\pv_i^\ct\pu_k\big|^2 =\left\{ \begin{array}{ll}
  1 &  \text{if } \pv_i=\pu_k\\
  0 &  \text{otherwise }
  \end{array} \right. .
$$ 
We can see that  $|\pv_i^\ct\pu_k|^2$'s are i.i.d. Bernoulli with $1/p$, thus yielding $\mathbb{E}\left[ |\pv_i^\ct\pu_k|^2\right]=1/p$.  
Using the independence assumption and the law of total expectation again, we have  
\begin{align}  
   \mathbb{E}\bigg[\sum_i d_{ii}|\pv_i^\ct\pu_k|^2\bigg] &= \mathbb{E}\Bigg[\mathbb{E}\bigg[\sum_i d_{ii}|\pv_i^\ct\pu_k|^2\Big|\pD\bigg]\Bigg]  
   = \frac{\mathbb{E}[\trace\pD]}{p}. \nonumber
\end{align}  
Noticing that the expectation of the remaining terms in \eqref{eq:Pre-11} equals to zero, we have  \eqref{eq:Pre-1c}.
\end{IEEEproof}

\section{Proof of Lemma \ref{lem-4}}
\label{app-1b}

Let $\tilde{\pw}_{\ell\ell_k}$ denote the MMSE estimate of ${\pw}_{\ell\ell_k}$ from $\bar{\sv}_{\ell_k}$ instead of $\sv_{\ell_k}$ given the knowledge of $\pU_{\ell}$. We only prove the random partial unitary correlation model and omit the random partial Fourier model because one can follow the same footsteps. The estimate $\tilde{\pw}_{\ell\ell_k}$ is given by  
\begin{align} \label{eq:Pre-13}
   &\tilde{\pw}_{\ell\ell_k} = \Lambdam_{\ell\ell_k}\pU_{\ell\ell_k}^\ct\bigg(\pR_{\ell\ell_k} + \sum_{\ell'\neq\ell} \pR_{\ell\ell'_k} +\rho_p^{-1}\pI_M\bigg)^{-1} \bar{\sv}_{\ell_k} .
\end{align} 
Letting $\pR_{-\ell}\triangleq\sum_{\ell'\neq\ell} \pR_{\ell\ell'_k}$, we have 
\begin{align} \label{eq:Pre-14}
   &\big(\rho_p^{-1}\pI_M +\pR_{\ell\ell_k} +\pR_{-\ell} \big)^{-1}  \nonumber \\ &= \Big(\pI_M + \big(\rho_p^{-1}\pI_M +\pU_{\ell\ell_k}\Lambdam_{\ell\ell_k} \pU_{\ell\ell_k}^\ct  \big)^{-1}\pR_{-\ell} \Big)^{-1} \times  \nonumber \\ &\ \ \ \ \ \ \ \  \big(\rho_p^{-1}\pI_M +\pR_{\ell\ell_k} \big)^{-1} \nonumber \\
   &\overset{(a)}{=} \Big(\pI_M + \big(\rho_p\pI_M -\rho_p\pU_{\ell\ell_k}(\Lambdam_{\ell\ell_k}^{-1}+\rho_p\pI_{r_{\ell\ell_k}})^{-1}\rho_p\pU_{\ell\ell_k}^\ct \big)  \times \nonumber \\ & \ \ \ \ \ \ \ \  \pR_{-\ell} \Big)^{-1} \big(\rho_p^{-1}\pI_M +\pR_{\ell\ell_k} \big)^{-1} \nonumber \\
   &\overset{(b)}{\simeq} \big(\pI_M + \rho_p\pR_{-\ell} \big)^{-1} \big(\rho_p^{-1}\pI_M +\pR_{\ell\ell_k} \big)^{-1} \nonumber \\
   &\overset{(c)}{\simeq} \prod_{\ell'\neq\ell}\big(\pI_M + \rho_p\pR_{\ell\ell'_k} \big)^{-1} \big(\rho_p^{-1}\pI_M +\pR_{\ell\ell_k} \big)^{-1} 
\end{align} 
where $(a)$ follows from the matrix inversion lemma, $(b)$ follows from the fact that $\pU_{\ell\ell_k}^\ct \pR_{-\ell}= \pU_{\ell\ell_k}^\ct \sum_{\ell'\neq\ell} \pU_{\ell\ell'_k}\Lambdam_{\ell\ell'_k} \pU_{\ell\ell'_k}^\ct\simeq \p0$ by Lemma \ref{lem-2} as $M\rightarrow\infty$, and we also used $\pU_{\ell\ell_k}^\ct\pU_{\ell\ell_k}=\pI_{r_{\ell\ell_k}}$. In $(c)$, we repeated the same decomposition upon $\big(\rho_p^{-1}\pI_M +\pR_{\ell j_k} +\pR_{-\{\ell,j\}} \big)^{-1}$ for all $j\neq \ell$, where $\pR_{-\{\ell,j\}}\triangleq\sum_{\ell'\neq\{\ell,j\}} \pR_{\ell\ell'_k}$. 

Substituting \eqref{eq:Pre-14} into \eqref{eq:Pre-13}, we have
\begin{align} 
   \tilde{\pw}_{\ell\ell_k} &\simeq \Lambdam_{\ell\ell_k}\pU_{\ell\ell_k}^\ct \prod_{\ell'\neq\ell}\big(\pI_M + \rho_p\pR_{\ell\ell'_k} \big)^{-1} \big(\rho_p^{-1}\pI_M +\pR_{\ell\ell_k} \big)^{-1} \bar{\sv}_{\ell_k} \nonumber \\
   &\overset{(a)}{\simeq} \Lambdam_{\ell\ell_k}\pU_{\ell\ell_k}^\ct \big(\rho_p^{-1}\pI_M +\pR_{\ell\ell_k} \big)^{-1} \bar{\sv}_{\ell_k} \nonumber \\
   &\overset{(b)}{\simeq} \Lambdam_{\ell\ell_k}\big(\rho_p^{-1}\pI_{r_{\ell\ell_k}} +\Lambdam_{\ell\ell_k} \big)^{-1} \sv_{\ell_k} \nonumber \\
   &\overset{(c)}{\simeq} \hat{\pw}_{\ell\ell_k} \nonumber 
\end{align} 
where $(a)$ follows from the matrix inversion lemma and Lemma \ref{lem-2}, $(b)$ follows from the matrix inversion lemma again. In $(c)$, we applied Lemma \ref{lem-1b} to  $\tilde{\pR}_{\ell_k{\ell'}\!\!_{k'}}$ in \eqref{eq:Pre-3} to have $$\tilde{\pR}_{\ell_k{\ell'}\!\!_{k'}}-\frac{\trace\Lambdam_{\ell\ell'_k}}{M}\;\pI_{r_{\ell\ell_k}} \overset{a.s.}{\longrightarrow} 0$$
and then $\tilde{\pR}_{\ell_k{\ell'}\!\!_{k'}}$ vanishes as in \eqref{eq:MR-9c}. 
The desired result then follows from the well-known fact that the MMSE estimate $\tilde{\pw}_{\ell\ell_k}$ is a sufficient statistic for the estimation of ${\pw}_{\ell\ell_k}$ (equivalently, ${\ph}_{\ell\ell_k}$) from $\bar{\sv}_{\ell_k}$ (e.g., see \cite{For04}).

\section{Proof of Theorem \ref{thm-a1}}
\label{app-4}

The complete proof is lengthy but somehow straightforward after realizing that all the assumptions in \cite[Thm. 1]{Cou11a} and \cite[Thm. 1]{Wag12} except the uniformly bounded spectral norm assumption apply to our case. For compactness, therefore, we only describe some essential parts without all the details.

Let $\tilde{\Psim}_i={\sqrt{b_i}}\Psim_i$, and $\tilde{\Thetam}_i=\tilde{\Psim}_i\tilde{\Psim}_i^\ct =b_i\Thetam_i$. 
To make sure that the random variable $m_N(z)$  converges to its deterministic equivalent, we need the generalized trace lemma in Corollary \ref{corol-a1} to hold. Following the same line of arguments in \cite[Thm. 1]{Cou11a} and \cite[Thm. 1]{Wag12}, we arrive at obtaining an asymptotic approximation of $\py_i^\ct\Psim_i^\ct\big(\pB_{[i]} -z\pI_N\big)^{-1}\Psim_i\py_i $,
where $\pB_{[i]}=\big(\pB_N -\Psim_i\py_i\py_i^\ct\Psim_i^\ct\big)^{-1} $. Although the $\Psim_i^\ct\big(\pB_{[i]} -z\pI_N\big)^{-1}\Psim_i$ may not have a uniformly bounded spectral norm, we can see that $\limsup_{N\rightarrow \infty}\|\tilde{\Psim}_i^\ct \big(\pB_{[i]} -z\pI_N\big)^{-1}\tilde{\Psim}_i\| <\infty$ due to assumption 3) and the fact that $\pB_{[i]}$ is Hermitian and $z\in \mathbb{C}\setminus\mathbb{R}^+$. Then, it immediately follows from Corollary \ref{corol-a1} that 
\begin{align} \label{eq:ED-5}
     {\py}_i^\ct\tilde{\Psim}_i^\ct \big(\pB_{[i]} -z\pI_N\big)^{-1}\tilde{\Psim}_i{\py}_i -\frac{1}{N}\trace\;\tilde{\Thetam}_i \big(\pB_{[i]} -z\pI_N\big)^{-1} \overset{a.s.}{\longrightarrow} 0 .
\end{align}
For $z\in \mathbb{C}\setminus\mathbb{R}^+$, we further obtain by the rank-1 perturbation lemma \cite[Lem. 2.1]{Bai07}  
\begin{align} 
     \frac{1}{N}\trace\;\tilde{\Thetam}_i \big(\pB_{[i]} -z\pI_N\big)^{-1}  -\frac{1}{N}\trace\;\tilde{\Thetam}_i \big(\pB_N -z\pI_N\big)^{-1} \overset{a.s.}{\longrightarrow} 0   \nonumber .
\end{align}
Notice that the rank-1 perturbation lemma does not depend on the asymptotic properties of the perturbation so that $\Psim_i\py_i\py_i^\ct\Psim_i^\ct$ may even have a large Euclidean norm.

It is important to guarantee that the fixed-point algorithm in (\ref{eq:ED-4}) converges. We already showed that $\frac{1}{ N}\trace\;\tilde{\Thetam}_i \big(\pB_N -z\pI_N\big)^{-1}<\infty, \forall i,$. For some finite constant $b^*>0$, therefore, $\frac{1}{ N}\trace\;\tilde{\Thetam}_i \big(b^*\pB_N -b^*z\pI_N\big)^{-1}$ is further required not to trivially go to zero as well.   
Letting $\pC=\tilde{\Thetam}_i$ and $\pD=\big(b^*\pB_N -b^*z\pI_N\big)^{-1}$, we have
\begin{align} 
   \frac{1}{N}\trace\;\tilde{\Thetam}_i \big(b^*\pB_N -b^*z\pI_N\big)^{-1}  
   &= \frac{1}{N}\trace\;\pC\pD  \nonumber\\
   &\overset{(a)}{\ge} \frac{1}{N}\sum_{i=1}^N\lambda_i(\pC)\lambda_{N-i+1}(\pD)  \nonumber\\
   &\ge \frac{1}{N}\sum_{i=1}^N\lambda_i(\pC)\lambda_{N}(\pD)  \nonumber\\
   &= \frac{\trace\pC}{N}\lambda_N(\pD)  \nonumber\end{align} 
where $(a)$ follows from \cite[H.1.h.]{Mar11}. To make sure that $\frac{1}{N}\trace\;\tilde{\Thetam}_i \big(b^*\pB_N -b^*z\pI_N\big)^{-1}>0$, we just need to let $b^*=b_\text{min}$ (equivalently, $\beta_i=\frac{b_\text{min}}{b_i}$), where $b_\text{min}=\min_{i\in\Nc} b_i$ with $\Nc=[1:n]$, by noticing that   
\begin{align} 
     \lambda_1(\pD^{-1}) &= b_\text{min}\lambda_1\bigg(\sum_{j\in \Nc\setminus i} \px_j\px_j^\ct+\pA_N-z\pI\bigg)  \nonumber \\
     &\le \sum_{j\in \Nc\setminus i}\lambda_1( b_\text{min}\px_j\px_j^\ct)+\lambda_1(b_\text{min}\pA_N)-b_\text{min}z < \infty \nonumber
\end{align}
where the first inequality follows from \cite{Fan51}, and the last step comes from the fact that $b_\text{min}\max_{j\in \Nc}\lambda_1( \px_j\px_j^\ct) < \infty$ due to Corollary \ref{corol-a1} and assumption 3). 
Then, one can show by the same arguments in \cite[Thm. 1]{Cou11a} and \cite[Thm. 1]{Wag12} that (\ref{eq:ED-4}) converges to a fixed point. 


It should be further noticed that $\frac{1}{ N}\trace\;{\pQ}_N \big(\pB_N -z\pI_N\big)^{-1}$ may trivially go to zero or infinity. This can be similarly avoided by appropriate scaling for ${\pQ}_N$ under assumption 5). 
The remainder of the proof then follows from the steps in \cite{Cou11a,Wag12}, and we eventually have (\ref{eq:ED-2}).

\section{Proof of Lemma \ref{thm-1}}
\label{app-2}

In what follows, we will derive the deterministic equivalents of each term in (\ref{eq:MR-9}) for the MMSE detector with $\pv_{\ell_k}=\pv_{\ell_k}^\mmse$ in (\ref{eq:MR-15}) conditioned on $\underline{\pU}$. {In the sequel, the superscript $^\mmse$ is omitted for notational convenience.} For channel covariance, we consider only the random partial unitary model in Sec. \ref{sec:CM}; A proof for the random partial Fourier model is straightforwardly given by using Corollary \ref{corol-a0} and Lemma \ref{lem-2} and Lemma \ref{lem-1e} instead of Lemmas \ref{lem-a1}, \ref{lem-2}, and \ref{lem-1b}.

\subsection{Signal Power}

For the asymptotic approximation of $|\pv_{\ell_k}^\ct\hat{\pw}_{\ell\ell_k}|^2$, we begin with 
\begin{align} \label{eq:MR-20}
   \pv_{\ell_k}^\ct\hat{\pw}_{\ell\ell_k} &= \hat{\pw}_{\ell\ell_k}^\ct\Upsilonm _{\ell_k}\hat{\pw}_{\ell\ell_k} \nonumber \\
   &\overset{(a)}{=} \frac{{\hat{\pw}_{\ell\ell_k}^\ct\Upsilonm _{-\ell\ell_k}\hat{\pw}_{\ell\ell_k}}}{1+\hat{\pw}_{\ell\ell_k}^\ct\Upsilonm _{-\ell\ell_k}\hat{\pw}_{\ell\ell_k}} \nonumber \\
   &\overset{(b)}{\simeq} \frac{\frac{1}{r_{\ell\ell_k}}\trace\; \tilde{\Phim}_{\ell\ell_k}\tilde{\Upsilonm} _{\ell_k}}{1+\frac{1}{r_{\ell\ell_k}}\trace\; \tilde{\Phim}_{\ell\ell_k}\tilde{\Upsilonm} _{\ell_k}} \nonumber \\
   &\overset{(c)}{\simeq} 
   \frac{\delta_{\ell_k}}{1+\delta_{\ell_k}}  
\end{align}
where $\Upsilonm _{-\ell\ell_k}=\big(\Upsilonm _{\ell_k}^{-1}-\hat{\pw}_{\ell\ell_k}^\ct \hat{\pw}_{\ell\ell_k}\big)^{-1}$, $\tilde{\Phim}_{\ell\ell_k} = \frac{r_{\ell\ell_k}}{M}{\Phim}_{\ell\ell_k}$, $\tilde{\Upsilonm} _{\ell_k}=M{\Upsilonm} _{\ell_k}$. $(a)$ follows from the matrix inversion lemma \cite[Eq. (2.2)]{Sil95}, $(b)$ is from Corollary \ref{corol-a1} and the rank-1 perturbation lemma
\cite[Lem. 2.1]{Bai07}, and $(c)$ is from Theorem \ref{thm-a1}.
In $(b)$, we used the fact that the spectral norm of $\tilde{\Phim}_{\ell\ell_k}$ is always uniformly bounded, whereas  ${\Phim}_{\ell\ell_k}$ is not necessarily the case, and that $\liminf_{M\rightarrow \infty}\frac{1}{r_{\ell\ell_k}}{\Upsilonm} _{\ell_k}>0 $ and $\limsup_{M\rightarrow \infty}\frac{1}{r_{\ell\ell_k}}{\Upsilonm} _{\ell_k}<\infty$. 
{Notice that the step $(b)$ is invalid for the case of the non-orthogonal pilot $\sv_{\ell_k}'$ in \eqref{eq:SC2-1} over the network since $\check{\pw}_{\ell\ell_k}$ in \eqref{eq:SC2-3} and $\Upsilonm _{-\ell\ell_k}$, where $\hat{\pw}_{\ell\ell_k}$ is replaced with $\check{\pw}_{\ell\ell_k}$,  are not independent any longer. This is also the  case with the remaining terms including $\Upsilonm _{-\ell\ell_k}$.}

It follows from (\ref{eq:MR-20}) and the continuous mapping theorem \cite[Thm. 2.3]{Vaa00} that, for any $\epsilon>0$, there exists $r_0>0$ such that, for all $r_{\ell\ell_k} \ge r_0$, $\Big||\pv_{\ell_k}^\ct\hat{\pw}_{\ell\ell_k}|^2 - \frac{\delta_{\ell_k}^2}{(1+\delta_{\ell_k})^2}\Big| < \epsilon$. Then, using the dominated convergence theorem \cite[Thm.  16.4]{Bil95}, we obtain
 \begin{align} \label{eq:MR-27}
   |\pv_{\ell_k}^\ct\hat{\pw}_{\ell\ell_k}|^2 - \frac{\delta_{\ell_k}^2}{(1+\delta_{\ell_k})^2} \xrightarrow[M\rightarrow\infty]{} 0.
\end{align}

\subsection{Noise Term}

Noticing that $\mathbb{E} \big[|\pv_{\ell_k}^\ct\pz_{\ell_k}|^2\big|\; \hat{\pw}_{\ell}\big]=\|\pv_{\ell_k}\|_2^2$ and applying \cite[Eq. (2.2)]{Sil95} twice and Theorem \ref{thm-4}, we have 
 \begin{align} 
   \|\pv_{\ell_k}\|_2^2 &= \frac{\hat{\pw}_{\ell\ell_k}^\ct\Upsilonm _{-\ell\ell_k}^2\hat{\pw}_{\ell\ell_k}}{\big(1+\hat{\pw}_{\ell\ell_k}^\ct\Upsilonm _{-\ell\ell_k}\hat{\pw}_{\ell\ell_k}\big)^2}  \nonumber \\
   &\simeq \frac{\frac{1}{r_{\ell\ell_k}M}\trace\; \tilde{\Phim}_{\ell\ell_k}\tilde{\Upsilonm} _{\ell_k}^2}{\big(1+\frac{1}{r_{\ell\ell_k}}\trace\; \tilde{\Phim}_{\ell\ell_k}\tilde{\Upsilonm} _{\ell_k}\big)^2} \nonumber \\
   &\simeq  \frac{\mu_{\ell_k}}{M(1+\delta_{\ell_k})^2} .
\end{align}

For the channel estimation error, we have 
 \begin{align} 
   \pv_{\ell_k}^\ct\pn_{\ell\ell_k} &=  \frac{\hat{\pw}_{\ell\ell_k}^\ct\Upsilonm _{-\ell\ell_k}\pn_{\ell\ell_k}}{1+\hat{\pw}_{\ell\ell_k}^\ct\Upsilonm _{-\ell\ell_k}\hat{\pw}_{\ell\ell_k}}  \simeq 0 \nonumber
\end{align}
which follows from \cite[Eq. (2.2)]{Sil95} and Lemma \ref{lem-2}. Applying the continuous mapping theorem and the dominated convergence theorem again leads to  
 \begin{align} \label{eq:MR-28}
\mathbb{E} \big[|\pv_{\ell_k}^\ct\nv_{\ell\ell_k}|^2 \big|\; \hat{\pw}_{\ell}\big] \overset{}{\longrightarrow} 0.
\end{align}

\subsection{Pilot Contamination and Interference Terms}

It follows from \cite[Eq. (2.2)]{Sil95} that
\begin{align} \label{eq:MR-21}
   \big|\pv_{\ell_k}^\ct\pw_{\ell_k{\ell'}\!\!_{k'}}\big|^2 
   &= \frac{\hat{\pw}_{\ell\ell_k}^\ct\Upsilonm _{-\ell\ell_k}{\pw}_{\ell_k{\ell'}\!\!_{k'}}{\pw}_{\ell_k{\ell'}\!\!_{k'}}^\ct\Upsilonm _{-\ell\ell_k}\hat{\pw}_{\ell\ell_k}}{1+\hat{\pw}_{\ell\ell_k}^\ct\Upsilonm _{-\ell\ell_k}\hat{\pw}_{\ell\ell_k}} 
\end{align}
where $\pw_{\ell_k{\ell'}\!\!_{k'}}$ is given by \eqref{eq:SM-5b}.
Since only the effective channels having the same user (pilot) index $k$ are correlated with each other in our setting, we address the following two cases for the numerator in the RHS of (\ref{eq:MR-21}).

Let us first consider the case of $\ell'\neq\ell$ and $k'=k$, which takes the major portion of pilot contamination and causes a \emph{nonvanishing}\footnote{More precisely, the pilot contamination component is nonvanishing unless ${\alpha_{\ell\ell_k}}\rightarrow 0$.} interference power  in the limit of large $M$. We can have
\begin{align} 
   &\hat{\pw}_{\ell\ell_k}^\ct\Upsilonm _{-\ell\ell_k}{\pw}_{\ell_k{\ell'}\!\!_{k}}  \nonumber \\ 
   &= \bigg(\pw_{\ell\ell_k} +\sum_{j\neq\ell}\pw_{\ell_k{j}_{k}} +{\frac{1}{\sqrt{\rho_p}}}\pz_{\ell_k}\bigg)^\ct 
   \Xim_{\ell\ell_k}\Lambdam_{\ell\ell_k} \Upsilonm _{-\ell\ell_k}\pw_{\ell_k{\ell'_k}}  \nonumber \\
   &\overset{(a)}{\simeq} \pw_{\ell_k{\ell'_k}}^\ct \Xim_{\ell\ell_k}\Lambdam_{\ell\ell_k} \Upsilonm _{-\ell\ell_k}\pw_{\ell_k{\ell'_k}}   \nonumber \\
   &= \pw_{\ell{\ell'}\!\!_{k'}}^\ct \underbrace{\pU_{\ell{\ell'}\!\!_{k'}}^\ct \pU_{\ell\ell_k}\Xim_{\ell\ell_k}\Lambdam_{\ell\ell_k} \Upsilonm _{-\ell\ell_k}\pU_{\ell\ell_k}^\ct\pU_{\ell{\ell'}\!\!_{k'}}}_{\overset{(b)}{\simeq} \ \frac{1}{M}\trace \big(\Xim_{\ell\ell_k}\Lambdam_{\ell\ell_k} \Upsilonm _{-\ell\ell_k}\big) {\hbox{\boldmath$I$}}_{r_{\ell{\ell'}\!\!_{k'}}}} \pw_{\ell{\ell'}\!\!_{k'}}   \label{eq:MR-23b} \\
   &\overset{(c)}{\simeq} \frac{1}{M^2} \trace\Lambdam_{\ell\ell'_k} \trace\;\Xim_{\ell\ell_k}\Lambdam_{\ell\ell_k} \tilde{\Upsilonm} _{\ell_k}  \nonumber \\
   &\overset{(d)}{\simeq} \frac{r_{\ell\ell_k}}{M} \nu_{\ell\ell'_k} \nonumber \\
   &=\; {\alpha_{\ell\ell_k}} \nu_{\ell\ell'_k} \label{eq:MR-23}
\end{align}
where $(a)$ follows from Lemma \ref{lem-2} for $j=\ell'$,  $(b)$ from Lemma \ref{lem-1b}, and $(c)$  from Corollary \ref{corol-a1}. As a matter of fact, we used the almost sure convergence in Lemma \ref{lem-a10} to obtain $(c)$ since \eqref{eq:MR-23b} is a mixture of random vector $\pw_{\ell{\ell'}\!\!_{k'}}$ and random matrix product $\pU_{\ell{\ell'}\!\!_{k'}}^\ct \pU_{\ell\ell_k}$ independent of each other.
For the rest of this work, we use Lemma \ref{lem-a10} in the same kind of mixtures of random variables. 
Noticing that 
$\limsup_{M\rightarrow \infty}\max\big\{\frac{1}{M}\trace\Lambdam_{\ell\ell'_k}, \|\tilde{\Upsilonm} _{\ell_k}\|_2, \frac{1}{r_{\ell\ell_k}}\trace{\Xim_{\ell\ell_k}\Lambdam_{\ell\ell_k}}\big\}<\infty$ by Assumption \ref{as-2} and using the trace inequality \cite{Hor99} that $\trace{\pA\pB}\le \|\pA\|_2\trace{\pB}$ for positive semidefinite matrices $\pA$ and $\pB$ and also the fact that the spectral norm is a submultiplicative norm, we have $\limsup_{M\rightarrow \infty}\nu_{\ell\ell'_k}<\infty$. Similarly, $\liminf_{M\rightarrow \infty}\nu_{\ell\ell'_k}>0$. In $(d)$ we used Theorem \ref{thm-a1}.

For the residual interference term, where $k'\neq k$,  we can rewrite by Corollary \ref{corol-a1} the numerator in the RHS of (\ref{eq:MR-21}) as 
\begin{align} \label{eq:MR-22}
   \hat{\pw}_{\ell\ell_k}^\ct\Upsilonm _{-\ell\ell_k}&{\pw}_{\ell_k{\ell'}\!\!_{k'}}{\pw}_{\ell_k{\ell'}\!\!_{k'}}^\ct\Upsilonm _{-\ell\ell_k}\hat{\pw}_{\ell\ell_k} \nonumber \\
   &\simeq {\frac{1}{r_{\ell\ell_k}M}\trace\; \tilde{\Phim}_{\ell\ell_k}\tilde{\Upsilonm} _{\ell_k}{\pw}_{\ell_k{\ell'}\!\!_{k'}}{\pw}_{\ell_k{\ell'}\!\!_{k'}}^\ct\tilde{\Upsilonm} _{\ell_k}}  \nonumber \\
   &= {\pw}_{\ell_k{\ell'}\!\!_{k'}}^\ct{\Upsilonm} _{\ell_k}{\Phim}_{\ell\ell_k}{\Upsilonm} _{\ell_k}{\pw}_{\ell_k{\ell'}\!\!_{k'}}  .
\end{align}
Let $\Upsilonm _{-{\ell_{kk'}}}=\big(\Upsilonm _{-\ell\ell_k}^{-1} -{\hat{\pw}}_{\ell_k{\ell}_{k'}}{\hat{\pw}}_{\ell_k{\ell}_{k'}}^\ct \big)^{-1}$, then by  \cite[Lem. 2]{Hoy13} we have  
$$\Upsilonm _{-\ell\ell_k} =\Upsilonm _{-{\ell_{kk'}}}-\frac{\Upsilonm _{-{\ell_{kk'}}}\hat{\pw}_{\ell_k{\ell}_{k'}}\hat{\pw}_{\ell_k{\ell}_{k'}}^\ct\Upsilonm _{-{\ell_{kk'}}}}{1+\hat{\pw}_{\ell_k{\ell}_{k'}}^\ct\Upsilonm _{-{\ell_{kk'}}}\hat{\pw}_{\ell_k{\ell}_{k'}}}.$$ 
We can further rewrite (\ref{eq:MR-22}) as
\begin{align} \label{eq:MR-24}
   &{\pw}_{\ell_k{\ell'}\!\!_{k'}}^\ct {\Upsilonm} _{\ell_k}{\Phim}_{\ell\ell_k}{\Upsilonm} _{\ell_k}{\pw}_{\ell_k{\ell'}\!\!_{k'}} 
   = {\pw}_{\ell_k{\ell'}\!\!_{k'}}^\ct{\Upsilonm} _{-{\ell_{kk'}}}{\Phim}_{\ell\ell_k}{\Upsilonm} _{-{\ell_{kk'}}}{\pw}_{\ell_k{\ell'}\!\!_{k'}} \nonumber \\
   &+\frac{|{\hat{\pw}_{\ell_k{\ell}_{k'}}^\ct\Upsilonm _{-{\ell_{kk'}}}{\pw}_{\ell_k{\ell'}\!\!_{k'}}}|^2\;\hat{\pw}_{\ell_k{\ell}_{k'}}^\ct\Upsilonm _{-{\ell_{kk'}}}{\Phim}_{\ell\ell_k}{\Upsilonm} _{-{\ell_{kk'}}}\hat{\pw}_{\ell_k{\ell}_{k'}}}{(1+\hat{\pw}_{\ell_k{\ell}_{k'}}^\ct\Upsilonm _{-{\ell_{kk'}}}\hat{\pw}_{\ell_k{\ell}_{k'}})^2} \nonumber \\
   & -2\Re \left\{\frac{\hat{\pw}_{\ell_k{\ell}_{k'}}^\ct\Upsilonm _{-{\ell_{kk'}}}{\pw}_{\ell_k{\ell'}\!\!_{k'}}{\pw}_{\ell_k{\ell'}\!\!_{k'}}^\ct\Upsilonm _{-{\ell_{kk'}}}{\Phim}_{\ell\ell_k}{\Upsilonm} _{-{\ell_{kk'}}}\hat{\pw}_{\ell_k{\ell}_{k'}}}{1+\hat{\pw}_{\ell_k{\ell}_{k'}}^\ct\Upsilonm _{-{\ell_{kk'}}}\hat{\pw}_{\ell_k{\ell}_{k'}}}\right\}.
\end{align}
We can further have the asymptotic approximations
\begin{align} 
   {\pw}_{\ell_k{\ell'}\!\!_{k'}}^\ct{\Upsilonm} _{-{\ell_{kk'}}}{\Phim}_{\ell\ell_k}{\Upsilonm} _{-{\ell_{kk'}}}{\pw}_{\ell_k{\ell'}\!\!_{k'}} 
   &\overset{(a)}{\simeq} \frac{\trace\Lambdam_{\ell{\ell'}\!\!_{k'}}}{M^2}{\frac{1}{r_{\ell\ell_k}}\trace\; \tilde{\Phim}_{\ell\ell_k}\tilde{\Upsilonm} _{\ell_k}^2}  \nonumber \\
   &= \frac{1}{M}\;\mu_{\ell_k{\ell'}\!\!_{k'}} \nonumber
\end{align}
where  $(a)$ follows from Lemma \ref{lem-1b}, and  
\begin{align} \label{eq:MR-29}
   &{\hat{\pw}_{\ell_k{\ell}_{k'}}^\ct\Upsilonm _{-{\ell_{kk'}}}{\pw}_{\ell_k{\ell'}\!\!_{k'}}} \nonumber \\
   &\simeq   {\pw}_{\ell_{k'}{\ell'}\!\!_{k'}}^\ct\Xim_{{\ell}_{k'}}\Lambdam_{{\ell}_{k'}}\pU_{{\ell}_{k'}}^\ct\pU_{\ell\ell_k} \Upsilonm _{-{\ell_{kk'}}}{\pw}_{\ell_k{\ell'}\!\!_{k'}}
  \nonumber \\
   &\simeq \frac{1}{M} \trace\; \underbrace{\pU_{\ell\ell_k}^\ct\pU_{{\ell'}\!\!_{k'}} \Lambdam_{\ell{\ell'}\!\!_{k'}}\pU_{{\ell'}\!\!_{k'}}^\ct\pU_{{\ell}_{k'}}}_{\xrightarrow[a.s.]{(a)}\ \p0}\Xim_{{\ell}_{k'}}\Lambdam_{{\ell}_{k'}}\underbrace{\pU_{{\ell}_{k'}}^\ct\pU_{\ell\ell_k}}_{\xrightarrow[a.s.]{(b)}\ \p0} \Upsilonm _{-{\ell_{kk'}}}   \nonumber \\ &\simeq  0 
\end{align} 
where $(a)$ and $(b)$ follow from a direct consequence of Lemma \ref{lem-1b} due to the fact that $\pU_{\ell\ell_k}$ and $\pU_{{\ell}_{k'}}$ are mutually independent. It is important to {note that this beneficial effect of interference elimination is enabled by spatial despreading.} Combining the above asymptotic approximations into (\ref{eq:MR-24}) yields 
\begin{align} \label{eq:MR-25}
   {\pw}_{\ell_k{\ell'}\!\!_{k'}}^\ct{\Upsilonm} _{\ell_k}{\Phim}_{\ell\ell_k}{\Upsilonm} _{\ell_k}{\pw}_{\ell_k{\ell'}\!\!_{k'}} 
   &\simeq \frac{1}{M}{\mu_{\ell_k{\ell'}\!\!_{k'}}}.
\end{align}

Plugging (\ref{eq:MR-23}) and (\ref{eq:MR-25}) into (\ref{eq:MR-21}) and noticing that (\ref{eq:MR-21}) is bounded away from infinity, we have by the dominated convergence theorem 
\begin{align} \label{eq:MR-26}
   \sum_{(\ell',k')\neq(\ell,k)}\big|\pv_{\ell_k}^\ct\pw_{\ell_k{\ell'}\!\!_{k'}}\big|^2 
   &\simeq \sum_{\ell'\neq\ell}\left(\frac{r_{\ell\ell_k}}{M}\right)^2 \frac{|\nu_{\ell\ell'_k} |^2}{(1+\delta_{\ell_k})^2}  \nonumber \\
   &\ \ \ +\sum_{\ell'\neq\ell, k'\neq k}\frac1{M}\frac{\mu_{\ell_k{\ell'}\!\!_{k'}}}{(1+\delta_{\ell_k})^2} .
\end{align}

Combining all the results in (\ref{eq:MR-27}) and (\ref{eq:MR-28}) into (\ref{eq:MR-26}), we arrive at (\ref{eq:MR-9b}).

\bibliographystyle{IEEEtran}
\bibliography{bib_nam}

\begin{IEEEbiographynophoto}{Junyoung Nam}  (M'17)
received the B.Sc. in Statistics from Inha University, Incheon, Korea, in 1997 and the M.Sc. and the Ph.D. degrees in Electrical Engineering (Information and Communication) from Korea Advanced Institute of Science and Technology (KAIST), Daejeon, Korea, in 2008 and 2015, respectively.    
He was with the Communications R\&D Center, Samsung Electronics, Seoul, Korea, from 1997 to 2001, and also with the Communications Lab., Seodu InChip, Seoul, from 2001 to 2006. In Summer 2006, he joined Electronics and Telecommunications Research Institute (ETRI), Deajeon, Korea, where he was a principal member of research staff. From January to June 2017, he was with the department of Wireless Communications and Networks, Fraunhofer Heinrich Hertz Institute (HHI), Berlin, Germany. In Summer 2017, he joined Intel Labs, Santa Clara, CA, where he was a senior research scientist. Since March 2019, he has been with Qualcomm, San Jose, CA, where he is currently a senior staff system engineer. 
Dr. Nam was a 3GPP RAN1 delegate with more than 30 patents and 10 standards contributions. His inventions led to a core part of MIMO in 3GPP Release 13/14 and 5G NR. His research interests are in the areas of wireless communications, information theory, advanced cellular system design, and compressed sensing.  
\end{IEEEbiographynophoto} 

\begin{IEEEbiographynophoto}{Giuseppe Caire }  (S'92--M'94--SM'03--F'05) 
was born in Torino in 1965. He received the B.Sc. in Electrical Engineering  from Politecnico di Torino in 1990, 
the M.Sc. in Electrical Engineering from Princeton University in 1992, and the Ph.D. from Politecnico di Torino in 1994. 
He has been a post-doctoral research fellow with the European Space Agency (ESTEC, Noordwijk, The Netherlands) in 1994-1995,
Assistant Professor in Telecommunications at the Politecnico di Torino, Associate Professor at the University of Parma, Italy, 
Professor with the Department of Mobile Communications at the Eurecom Institute,  Sophia-Antipolis, France,
a Professor of Electrical Engineering with the Viterbi School of Engineering, University of Southern California, Los Angeles,
and he is currently an Alexander von Humboldt Professor with the Faculty of Electrical Engineering and Computer Science at the
Technical University of Berlin, Germany.

He received the Jack Neubauer Best System Paper Award from the IEEE Vehicular Technology Society in 2003,  the
IEEE Communications Society \& Information Theory Society Joint Paper Award in 2004 and in 2011, the 
Leonard G. Abraham Prize for best IEEE JSAC paper in 2019,  the Okawa Research Award in 2006,
the Alexander von Humboldt Professorship in 2014, the Vodafone Innovation Prize in 2015, and an ERC Advanced Grant in 2018.
Giuseppe Caire is a Fellow of IEEE since 2005.  He has served in the Board of Governors of the IEEE Information Theory Society from 2004 to 2007,
and as officer from 2008 to 2013. He was President of the IEEE Information Theory Society in 2011. 
His main research interests are in the field of communications theory, information theory, channel and source coding
with particular focus on wireless communications.   
\end{IEEEbiographynophoto}

\begin{IEEEbiographynophoto}{M{\'e}rouane Debbah}  (S’01–M’04–SM’08–F’15) 
received the M.Sc. and Ph.D. degrees from the Ecole Normale SupérieureParis-Saclay, France. He was with Motorola Labs, Saclay, France, from 1999 to2002, and also with the Vienna Research Center for Telecommunications, Vienna,Austria, until 2003. From 2003 to 2007, he was an Assistant Professor with theMobile Communications Department, Institut Eurecom, Sophia Antipolis, France.In 2007, he was appointed Full Professor at CentraleSupelec, Gif-sur-Yvette,France. From 2007 to 2014, he was the Director of the Alcatel-Lucent Chair onFlexible Radio. Since 2014, he has been a Vice-President of the Huawei FranceResearch Center and the Director of the Mathematical and Algorithmic SciencesLab. He has managed 8 EU projects and more than 24 national and internationalprojects. His research interests lie in fundamental mathematics, algorithms,statistics, information, and communication sciences research. He is an IEEEFellow, a WWRF Fellow, and a Membre émérite SEE. 

He was a recipient of the ERCGrant MORE (Advanced Mathematical Tools for Complex Network Engineering) from2012 to 2017. He was a recipient of the Mario Boella Award in 2005, the IEEEGlavieux Prize Award in 2011, the Qualcomm Innovation Prize Award in 2012 andthe 2019 IEEE Radio Communications Committee Technical Recognition Award. Hereceived 20 best paper awards, among which the 2007 IEEE GLOBECOM Best PaperAward, the Wi-Opt 2009 Best Paper Award, the 2010 Newcom++ Best Paper Award,the WUN CogCom Best Paper 2012 and 2013 Award, the 2014 WCNC Best Paper Award,the 2015 ICC Best Paper Award, the 2015 IEEE Communications Society Leonard G.Abraham Prize, the 2015 IEEE Communications Society Fred W. Ellersick Prize,the 2016 IEEE Communications Society Best Tutorial Paper Award, the 2016European Wireless Best Paper Award, the 2017 Eurasip Best Paper Award, the 2018IEEE Marconi Prize Paper Award, the 2019 IEEE Communications Society YoungAuthor Best Paper Award and the Valuetools 2007, Valuetools 2008, CrownCom2009, Valuetools 2012, SAM 2014, and 2017 IEEE Sweden VT-COM-IT Joint Chapterbest student paper awards. He is an Associate Editor-in-Chief of the journalRandom Matrix: Theory and Applications. He was an Associate Area Editor andSenior Area Editor of the IEEE TRANSACTIONS ON SIGNAL PROCESSING from 2011 to2013 and from 2013 to 2014, respectively.
\end{IEEEbiographynophoto}

\begin{IEEEbiographynophoto}{H.Vincent Poor}   (S’72--M’77--SM’82--F’87) 
received the Ph.D. degree in electrical engineering and computerscience from Princeton University in 1977. From 1977 until 1990, he was on the faculty of the University ofIllinois at Urbana-Champaign. Since 1990 he has been on the faculty at Princeton,where he is the Michael Henry Strater University Professor of ElectricalEngineering. During 2006 to 2016, he served as Dean of Princeton’s School ofEngineering and Applied Science. He has also held visiting appointments atseveral other institutions, most recently at Berkeley and Cambridge. Hisresearch interests are in the areas of information theory and signal processing,and their applications in wireless networks, energy systems and related fields.Among his publications in these areas is the recent book Information Theoretic Security and Privacy of Information Systems (CambridgeUniversity Press, 2017). 

Dr. Poor is a member of the National Academy ofEngineering and the National Academy of Sciences, and is a foreign member ofthe Chinese Academy of Sciences, the Royal Society, and other national andinternational academies. Recent recognition of his work includes the 2017 IEEE Alexander Graham BellMedal, the2019 ASEE Benjamin Garver Lamme Award, a D.Sc. honoris causa from Syracuse University, awarded in 2017, and a D.Eng.honoris causa from the University of Waterloo, awarded in 2019.
\end{IEEEbiographynophoto}

\end{document}